\documentclass[fleqn,usenatbib]{mnras}

\usepackage{newtxtext,newtxmath}

\usepackage[T1]{fontenc}
\usepackage{xcolor}
\DeclareRobustCommand{\VAN}[3]{#2}
\let\VANthebibliography\thebibliography
\def\thebibliography{\DeclareRobustCommand{\VAN}[3]{##3}\VANthebibliography}

\usepackage{graphicx}	
\usepackage{amsmath}	
\usepackage{deluxetable}
\usepackage{lineno}

\usepackage{float}
\usepackage{pdflscape}
\usepackage{afterpage}
\usepackage{capt-of}

\usepackage{multirow}
\usepackage{ulem}

\newcommand{\catname}[1]{SCOTCH}

\usepackage{datatool}
\usepackage{etoolbox}

\DTLnewdb{bibnotes}

\makeatletter
\patchcmd{\@lbibitem}%
  {\item[\hfil\NAT@anchor{#2}{\NAT@num}]}%
  {%
    \item[\hfil\NAT@anchor{#2}{\NAT@num}]%
    \DTLforeach[\DTLiseq{\mylabel}{#2}]{bibnotes}{\mylabel=mylabel,\mynote=mynote}{{\mynote}}
  }{}{\message{^^JPatching failed^^J}}%

\title{The Simulated Catalogue of Optical Transients and Correlated Hosts (SCOTCH)}

\author[Lokken, Gagliano et al.]{\href{https://orcid.org/0000-0001-5917-955X}{Martine Lokken} $^{1,2,3}$\thanks{E-mail: m.lokken@mail.utoronto.ca}, \href{https://orcid.org/0000-0003-4906-8447}{Alexander~Gagliano}$^{4,5,6}$\thanks{E-mail: gaglian2@illinois.edu}, \href{https://orcid.org/0000-0001-6022-0484}{Gautham Narayan}$^{4,5}$, \href{https://orcid.org/0000-0002-0965-7864}{Ren\'ee Hlo\v{z}ek}$^{1,3}$, \href{https://orcid.org/0000-0003-3221-0419}{Richard Kessler}$^{7,8}$, \newauthor{\href{https://orcid.org/0000-0002-2495-3514}{John Franklin Crenshaw}$^{9}$, \href{https://orcid.org/0000-0001-5473-6871}{Laura Salo}$^{10}$, \href{https://orcid.org/0000-0002-6164-9044}{Catarina S. Alves}$^{11}$, \href{https://orcid.org/0000-0003-0038-5468}{Deep Chatterjee}$^{4,5}$, \href{https://orcid.org/0000-0001-8788-1688}{Maria Vincenzi}$^{12}$,}
\newauthor{\href{https://orcid.org/0000-0002-8676-1622}{Alex I. Malz}$^{13}$, The LSST Dark Energy Science Collaboration}
\\
$^{1}$David A. Dunlap Department of Astronomy and Astrophysics, University of Toronto, 50 St. George Street, Toronto, Ontario, M5S 3H4 Canada\\
$^{2}$Canadian Institute for Theoretical Astrophysics, University of Toronto, 60 St. George St., Toronto, ON M5S 3H4, Canada\\
$^{3}$Dunlap Institute of Astronomy \& Astrophysics, 50 St. George St., Toronto, ON M5S 3H4, Canada\\
$^{4}$Department of Astronomy, University of Illinois at Urbana-Champaign, 1002 W. Green St., IL 61801, USA\\
$^{5}$Center for Astrophysical Surveys, National Center for Supercomputing Applications, Urbana, IL, 61801, USA\\
$^{6}$National Science Foundation Graduate Research Fellow\\
$^{7}$Department of Astronomy and Astrophysics, University of Chicago, Chicago, IL 60637, USA\\
$^{8}$Kavli Institute for Cosmological Physics, University of Chicago, Chicago, IL 60637, USA\\
$^{9}$DIRAC Institute and Department of Physics, University of Washington, Seattle, WA 98195, USA\\
$^{10}$School of Physics and Astronomy, University of Minnesota, 116 Church Street S.E., Minneapolis, MN 55455, USA\\
$^{11}$Department of Physics \& Astronomy, University College London, Gower Street, London WC1E 6BT, UK\\
$^{12}$Department of Physics, Duke University Durham, NC 27708, USA\\
$^{13}$McWilliams Center for Cosmology, Department of Physics, Carnegie Mellon University}

\graphicspath{{./}{figures/}}
\date{Accepted XXX. Received YYY; in original form ZZZ}

\pubyear{2022}

\begin{document}
\label{firstpage}
\pagerange{\pageref{firstpage}--\pageref{lastpage}}
\maketitle

\begin{abstract}
As we observe a rapidly growing number of astrophysical transients, we learn more about the diverse host galaxy environments in which they occur. Host galaxy information can be used to purify samples of cosmological Type~Ia supernovae, uncover the progenitor systems of individual classes, and facilitate low-latency follow-up of rare and peculiar explosions. In this work, we develop a novel data-driven methodology to simulate the time-domain sky that includes detailed modeling of the probability density function for multiple transient classes conditioned on host galaxy magnitudes, colours, star formation rates, and masses. We have designed these simulations to optimize photometric classification and analysis in upcoming large synoptic surveys. We integrate host galaxy information into the \texttt{SNANA} simulation framework to construct the Simulated Catalogue of Optical Transients and Correlated Hosts (\catname\;), a publicly-available catalogue of 5 million idealized transient light curves \textcolor{black}{in LSST passbands} and their host galaxy properties over the redshift range $0<z<3$. This catalogue includes supernovae, tidal disruption events, kilonovae, and active galactic nuclei. Each light curve consists of true top-of-the-galaxy magnitudes \textcolor{black}{sampled with high ($\lesssim$2 day) cadence}. In conjunction with \catname\;, we also release an associated set of tutorials and transient-specific libraries to enable simulations of arbitrary space- and ground-based surveys. Our methodology is being used to test critical science infrastructure in advance of surveys by the Vera C. Rubin Observatory and the Nancy G. Roman Space Telescope. 

\end{abstract}



\section{Introduction} \label{sec:intro}

An era of rapid and large-scale astronomical data collection is underway. Surveys are \textcolor{black}{now} detecting transient events across large swaths of the sky, e.g. the Zwicky Transient Facility \cite[ZTF;][]{ZTF2019}, ASAS-SN \citep{Shappee2014}, Gaia \citep{Gaia2016}, and MeerKAT \citep{Jonas}. 
\textcolor{black}{Entirely novel classes of events,} including those at higher redshifts, \textcolor{black}{will soon be revealed} as telescopes such as the Vera C. Rubin Observatory \citep{LSST_Science_book2019}, the Nancy Grace Roman Space Telescope \citep{Spergel2015}, and the James Webb Space Telescope \citep{Gardner2006} begin collecting data. Notable among these, the Rubin Observatory (set to begin science operations in Chile in 2024) will undertake the Legacy Survey of Space and Time (LSST), expected to observe millions of transients out to and beyond $z\sim1$ as it repeatedly scans 18,000 square degrees of the sky over ten years. 

The science goals of these large-scale transient surveys are manifold, but include measuring the expansion history of the universe, improving our understanding of progenitor physics, and discovering faster and fainter transients that have eluded prior searches. Accurate classification will be paramount for each of these goals. In the majority of cases, transients discovered in upcoming surveys must be classified without spectroscopic information. Algorithms that classify transients based on photometry alone have proliferated in recent years, some based on traditional template-matching methods and others employing novel machine learning algorithms \citep[e.g.,][]{Villar2020,moller2020, Qu2021AJ, Alves2022}. 

Multiple studies have demonstrated correlations between the class of a transient and the properties of the galaxy where it occurs (deemed its ``host galaxy''). Where light curve sampling is sparse, host galaxy data can be used to more easily distinguish transients of different classes.
The \textcolor{black}{occurrence rate} of Type~Ia supernovae (SNe~Ia) is directly proportional to both the stellar mass and the star formation rate of its host galaxy, although a non-negligible subset has also been discovered in passive galaxies \citep{Sullivan2006}. As the end-states of short-lived ($\sim50$ Myr) massive stars, Core-Collapse SNe (CCSNe) are most likely to occur in spiral galaxies undergoing periods of significant star formation \citep{2010Svensson_CCSNe}. Variations also arise in the host populations of stripped-envelope SNe, events characterized by a dearth of hydrogen (SNe~Ib) and helium (SNe~Ic) in spectral observations. Broad-lined SNe~Ic (SNe~Ic-BL), the only supernovae to be unambiguously associated with long-duration Gamma Ray Bursts \citep[LGRBs;][]{2016Japelj_LGRBs}, have host galaxies with lower average metallicity than their lower-energy counterparts \citep[SNe~Ic;][]{2020Modjaz_IcBL}. This suggests a potential relationship between progenitor metallicity and the formation of jets in the tumult of stellar death \citep{2013Graham_MetalAversion}. Further, SED fits to multi-band photometry of Type-I (Hydrogen-poor) Super-Luminous SN (SLSN-I) hosts have revealed that these events occur almost exclusively in metal-poor, low-mass, star-forming host galaxies \citep{2015Leloudas_SLSN, 2016Perley_SLSNe, 2016Angus_HSTSLSNe}. \textcolor{black}{In the pursuit of optimally-standardized light curves for cosmology, subtler correlations have also been identified between SN~Ia photometry and both host galaxy mass and star formation rate \citep{2010Lampeitl_SNIa, 2010Sullivan, 2019Barkhudaryan_91bglike, 2019Rose, 2021Brout}.}


At low redshifts, spatially resolved studies enabled by multi-wavelength photometry and Integral Field Unit spectroscopy have revealed local-scale correlations. From a sample of 519 host galaxies selected in the Sloan Digital Sky Survey \citep[SDSS;][]{2017Blanton_SDSS}, \citet{2012Kelly_CCSNe} found that SNe~Ic-BL and SNe~IIb occurred in extremely blue locations. Local correlations are weak for SNe~Ia \citep{2015Anderson_SNIa}: due to the advanced ages of their white-dwarf progenitors ($\sim$Gyrs), these systems can migrate substantially from formation to explosion. Nevertheless, the environment may alter the evolution of an event for the same progenitor: \textcolor{black}{the observational properties of SNe~Ia are correlated with local dust extinction, although this is itself linked to the global properties of the host galaxy \citep{2015Holwerda_SNIaDust,2021Popovic_Dust}}. Spatially-resolved host galaxy information can also be used to distinguish between SN explosion models in the absence of direct detections of a progenitor \citep{Raskin2008, Fruchter2006}.

Preliminary correlations have been identified in recent years for rare classes of events, including a preference of Rapidly-Evolving Transients \citep[RETs;][]{2014Drout_RETs, pursiainen2018} for hosts of intermediate mass between high-mass SN~Ia hosts and low-mass SLSN hosts \citep{2020Wiseman_DESHosts}. For observationally rare events \citep[e.g., kilonovae (KNe);][]{2017Smartt_KNe}, correlations remain as yet unconstrained, but may play a decisive role in maximizing the yield of future targeted searches. Upcoming surveys are well-poised to further extend our understanding of these class-specific correlations toward higher redshifts and rare subtypes.  However, only a few transient classifiers to date \citep[][]{FoleyMandel2013, Gagliano2021,2023Kisley} have incorporated host galaxy information. Efforts are stymied by the limited observational data available for training these algorithms; perhaps worse, extant samples are frequently biased toward low-redshift, bright host galaxies, uncharacteristic of the galaxies anticipated from upcoming surveys. 




In this transitional period when observed transient samples are limited to the local universe, simulations are vital for developing and testing classification algorithms that incorporate host galaxy information. Such algorithms will need to perform reasonably well on new faint and high-redshift data in the early days of LSST, until they can be iteratively tested and updated using the incoming data and the corresponding advancements in time-domain theory. This interim period requires high-redshift simulations that extrapolate bright, low-redshift existing measurements (such as transient rates and host correlations) to fainter magnitudes and into the universe's past. High-redshift simulations necessarily require assuming untested theoretical models for evolution, or assuming that trends at low redshift extend to high redshift.

We follow the latter assumption, and present the first large-scale simulation of optical transients \textit{and their host galaxies} that extrapolates existing data to high redshifts ($z\sim3$). This work extends the simulations generated for the Photometric LSST Astronomical Time-Series Classification Challenge \citep[PLAsTiCC;][]{Plasticcv12019}. 
PLAsTiCC data were generated by simulation code in the SuperNova ANAlysis package \citep[\texttt{SNANA}]{Kessler2009}, and the challenge represented the first large-scale effort to realistically simulate a broad diversity of transient phenomena. Simulated host-galaxy information, however, was limited: the dataset only includes the photometric redshifts of host galaxies assigned near the redshift of the transient. This work extends the PLAsTiCC framework with the inclusion of empirical and theoretical host-galaxy correlations for both common and rare transient classes. We embed these correlations using a series of host galaxy libraries (called \texttt{HOSTLIB}s within \texttt{SNANA}) and weight maps (\texttt{WGTMAP}s) \textcolor{black}{that describe the likelihood of a given galaxy to host a transient of a certain class within the simulation}.

\textcolor{black}{Our work} \textcolor{black}{is enabled by multiple previous efforts to consolidate the properties of transient host galaxy populations. The Gamma Ray Burst (GRB) Host Studies catalogue\footnote{\url{http://www.grbhosts.org/}} \citep[GHOstS;][]{2006GHOSTs} consolidates derived host galaxy information (including star formation rate, stellar mass, and redshift) for $\sim$100 GRBs, finding that these bursts occur preferentially in star-forming, low-mass galaxies. \cite{Gagliano2021} constructed the Galaxies Hosting Supernovae and other Transients (GHOST) Catalogue, which contains the optical photometric properties of $\sim16,500$ galaxies that have hosted observed SNe of all classes. At the time of writing, \cite{2022Qin_THEx} has released the Transient Host Exchange (THEx), a catalogue containing $>36,000$ host galaxies of many transient types and the photometric properties of their hosts spanning infrared to ultraviolet wavelengths.} 

There is precedent to the use of host galaxy libraries in supernova simulations. In \texttt{SNANA}, \texttt{HOSTLIB}s were first used to 
model Poisson noise in simulated bias corrections for SN~Ia distances for the Joint Lightcurve Analysis \citep{2013Kessler_JLA, 2014Betoule_JLA}. \texttt{HOSTLIB}s were similarly used to forecast constraints for the Nancy Grace Roman Space Telescope \citep{2018Hounsell_WFIRST}. The Pantheon \citep{2018Scolnic_Pantheon} and DES 3-year \citep{2019Abbott_DES, 2019Brout_Cosmology} SN~Ia-cosmology analyses further extended the scope of these files to account for correlations between SN properties and host galaxy stellar mass. An alternative machine learning-based simulation, \texttt{EmpiriciSN} \citep{Holoien2017}, employed a library of photometric host galaxy data in SDSS to generate synthetic light curves of SNe~Ia.

As part of the recent Dark Energy Survey (DES) cosmology analysis using photometrically-identified SNe~Ia, \citet{Vincenzi2021} expanded the \texttt{SNANA}-specific \texttt{HOSTLIB} framework to realistically simulate SN host galaxies by class using a set of physically-motivated \texttt{WGTMAP} files. To account for contamination from non-Ia classes, their simulation includes CCSNe (including SNe~II and Ib/c) and considers the dependence on host galaxy mass and star formation rate for each SN type. The resulting SN~Ia+CC~SN simulation of host properties and SN-host correlations shows excellent agreement with DES data, and this simulation is now in use as a training set for photometric classifiers (M\"oller et al. 2023, \textit{in prep.}).
%

In this work, we extend the results of \citet{Vincenzi2021} by simulating additional classes of supernovae and other non-SN transients, incorporating correlations across a wider range of host galaxy properties, and separately validating hosts of different classes across these properties. \textcolor{black}{We extrapolate information from archival events, mostly at low-redshift, to a high-redshift simulated sample of galaxies via associations in intrinsic --- rather than observed --- properties. In doing so, we encode correlations for each transient class to significantly higher redshifts ($z\sim3$) than the ranges spanned by current observations.} \textcolor{black}{Due to a lack of high-$z$ evidence, the most practical option is to assume no redshift evolution of the observed transient-host relationships in the local universe and smoothly extrapolate transient rate models measured at low redshifts. Therefore, the simulation is likely to become increasingly inaccurate at higher redshifts, and  \catname~\footnote{\url{https://zenodo.org/record/7563623\#.Y8_2my9h2yA}} by itself should not be used for precision cosmological forecasts or high-redshift transient analyses.} However, the catalogue can be used to test classifier algorithms and alert broker systems, examine the effects of contamination on cosmological constraints, and model selection biases in current and upcoming transient/host galaxy surveys. 

Our paper is organized as follows. We outline the overall methodology in \textsection \ref{sec:methodology_overview}. We introduce the archival catalogues used to construct our host galaxy sample in \textsection \ref{sec:data}. We describe
the \texttt{SNANA} simulation code that we use to generate our transient events and summarize the models underlying each event in \textsection \ref{sec:SNANA}. 
We present our methods in \textsection\ref{sec:methods} for combining observed and simulated catalogues into a library containing millions of host galaxies. We then describe our process for incorporating additional class-specific transient correlations using \texttt{WGTMAP}s in \textsection\ref{sec:host_correlations}. \textcolor{black}{We comment briefly on the inclusion of class-specific radial offsets in \textsection\ref{sec:offsets}.} We validate our results in \textsection \ref{sec:validation} and present the schema of our final catalogue in \textsection \ref{sec:cat_access}. We then conclude with a discussion of the limitations of our sample and areas for future research in \textsection \ref{sec:conclusions}.

\section{Methodology Overview}\label{sec:methodology_overview}
\begin{figure*}
    \centering
    \includegraphics[width=\linewidth]{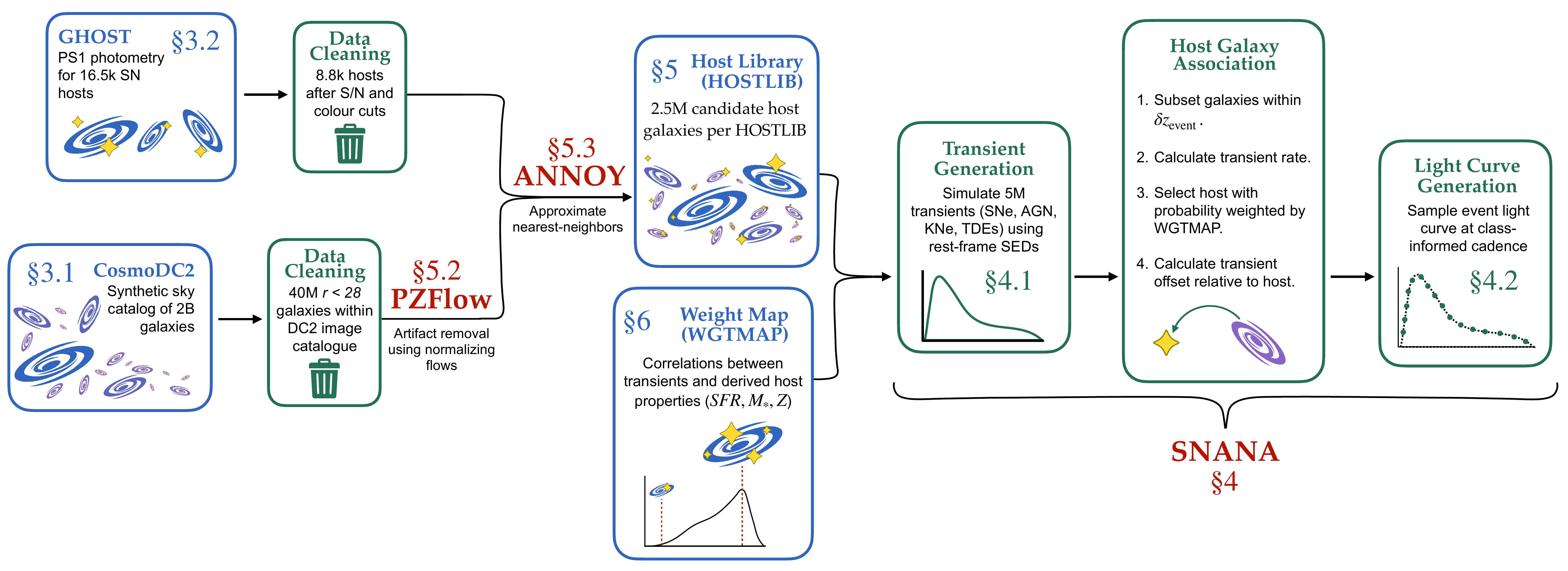}
    \caption{
    \textcolor{black}{Analysis flowchart with section labels. Intermediate catalogues and data products are given in blue text, major software packages in red text, and methodology in green text.}
    }
    \label{fig:flowchart}
\end{figure*}

 Because the methods we use to simulate the transients themselves \textcolor{black}{are nearly identical to those described in \citet{Plasticcv12019}, this paper focuses on our methodology for constructing a set of \texttt{HOSTLIBs} and \texttt{WGTMAPs} for the \texttt{SNANA} simulation code. This allows us to realistically associate a simulated host galaxy with each transient, given its class}. We provide a flowchart of this pipeline in Fig.~\ref{fig:flowchart}. The initial input is the CosmoDC2 synthetic galaxy catalogue \citep[][described in \textsection \ref{subsec:CosmoDC2}]{CosmoDC22019} with slight modifications \textcolor{black}{made via the \texttt{PZFlow} algorithm} \citep[][described in \textsection \ref{subsec:pzflow}]{PZFlow_JFC}. First, we create four large galaxy libraries: \textcolor{black}{for SNe,} by selecting synthetic galaxies matching the \textcolor{black}{rest-frame} colour and \textcolor{black}{absolute} magnitude distributions of three broad categories of transient hosts in observational data; \textcolor{black}{and for non-SNe, by generating a generic catalogue of galaxies which will be tailored to match observations at a later stage.} \textcolor{black}{For SNe}, we utilize the observational data in GHOST \citep{Gagliano2021}, a catalogue of supernova host galaxy properties described in additional detail in \textsection \ref{subsec:ghost}. Our process for selecting synthetic galaxies matching the statistical properties of the observational data, including the use of the \texttt{ANNOY}\footnote{\url{https://github.com/spotify/annoy}} algorithm, is described in detail in \textsection \ref{subsec:CosmoDC2_subsample}. The resulting \textcolor{black}{three} galaxy libraries include: Type~Ia hosts (\textbf{I}); Type~II, IIn, IIp hosts (\textbf{II}); and Type~Ib, Ic, IIb hosts (\textbf{III}). We construct a fourth library of randomly-selected CosmoDC2 galaxies to serve as the hosts of transients from classes that have little observational data, and for which the correlation between explosion and environment is unconstrained by GHOST. Next, for nine different transient classes \textcolor{black}{including SN and non-SN}, we encode the probability that an event of that class will occur in a host galaxy given its mass and star formation rate. These probabilities \textcolor{black}{are motivated by observations} and are described in \textsection \ref{sec:host_correlations}. Finally, the \texttt{SNANA} simulation (\textsection \ref{sec:SNANA}) is used to generate transient light curves and place each in a realistically-associated host galaxy, with host galaxies selected from one of the four libraries depending on the transient class and the probability of selection given by the class-specific functions described in \textsection \ref{sec:host_correlations}. 

\section{Galaxy Data}\label{sec:data}
\subsection{The DC2 Simulations}\label{subsec:CosmoDC2}
Our work aims to simulate transient-host associations to higher redshifts and fainter magnitudes than have been previously observed. We also wish to produce a catalogue containing valuable derived properties of transient hosts, including morphology, star formation rates and stellar masses. Observed galaxy catalogues are limited in these respects, so we make use of synthetic galaxy data from the LSST Dark Energy Science Collaboration (DESC).

DESC is in the midst of a series of data challenges in preparation for the Rubin Observatory. The goal of the data challenges is to produce realistic end-to-end simulations of LSST survey operations, data reduction, and analysis. For data challenge 2 (DC2), DESC produced a suite of simulated LSST-like datasets, including an extragalactic source catalogue and co-added images \citep{DESC2021}. We make use of the extragalactic catalogue, named the CosmoDC2 Synthetic Sky catalogue \citep{CosmoDC22019}, and its associated DC2 image data.

The CosmoDC2 catalogue was constructed from a dark matter only N-body simulation in a 4.225 Gpc$^3$ volume. \textcolor{black}{This volume is insufficient to probe to $z\sim3$ as needed for LSST, so to create lightcones, the volume was replicated once per axis. After the lightcone generation,} realistic galaxies with Sersic profiles and SEDs were associated with dark matter halos. The resultant galaxy catalogue, which is delivered in HEALPix\footnote{\url{http://healpix.sourceforge.net}} (Hierarchical Equal Area iso-Latitude Pixelization) format \citep{Gorski2005}, spans 440 deg$^2$ of the sky and cosmological distances ($z\simeq3$). The CosmoDC2 data has also undergone rigorous validation to ensure that it matches observed galaxy properties \citep{dc2val}. However, due to necessary compromises made to accommodate the needs of the diverse DESC working groups, the CosmoDC2 distributions of galaxy colours are far from perfect, particularly at higher redshifts ($z>0.5$). We discuss how these colour limitations affect the final SCOTCH catalogue in \textsection \ref{sec:validation}.

\begin{figure}
    \centering
    \includegraphics[width=\columnwidth]{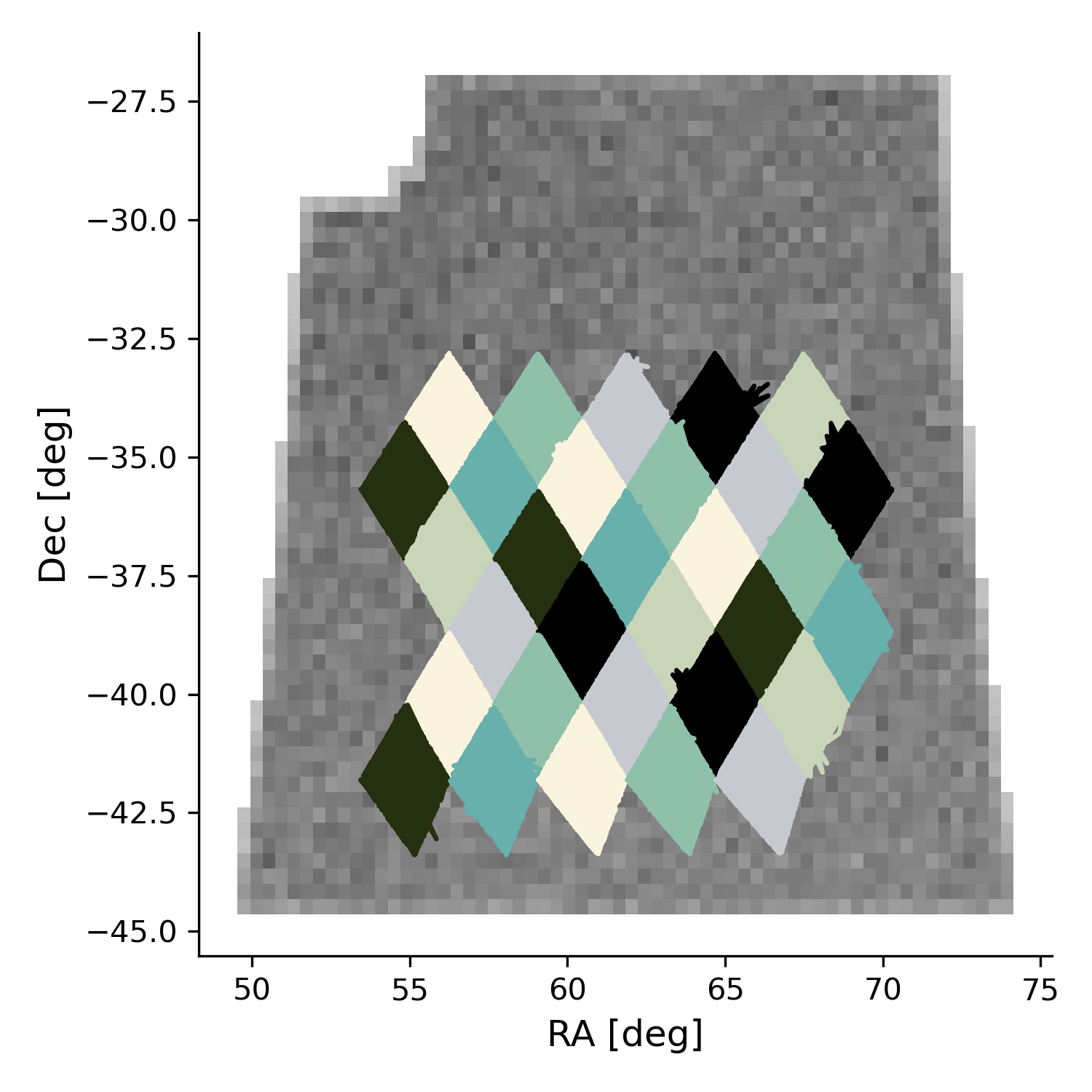}
    \caption{Selected HEALPixels from CosmoDC2 (diamonds) overlaid on the grayscale footprint of the DC2 image catalogues. We cross-match CosmoDC2 and the DC2 image catalogues for the image properties of extragalactic objects.}
    \label{fig:healpixels_objcat}
\end{figure}

We use the \texttt{GCRCatalogs}\footnote{\url{https://github.com/LSSTDESC/gcr-catalogs}} analysis tool \citep{2018Mao} to select CosmoDC2 galaxies within 31 neighbouring HEALPixels \textcolor{black}{from an $N_{\mathrm{side}}=32$ pixelisation, corresponding to 104 square degrees of total area.} The selected HEALPixels are chosen to completely overlap with the DC2 image catalogues \citep{DC2skycatalogs2021}, which contain image properties for CosmoDC2 galaxies as LSST would observe them over 5 years, corresponding to a co-add depth of ${r=28}$ (where $r$ is the apparent AB magnitude in the LSST filter). The overlap in coverage between these HEALPixels and the image catalogue is shown in Fig.~\ref{fig:healpixels_objcat}. To ensure the generality of our final catalogue, the sky coverage displayed in the figure is not maintained in SCOTCH (we discuss this in additional detail in \textsection\ref{subsec:SNANA_descript}).

Within these HEALPixels, we select CosmoDC2 galaxies with $r<28$, which removes the majority of galaxies that were not detected within the DC2 5-year image co-add catalogues. We further refine this sample by removing objects with unreported image properties. \textcolor{black}{After this cut, $\sim98\%$ of galaxies have an average $r$-band surface brightness within the half-light radius of less than 26 mag/arcsec$^2$.} This selection yields a sample of $\sim40$ million galaxies, which represent a realistic magnitude-limited galaxy sample.  In Section~\ref{subsec:CosmoDC2_subsample}, we describe the process of selecting galaxies that match observed properties of transient hosts; this further reduces the sample to $\sim10$ million galaxies.

\subsection{The GHOST Catalogue of Transient-Host Galaxies}\label{subsec:ghost}
The GHOST catalogue \citep{Gagliano2021} consists of $\sim16,500$ archival SNe consolidated from the Transient Name Server (TNS) and the Open SN Catalogue (OSC), as well as the observed properties of their host-galaxies from the first data release of Pan-STARRS (PS1; \citealt{chambers2016pan}). Host galaxies within the catalogue were identified and validated through a combination of the established Directional-Light Radius method \citep[DLR;][]{gupta2016host} and the Gradient Ascent method \citep{Gagliano2021}. Although this GHOST sample contains realistic statistical correlations between transients and their host galaxies, it does not contain a representative sample of events anticipated from upcoming surveys. First, the catalogue comprises a small number of low-redshift ($z<0.2$) explosions. LSST will dramatically enhance our current discovery rate for luminous SNe, motivating the need to scale the sample sizes in the GHOST catalogue for every class.  
Further, GHOST does not include non-SN transients such as Tidal Disruption Events (TDEs) and Active Galactic Nuclei (AGN), classes whose discovery and characterization are key science drivers for LSST. 
Finally, because the catalogue contains only the photometric properties of SN host galaxies, it lacks an explicit correlation between a transient and the stellar-mass and star formation rate of its host galaxy. 

The CosmoDC2 catalogue provides photometry in LSST ($ugrizY$) and SDSS ($griz$) filters, while GHOST galaxy photometry is provided only in the PS1 ($grizy$) filter system. To accurately match galaxy photometry between these catalogues, we convert the GHOST photometry from the PS1 to the SDSS system. We use the approximate quadratic conversions defined in \cite{2012Tonry_PS1System} and k-correct the apparent magnitude in each band using the analytic approximations\footnote{\url{http://kcor.sai.msu.ru/getthecode/\#python}} described in \citet{2010Chilingarian_kCorr} and \citet{2012Chilingarian_Colorcolor}\textcolor{black}{, which were computed according to the SEDs of simple stellar populations ranging from 25 Myr to 16.5 Gyr and $-2.5 < [Fe/H] < +1.0$ dex fit to 170,533 galaxies within SDSS Data Release 7}. 
This transformation allows us to accurately match GHOST galaxies to those in CosmoDC2. 
Next, we use the spectroscopic redshift of each galaxy's associated transient (or the spectroscopic redshift of the host, if the former was unreported on TNS) to calculate the absolute rest-frame brightness of each galaxy in SDSS pass-bands. 
We use these to compute rest-frame $g-r$, $r-i$, and $i-z$ colours. 
To ensure that only galaxies with high-SNR photometry are considered, we apply stringent quality cuts to the galaxies in the GHOST \textcolor{black}{sample based on the values of the PS1 \texttt{primaryDetection} and \texttt{objInfoFlag} flags, as well as the \texttt{InfoFlag1}, \texttt{InfoFlag2}, and \texttt{InfoFlag3} values in each band.\footnote{A full list of the imposed quality cuts can be found at the GitHub repository for this work.}. While this may limit our observed sample to only low-redshift events, it ensures that the photometric correlations we encode from this regime are accurate.} Our final sample after these cuts consists of \textcolor{black}{8,800} SN host galaxies.

\section{Generating Transients with the \texttt{SNANA} Simulation}\label{sec:SNANA}

\subsection{The SNANA Framework for Generating Transients}\label{subsec:SNANA_descript}
\textcolor{black}{Our methodology is shaped by the existing framework of the \texttt{SNANA} simulation code \citep{Kessler2009}. Therefore, although the use of \texttt{SNANA} constitutes the final stage of our pipeline as displayed in Fig.~\ref{fig:flowchart}, we first describe the \texttt{SNANA} simulation and introduce the simulation-specific terminology before detailing our methods for host galaxy association.} This simulation reads user-defined transient model SEDs (rest-frame), redshift-dependent rates, and realistic observing conditions, 
and outputs transient events with realistic light curves and spatial distributions as a function of redshift. The \texttt{SNANA} simulation was used to generate nine distinct classes of extragalactic transients for PLAsTiCC; Figure~1 from \citet{Plasticcv12019} shows example output light curves in \textit{ugrizY} bands from each model. In this paper, we distinguish SNe~Ib from SNe~Ic and SNe~IIn from SNe~II to simulate 11 classes: 
AGN, KN, TDE, SLSN-I, SN~Ib, SN~Ic, SN~II, SN~IIn, SN~Iax, SN~Ia-91bg, and SN~Ia, with the addition of SN~IIb and SN~Ic-BL and with some updates to the PLAsTiCC models as described in \textsection \ref{sec:transient_models}. \textcolor{black}{We implement a cosmology of $\Omega_\mathrm{M}=0.3$, $\Omega_\Lambda=0.7$, $w_0=-1$, and $H_0=70$ km s$^{-1}$Mpc$^{-1}$. We note that, while these values are slightly different from the \textit{Planck} \citep{Planck2020A&A...641A...6P} parameters used in the Outer Rim simulation to generate the CosmoDC2 catalogue, the small inconsistencies that this difference introduces are far sub-dominant to the shifts that we apply to galaxy redshifts in \textsection\ref{subsec:pzflow}.}
The transient models differ in the distributions of emission across bands, the shape of the light curves, and the peak brightness. For example, most of the SN~Ib/c model flux is emitted at redder wavelengths; the KN model exhibits a characteristic timescale shorter than most other transients in our sample; and simulated SLSNe-I emit significantly higher flux than the other models.

The \texttt{SNANA} simulation includes the option of associating each simulated transient to a host galaxy from a user-input catalogue (the \textit{host library} or `\texttt{HOSTLIB}') that includes redshift and other galaxy properties (\textsection \ref{sec:methods}). After each transient is simulated, the algorithm selects a \texttt{HOSTLIB} galaxy that lies within $\delta z$ of the transient. We set $\delta z$ to correspond \textcolor{black}{approximately} to 100 comoving Mpc, which ranges from 0.02 at $z\sim0$ to 0.1 at $z=3$ given the 2018 cosmological parameters from the Planck satellite \citep{Planck2020A&A...641A...6P}.
Host galaxy selection is further influenced by a user-input \textit{weight map} (\texttt{WGTMAP}) 
that defines the probability of selecting a galaxy as a function of its properties (\textsection \ref{sec:host_correlations}). \textcolor{black}{The \texttt{WGTMAP} framework is flexible, such that the probability can depend on different galaxy properties for different classes---in other words, the parameter space over which the \texttt{WGTMAP} probabilities are defined is not fixed but rather user-defined.}

\begin{table}
    \centering
    \begin{tabular}{c|c|c|c}
      \hline 
        HOSTLIB & WGTMAP & Class & Cadence (Days)\\
    \hline
    \hline
        SN~Ia Hosts & $P_{\textrm{Ia}}$ & SN~Ia & 2.0\\
    \hline
        SN~II Hosts & $P_{\textrm{II}}$ & SN~II/IIn/IIP/IIL & 2.0 \\
        \hline
         & $P_{\textrm{Ibc}}$ & SN~Ib/Ic & 2.0 \\
        H-Poor SN Hosts & $R_{\textrm{Ic-BL}}$ & SN~Ic-BL & 2.0 \\
        & $P_{\textrm{SLSNI}}$ & SLSN-I & 2.0 \\
        & $P_{\textrm{II}}$ & SN~IIb  & 2.0\\
        \hline
         & $P_{\textrm{AGN}}$ & AGN & 0.1, 2.0 \\
        Rand. DC2 Subset & $R_{\textrm{KN}}$ & KN & 0.1--2.0 \\
        & $P_{\textrm{TDE}}$ & TDE & 0.1--2.0\\
    \hline
    \end{tabular}
    \caption{Summary of the {\tt HOSTLIB}, {\tt WGTMAP}, \textcolor{black}{and cadence used for each transient class in our simulation. The first three {\tt HOSTLIB}s contain CosmoDC2 galaxies that extrapolate distributions of observed hosts in GHOST; the fourth is a representative sample from CosmoDC2. $P_{\mathrm{class}}$ refers to the \textcolor{black}{probability} of each transient class occurring given the properties of a galaxy; these functions are defined in \textsection \ref{sec:host_correlations} and implemented via the {\tt WGTMAP}s. The cadence refers to the time interval between light curve samples in the simulation.}}
    \label{tbl:hostlib_wgtmap_summary}
\end{table}

Following Fig.~\ref{fig:flowchart}, the simulation begins by generating a 
transient SED and \textcolor{black}{drawing a redshift according to the class-specific rate model. The rate models are detailed in \textsection~\ref{sec:transient_models} and are entirely responsible for the redshift distributions of events in the final catalogue. This may be counterintuitive, as one might expect that the features of the synthetic host galaxy population 
should determine the transient event rates; however, myriad uncertainties in the transient-host connections make such an approach far more challenging to match observed event rates at low redshifts.} 

Next, the transient is matched to a host galaxy. \textcolor{black}{The surface brightness profile of each galaxy in cosmoDC2 is a weighted combination of two Sersic profiles: that of an exponential disk ($n_0$), and that of a deVaucouleurs bulge ($n_1$). For each galaxy associated with a supernova, one of these profiles is selected using their relative weights. Next, a radius and azimuthal angle are selected (where the radius selection is weighted by the Sersic profile), and this determines the position of the transient with respect to the host centre. We further encode simple models for the radial offsets of TDEs, AGN, and KNe as described in \textsection~\ref{sec:offsets}.}

\textcolor{black}{After the transient offset is computed, the} host galaxy redshift is set to the redshift of the transient and the host's apparent magnitudes are updated to reflect the small change in distance (100 comoving Mpc or less). With this framework, one host galaxy from the \texttt{HOSTLIB} can be used multiple times, \textcolor{black}{which is often necessary given the desired number of transients (5M) that we aim to simulate, the similar size of the \texttt{HOSTLIB}s, and the different redshift distributions for the transient model and the \texttt{HOSTLIB}s. In such a case, the single input galaxy} is mapped to multiple host galaxies with slightly different redshifts and apparent magnitudes in the final catalogue. \textcolor{black}{Due to this duplication, it is necessary to remove on-sky position information from the galaxies such that duplicated galaxies do not appear at the same sky location. Unfortunately, this loses the initial positioning within realistic large-scale structure from CosmoDC2. We leave an accurate embedding of transients within the cosmic web to future work. In our simulation, the host centres are arbitrarily set to be simulated at (RA,Dec)=(0,0) and positions are not reported in the final catalogue.}

The transient SED is modified for cosmic expansion, \textcolor{black}{weak lensing magnification}, redshift, \textcolor{black}{and host galaxy extinction. The weak lensing model aims to statistically match the average magnification effects as a function of redshift; \citet{Kessler2019} describes the model in further detail.} \textcolor{black}{We do not explicitly model host extinction from each galaxy's morphology; rather, we follow the statistical treatment in PLAsTiCC \citep{Plasticcv12019}. For the theoretical models (MOSFIT and KNe), extinction is parameterized by $A_V$. For most models, this parameter is drawn from a Galactic Line-of-Sight distribution with a Gaussian core of $\sigma=0.1$ and an exponential tail parameterized by $\tau=0.4$ \citep{2008Sako_SDSSII,2012Berstein_SNsims}. Dimming is then computed assuming the Milky Way color law from \cite{1999Fitzpatrick_Extinction}. For TDEs, only the exponential is included in order to simulate significantly higher extinction near the host galaxy centres where these transients occur. For KNe experiencing less host extinction at large offsets (see \textsection~\ref{sec:offsets}), the exponential tail is reduced by a factor of two. The other transient models have been constructed from observed photometry, which implicitly includes the effects of host-galaxy extinction. This formalism assumes no evolution in dust properties with redshift.}

\textcolor{black}{The final SED corresponds to what an observation would look like from a vantage point just outside the Milky Way, without Galactic extinction, because we have not simulated transient locations on the sky. This leaves flexibility for the user to place events within the sky footprint of a preferred survey and apply the corresponding line-of-sight extinction and noise models.} Next, the SED is converted to an apparent magnitude in the LSST photometric system ($ugrizY$). \textcolor{black}{We configure \texttt{SNANA} to report all light curves in true LSST magnitudes without applying any threshold for detection. Therefore, very faint magnitudes are included in the transient light curves. Because hosts \textit{are} restricted to $r<28$, catalogue users are encouraged to apply a limiting magnitude equal to or brighter than $r=28$ to the full catalogue to create consistency before use.}

\begin{figure*}
    \centering
    \includegraphics[width=\linewidth]{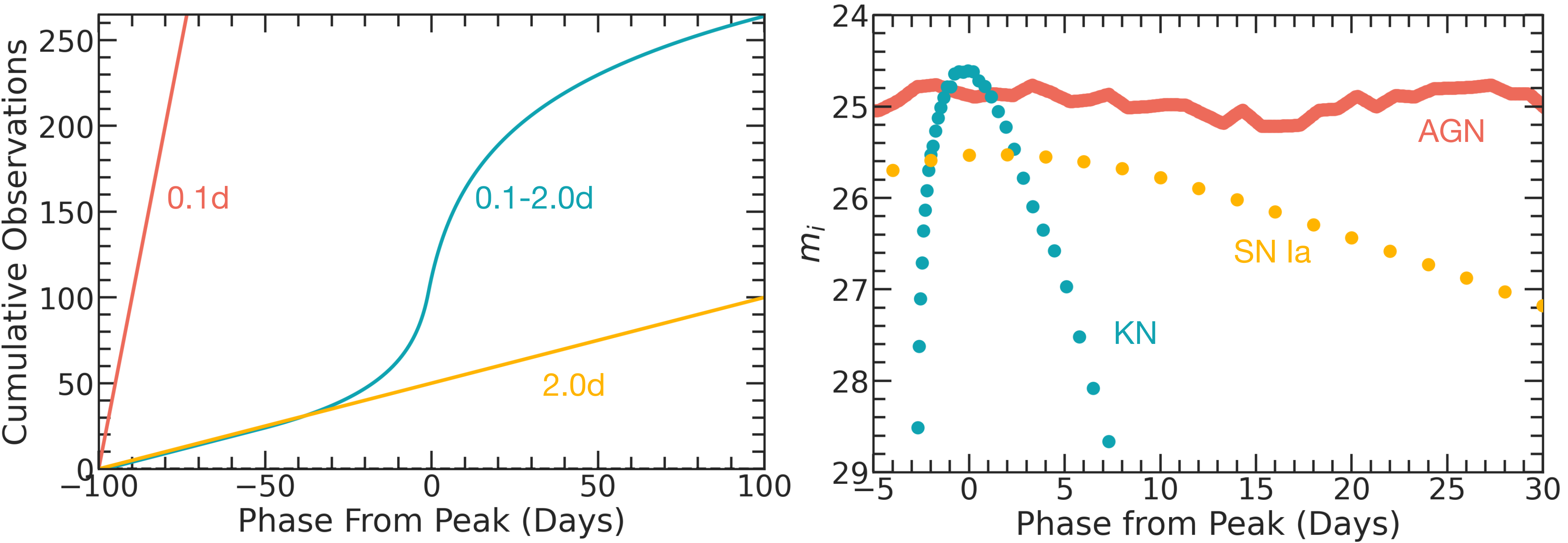}
    \caption{Left: Number of cumulative observations for the variable-cadence ($0.1$d$\leq\delta t\leq2.0$d) light curve sampling method for KNe \textcolor{black}{(turquoise)}, shown alongside the $\delta t=0.1$d (\textcolor{black}{red}) and the $\delta t=2.0$d (\textcolor{black}{yellow}) cadences.  The function describing the variable-cadence sampling rate is asymmetric in time to mimic the rapid rise and slow decay of transient phenomena. Right: Sample light curves for events using these cadences. Colours indicate the sampling method used.}
    \label{fig:sampling}
\end{figure*}

We sample most of our simulated light curves from 100 days preceding peak light to 100 days after peak. Due to the rapid rise and decline of KN photometry, we sample the light curve for only 10 days pre-peak and 60 days after-peak. For each transient class, we implement one of three cadence strategies. For SNe~Ia, CCSNe (II, IIn, IIP, and IIL), and H-poor SNe (Ib/c, Ic-BL, IIb, SLSN-I), we adopt a sampling of $\delta t=2.0$ days (so that each light curve has 100 observations each in $ugrizY$). Because of the rarity of high-cadence KN and TDE observations, we aim to preserve as much physical information about these light curves as possible while keeping file sizes manageable. For these events, we adopt a logarithmic sampling scheme, with $\delta t = 0.1$d at peak and increasing toward $\delta t = 2.0$d in either temporal direction. Because the rise of many transient light curves occurs on significantly shorter timescales than a decline powered by emission from radioactive decay, our cadence increases more slowly from peak light to late times than from peak light to pre-explosion epochs (Fig.~\ref{fig:sampling}). This sampling scheme generates $\sim260$ observations per pass-band. The rapidly-evolving stochasticity of AGN light curves cannot be accurately characterized by either of these cadences. For the AGN in our catalogue, we provide light curves at cadences of both $\delta t=0.1d$ (250 events) and $\delta t=2.0d$ (2,250 events). The choice in light curve can be made by the user, depending on their science goals. We show the cumulative number of observations for our three light curve cadences, along with sample light curves, in Fig.~\ref{fig:sampling}.

\textcolor{black}{In summary, the key features of the catalogue are as follows: top-of-the-galaxy observing conditions, near-perfect transient observing efficiency, host galaxies as faint as $r=28$ associated in a class-dependent manner, and physically-motivated transient-host separations.} We provide an overview of the classes of transients generated in this work, the {\tt HOSTLIB}s and {\tt WGTMAP}s used to associate them with hosts, and the sampling method for each class, in Table~\ref{tbl:hostlib_wgtmap_summary}.

\textcolor{black}{For completeness, we also summarize here the list of major assumptions and approximations used to construct the SCOTCH catalogue:} 
\begin{itemize}
\item \textcolor{black}{\textbf{Representative correlations from observed samples:} Limited observational data exists for many of the transient classes we simulate. Nevertheless, we assume that each data set presents a representative sample of that class's host population. With small sample sizes, this may cause host-transient correlations to be over-confidently inferred. }
\item \textcolor{black}{\textbf{sGRBs and KNe occur in the same host galaxies:} Given the extremely low number of KNe detected to date, and the intense interest from the community in detecting additional events, we choose to include KNe in our simulation framework with host-galaxy properties determined by a sample of well-localized sGRBs (\textsection~\ref{subsec:KN}).}
\item \textcolor{black}{\textbf{Galaxy populations have been shifted in redshift space:} We introduce redshift adjustments to host galaxies in several stages of the pipeline: when smoothing over unphysical tracks in colour-redshift space in CosmoDC2 (see \textsection~\ref{subsec:pzflow}) and when adjusting the host redshift to the transient redshift in \texttt{SNANA}. Thus, while the final host galaxy properties depend on redshift in a broadly similar way as in CosmoDC2, any realistic tight redshift dependencies that were encoded into the original simulation may be lost in \catname\,.}
\item \textcolor{black}{\textbf{No on-sky positions:} The hosts and their transients are not placed realistically on the sky, and thus the catalogue lacks realistic large-scale structure.}
\item \textcolor{black}{\textbf{Similar host-galaxy extinction parameters across cosmic time:} Dust parameters were determined from transient studies at low-$z$ (primarily SNe~Ia), and this work extends these properties to $z<3$. Deep host-galaxy studies will further clarify the dust properties of host galaxies at higher redshifts, which can be used to simulate an evolution in dust properties in future work.}
\item \textcolor{black}{\textbf{k-corrections linked to Simple Stellar Populations:} The k-correction equations used to calculate the rest-frame properties of host galaxies in GHOST \citep{Gagliano2021} were computed by SED fits to SDSS galaxies assuming simple stellar populations. This is done to obviate the need for detailed SED fitting, which would be a valuable independent study but is outside the scope of this work.}
\item \textcolor{black}{\textbf{No evolution in host-galaxy correlations:} We did not attempt to model any evolution of host-transient correlations with redshift, and therefore host populations at high-$z$ in the catalogue have similar distributions in intrinsic properties to hosts at low-$z$. Upcoming high-$z$ surveys will shed more light on any potential evolution in host galaxy populations.}

\item \textcolor{black}{\textbf{Malmquist-like bias:} The necessary $r<28$ cut on synthetic galaxies causes the hosts at higher redshift to be drawn from an intrinsically brighter population.}
\end{itemize}

\subsection{Transient Models}\label{sec:transient_models}


\begin{figure*}
    \centering
    \includegraphics[width=0.7\textwidth]{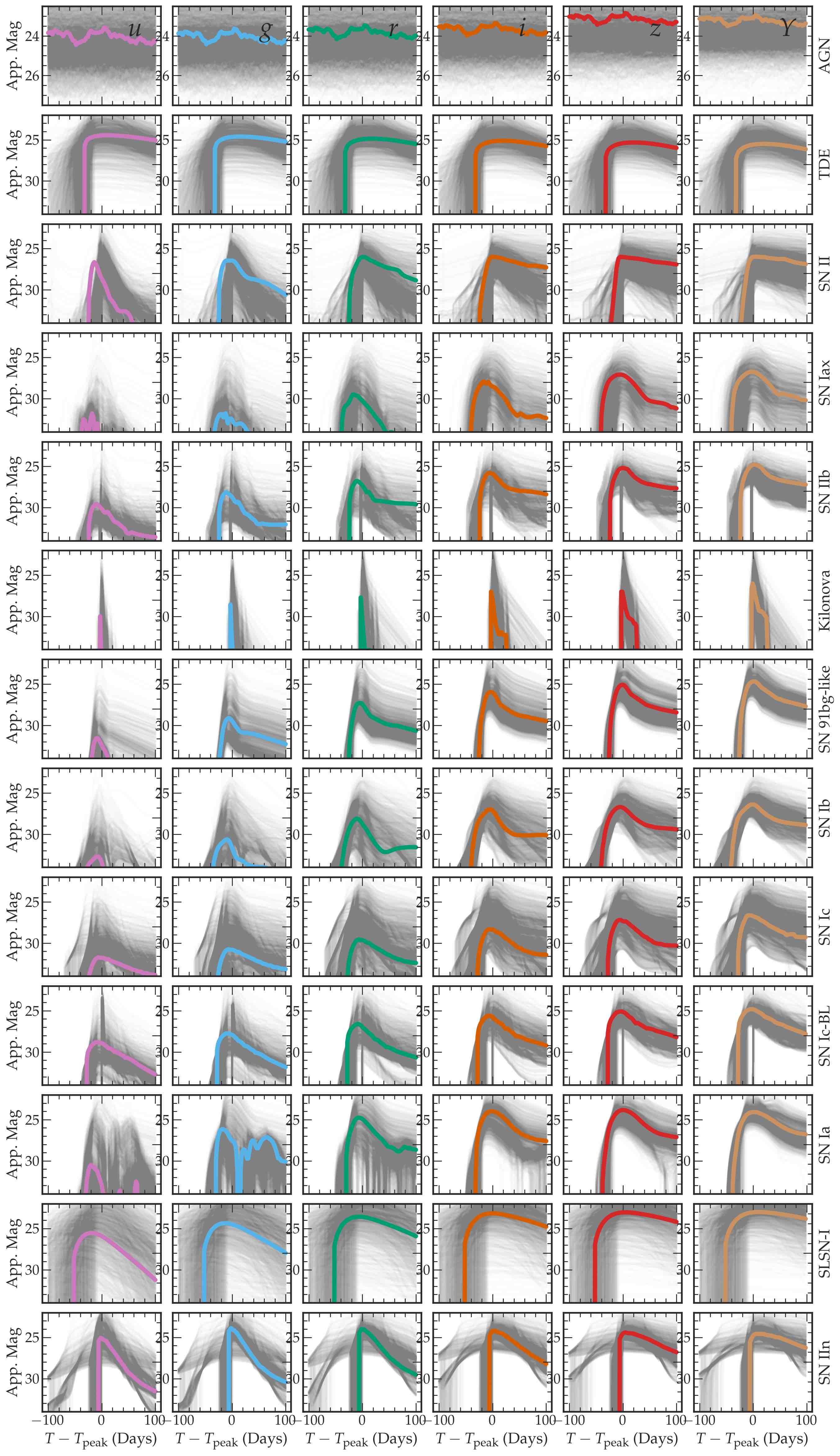}
    \caption{Light curves for the transient classes simulated in this work. Gray curves show a representative sample of events, and coloured curves show one example event for each class. Due to the rapid rise of the kilonova model, few KN light curves feature emission at epochs prior to peak light.}
    \label{fig:LCs}
\end{figure*}

We simulate each transient class with a similar setup as that of the PLAsTiCC challenge \citep[see][hereafter K19, and references therein]{Plasticcv12019}. Unless otherwise noted, the event rate of each class follows the same redshift dependency as in PLAsTiCC. \textcolor{black}{As detailed in K19, the redshift-dependent rates have been measured from existing observations, and thus become increasingly uncertain with redshift. We extrapolate the rates to $z=3$ for SCOTCH, but caution that the rate functions are unlikely to be trustworthy at high redshifts.} We are unable to simulate realistic \textit{volumetric} rates with the framework described, as our ideal `observing' conditions and wide redshift range would generate a prohibitively large catalogue. Instead, we choose to simulate 5 million events and divide the sample by class based on anticipated interest to users. Table \ref{tbl:class-models-numbers} displays the number of light curves and model used per class. The sky density of events at any given redshift, and the relative rates of different classes, are therefore inaccurate. Nevertheless, the volumetric rates predicted for each class by the respective rate model in K19 \textcolor{black}{can be calculated using information in the supplementary material, Appendix~B, and the corresponding functions provided in the GitHub repository for this work. This information can also be used to over- or under-sample the catalogue as needed to generate realistic mock observations.}

We simulate normal and peculiar SNe~Ia using the same models used in K19. Additionally, the core-collapse SN models suffixed with \texttt{NMF}, \texttt{Templates}, and \texttt{MOSFIT} in Table \ref{tbl:class-models-numbers} are identical to those detailed in K19. However, several features of our simulation are different from PLAsTiCC. We do not include the parametric SN~Ibc MOSFiT model, which was part of PLAsTiCC, because its parameters were drawn from unrealistic flat distributions and the models produce unphysical light curves. We also add the recent core-collapse spectrophotometric templates from \citet{Vincenzi2019CC}, which are suffixed by \texttt{HostXT\_V19} in Table~\ref{tbl:class-models-numbers}. The V19 models add realistic diversity to our core-collapse sample. The V19 models include templates for SNe~IIb and SNe~Ic-BL, which are both new since PLAsTiCC. SNe~Ic-BL are energetic Ic events observed with broad lines in their spectra, indicating enhanced absorption velocities of ${\sim}15,000$-$30,000$~km~s$^{-1}$ \citep{Modjaz2016}. Thus, the Ic-BL templates are distinguishable from the typical Ic model primarily through spectroscopy. As we only simulate photometric light curves for SCOTCH, the only relevant difference from SNe~Ic is that SN~Ic-BL light curves are brighter.



\begin{table}
    \centering
    \begin{tabular}{c|c|c}
    \hline
    \hline 
    Class & Model name & $N$\tabularnewline
    \hline 
    \hline
    \multirow{3}{*}{Ia (2.2M total)} & SALT2-Ia & 2M \tabularnewline
     & Iax & 100,000\tabularnewline
     & 91bg-like & 100,000 \tabularnewline
    \hline
    \hline
    \multicolumn{3}{c}{H-Rich Core-Collapse (1,999,000 total)}\tabularnewline
    \hline
    \hline 
    \multirow{3}{*}{SN II (1.9M total)} & SNII-Templates & 633,000 \tabularnewline
    & SNII-NMF & 633,000 \tabularnewline
    & SNII+HostXT\_V19 & 633,000 \tabularnewline
    \hline
    \multirow{2}{*}{SN IIn (100,000 total)} & SNIIn-\texttt{MOSFiT} & 50,000 \tabularnewline
    & SNIIn+HostXT\_V19 & 50,000 \tabularnewline
    
    \hline
    \hline
    
    \multicolumn{3}{c}{Stripped Envelope / H-Poor Core-Collapse (500,000 total)}\tabularnewline 
    \hline
    \hline 
    \multirow{2}{*}{Ib (100,000 total)} & SNIb-Templates & 50,000 \tabularnewline
     & SNIb+HostXT\_V19 & 50,000 \tabularnewline
    \hline 
    \multirow{2}{*}{Ic (100,000 total)} & SNIc-Templates & 50,000 \tabularnewline
     & SNIc+HostXT\_V19 & 50,000  \tabularnewline
    \hline 
    IcBL & SNIcBL+HostXT\_V19 & 100,000 \tabularnewline
    \hline 
    IIb & SNIIb+HostXT\_V19 & 100,000 \tabularnewline
    \hline 
    SLSN-I & SLSN-I-\texttt{MOSFiT} & 100,000 \tabularnewline
    \hline 
    \hline 
    \multicolumn{3}{c}{Non-SN Transients (301,000 total)}\tabularnewline 
    \hline
    \hline 
    \multirow{2}{*}{KN (100,000 total)} & Kasen 2017 & 50,000 \tabularnewline
     & Bulla 2019 & 50,000\tabularnewline
     \hline
    TDE & TDE\_\texttt{MOSFiT} & 101,000\tabularnewline
    \hline
    AGN & AGN-LSST & 100,000\tabularnewline
    \hline
    \end{tabular}
\caption{Simulated transients in the SCOTCH catalogue organized by class and model name. $N$, the number of transients simulated per class, sums to 5 million. }
\label{tbl:class-models-numbers}
\end{table}

The SLSN~I, TDE, and AGN models are similar to those in K19, the only difference being that we impose an upper redshift limit of $z=3$ to match the limit of CosmoDC2 galaxies within the \texttt{HOSTLIB}s.

PLAsTiCC included a set of kilonova SED models from \cite{Kasen_2017}. In this work, we incorporate the Kasen model as well as an SED model by \cite{bulla_2019}.
Both approaches parametrize the rapidly decompressing ejecta from the merger of two neutron stars that
are heated by radioactivity from r-process nucleosynthesis. The \citet{bulla_2019} model\footnote{We
consider the two-component model here; ``BULLA-BNS-M2-2COMP'' in the \texttt{SNANA} package data.} is parameterized by the mass of the ejecta, the half-opening angle of its lanthanide-rich component, and the
angle between the line of sight and the plane of merger. The \cite{Kasen_2017} models consider the
SED parameterized by the ejecta mass, the lanthanide fraction, and the ejecta velocity, without an observing angle dependence. In general, larger values of the ejecta mass lead to
brighter events, while increasing the lanthanide fraction leads to redder, longer-lived kilonovae. 
Both models are grid-based, i.e., an SED given on a grid of discrete model parameters. Recently, \cite{chatterjee_2021} incorporated the Bulla model into \texttt{SNANA}, and used both these kilonova models to create a classifier to filter kilonovae from other astrophysical transients using photometric and contextual information available from gravitational wave data products. The SED data for both the models are publicly available as part of the \texttt{SNANA} package.\footnote{\url{https://doi.org/10.5281/zenodo.7416122}}

\begin{figure}
    \centering
    \includegraphics[width=\linewidth]{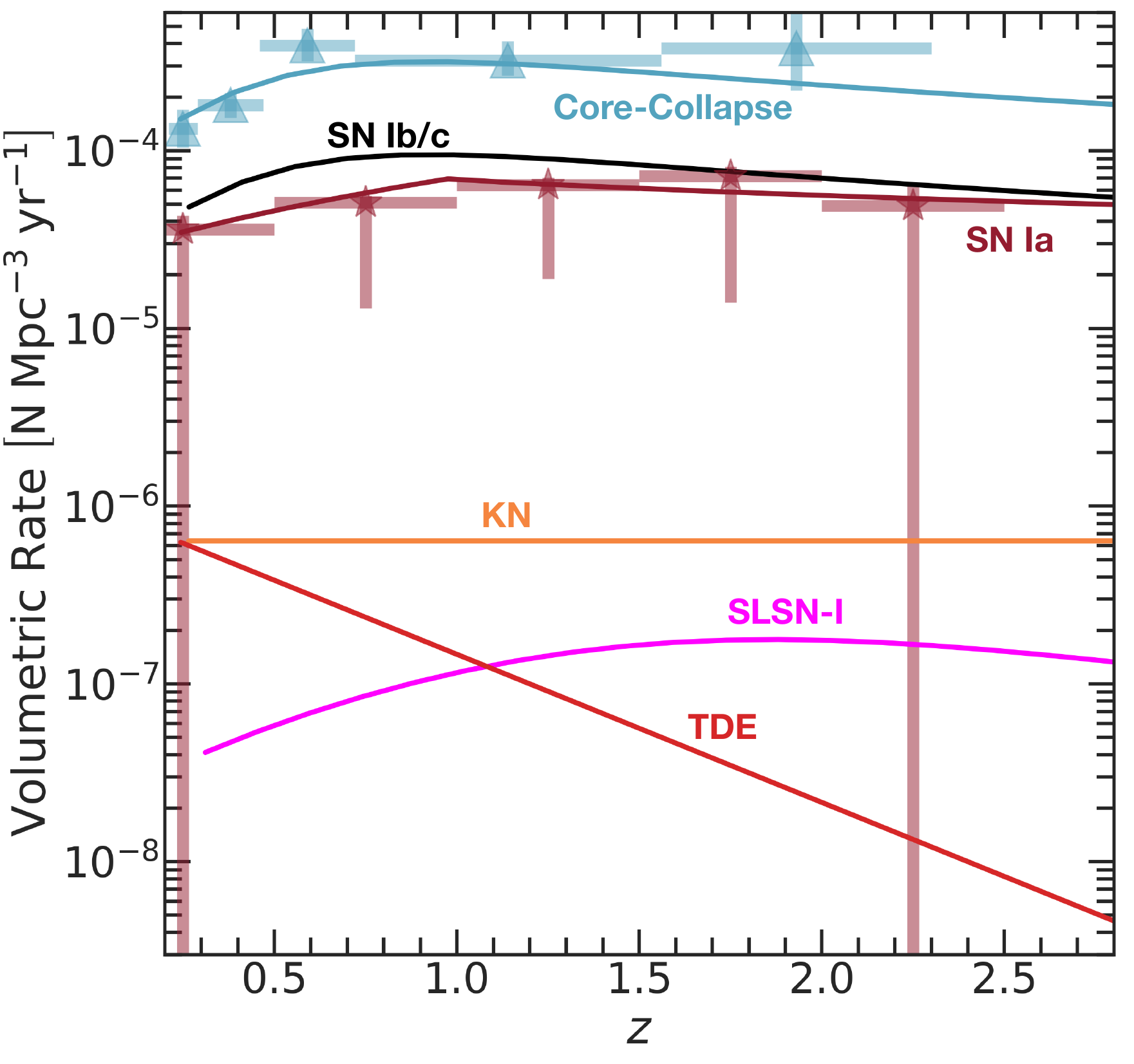}
    \caption{\textcolor{black}{Re-scaled volumetric rates in the SCOTCH catalogue as a function of redshift for multiple transient classes. The true catalogue rates (with the same redshift evolution but arbitrary total event numbers) have been re-scaled to match the predicted volumetric rates given the input model. Stars correspond to observed volumetric rates for SNe~Ia and core-collapse SNe, with statistical and systematic uncertainties plotted.}}
    \label{fig:volumetricRates}
\end{figure}

We present the distribution of light curves for a representative sample of each simulated transient class in Fig.~\ref{fig:LCs}. \textcolor{black}{We further present the volumetric rate evolution of transients within the catalogue, re-scaled to the match the model predictions using the values in Table~B1, in Fig.~\ref{fig:volumetricRates}. The re-scaling is necessary because the total event numbers in SCOTCH (Table~\ref{tbl:class-models-numbers}) are arbitrary. For comparison, we show the observed volumetric rates for SNe~Ia from \cite{2014Rodney_CANDELS} and CCSNe from \cite{2015Strolger_CANDELS}. We find agreement with the observed volumetric rates, with the deviations dominated by the theoretical fits to the data and not limitations in the simulation pipeline.} 

\section{ Host Library Creation (HOSTLIB)}
\label{sec:methods}
\subsection{Overview}
Our first step to realistically associate transients with hosts is to create \texttt{HOSTLIBs} from CosmoDC2 galaxies. In previous work, including \citet{Vincenzi2021}, \texttt{HOSTLIBs} were constructed of representative field galaxies and correlations with transient class were implemented solely via the \texttt{WGTMAP} schema. However, \texttt{WGTMAPs} that depend on more than a few host properties $N$ are prohibitively memory-intensive and computationally slow, as the \texttt{SNANA} code requires weights to be defined over an evenly-spaced grid across the full $N$-dimensional parameter space. To incorporate additional galaxy properties in our transient-host association, we generate \textit{tailored} \texttt{HOSTLIBs} for different transient categories containing correlations in host galaxy colour and magnitude. 

\subsection{Accounting for CosmoDC2 Artifacts with PZFlow}\label{subsec:pzflow}

\begin{figure}
    \centering
    \includegraphics[width=\columnwidth]{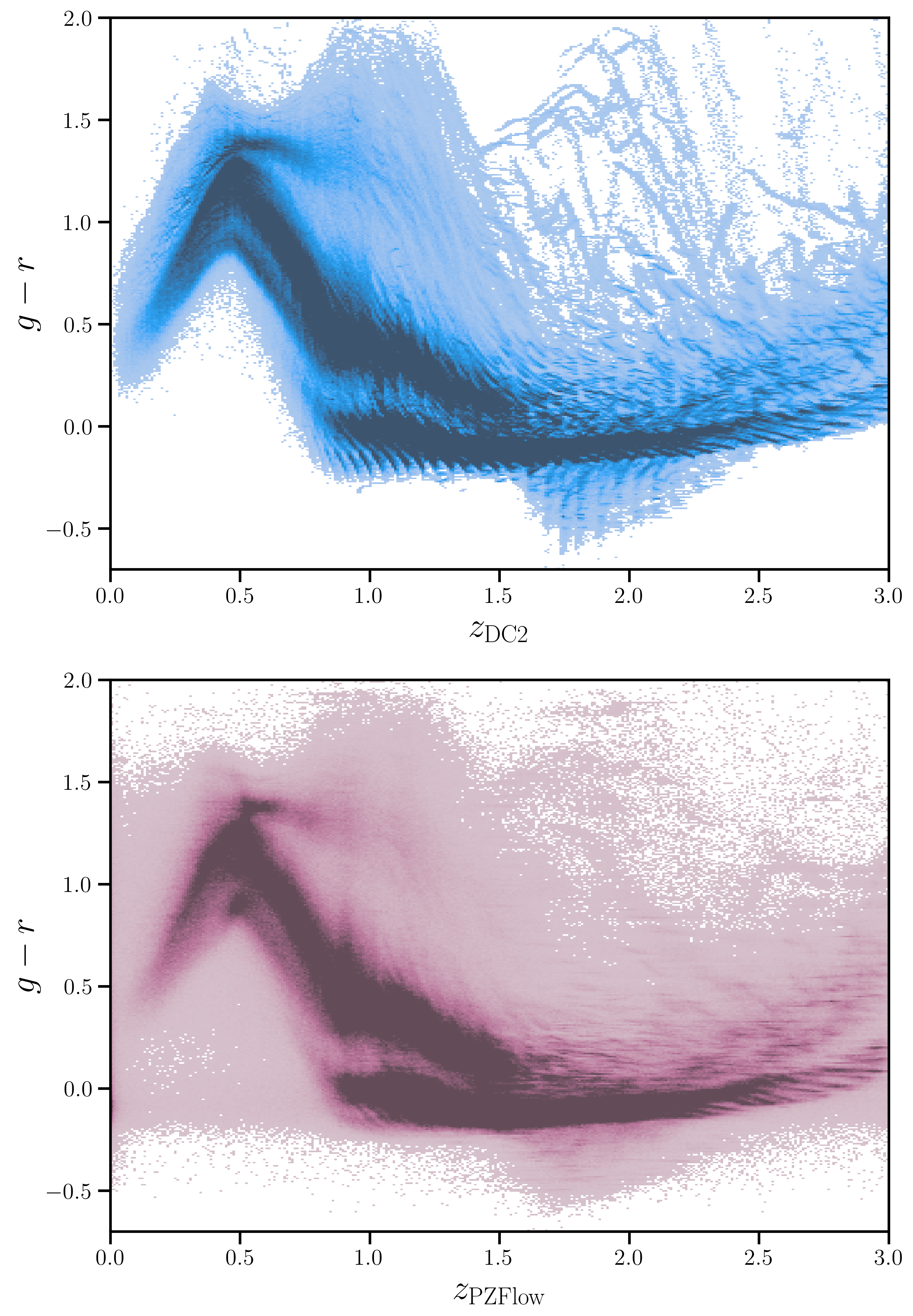}\\
    \caption{\textbf{Top:} \textcolor{black}{Observed} $g-r$ colour versus redshift for galaxies within the CosmoDC2 synthetic sky catalogue \citep{CosmoDC22019}. Discrete tracks visible in this space are an artifact of the limited number of SED models used to generate the derived properties of these galaxies.  \textbf{Bottom:} The same $g-r$ values plotted against redshifts drawn from a probability density function estimated using a normalizing flow \citep{PZFlow_JFC}. There is a one-to-one mapping between the galaxies in the top and bottom image; each galaxy's redshift is slightly shifted, which smears out many of the non-physical tracks.}
    \label{fig:pzflow_sfr}
\end{figure}

The distribution of CosmoDC2 galaxies in colour-redshift space contains non-physical structure \textcolor{black}{that worsens at higher redshifts.} \textcolor{black}{The problem is caused by a mismatch between the colour ranges in different components of the hybrid catalogue production pipeline for CosmoDC2 \citep{CosmoDC22019},} and manifests as discrete `tracks' in colour-redshift space across the full redshift range  (upper panel of Fig.~\ref{fig:pzflow_sfr}). These artifacts limit our ability to test upcoming machine-learning based photometric redshift estimators. Along each track, the dispersion about the colour-redshift relation is unrealistically low. Photometric redshift and classification algorithms trained using these data are likely to achieve unrealistically accurate results. 
As photo-$z$ estimation is an important component of upcoming large-sky transient and galaxy surveys, we mitigate this effect by smearing the streaky colour-redshift relationship for selected CosmoDC2 galaxies.

We use the code \texttt{PZFlow}\footnote{\url{https://github.com/jfcrenshaw/pzflow}} \citep{PZFlow_JFC} to realistically shift galaxy redshifts and smear the colour-redshift tracks. 
\texttt{PZFlow} uses normalizing flows \citep{Rezende2015} to efficiently model a multi-dimensional probability density function (PDF). 
The normalizing flow learns a bijection, a function which maps the PDF to a tractable latent distribution (such as a uniform distribution) in which we can easily sample points and calculate densities.
We train a normalizing flow to model the conditional PDF $p(z|\mathbf{m}, M_*, \text{SFR})$, where $\mathbf{m}$ is a vector of LSST $ugrizY$ apparent magnitudes.
The latent distribution for the flow is uniform over the range $[-5, 5]$.
The bijection consists of two layers:
the first layer shifts the redshift range $[0, 3]$ into the range of the latent uniform distribution, $[-5, 5]$;
the second layer is a Rational-Quadratic Neural Spline Coupling layer \citep[RQ-NSC;][]{durkan2019}, which transforms the redshift distribution into a uniform distribution.
The RQ-NSC performs this transformation using splines parameterized by a set of knots.
The number of spline knots, $K$, can be chosen to increase or decrease the resolution of information captured by the flow.
We choose a low value of $K=2$ so that the flow does not learn the discrete tracks in the CosmoDC2 colour-redshift space, and instead smooths over them.

\begin{figure*}
    \centering
    \includegraphics[width=\textwidth]{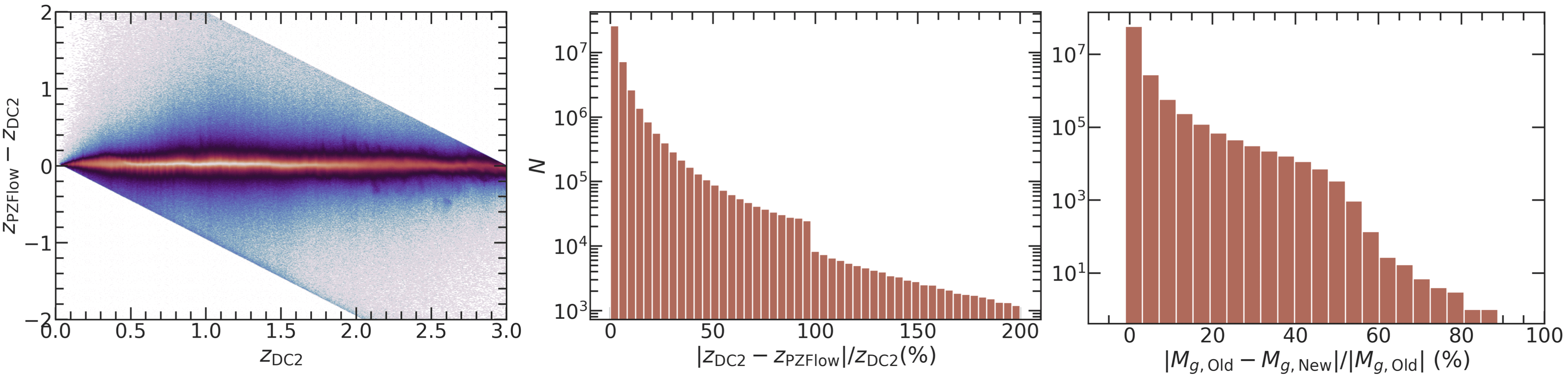}
    \caption{
    \textbf{Left:} Difference between PZFlow-sampled redshift and original CosmoDC2 redshift as a function of CosmoDC2 redshift for each galaxy in the sample. Shifts are minimal, and the discrete paths at top left and bottom right reflect the bounds of [0, 3] imposed in generating PZFlow samples. \textbf{Middle:} Histogram of fractional differences in galaxy redshift before and after the PZFlow-oversampling stage. \textbf{Right:} Histogram of fractional differences in absolute magnitude propagated from the PZFlow-shifted redshifts.}
    \label{fig:pzflow_shifts}
\end{figure*}

We train the normalizing flow on the 40M $r<28$ CosmoDC2 galaxies for 30 epochs using maximum likelihood estimation. Then, for each galaxy, we sample a new redshift $z_\text{PZFlow}$ from the normalizing flow. These new redshifts are consistent with the galaxies' photometry, stellar masses, and star formation rates, but are slightly perturbed from their original values (see left and middle panels of Fig.~\ref{fig:pzflow_shifts}).
For $92\%$ of the galaxies, the change in redshift is $\delta z < 0.1$. 
While $8\%$ of redshifts have more substantial perturbations, we stress that each perturbed redshift is consistent with that galaxy's observed and derived properties.
\textcolor{black}{It should not be surprising that, for some galaxies, a substantially different redshift remains consistent with its other properties \citep{salvator2019}, particularly if the original distribution contained discrete artifacts.}

We next adjust each galaxy's absolute magnitude to retain the same apparent magnitude given its shifted luminosity distance. The right panel of Fig.~\ref{fig:pzflow_shifts} shows the size of these adjustments. For $93\%$ of galaxies, the absolute magnitude change is $\delta M < 0.2$. We expect the small magnitude perturbations to lie within the realistic scatter of galaxy properties. 

We have perturbed the galaxy redshifts with the goal of removing the discrete tracks in colour-redshift space that would lead to overly-optimistic photo-$z$ estimates. We present the original and re-sampled redshift distributions in the upper and lower panels of Fig.~\ref{fig:pzflow_sfr}, respectively, in which only a fraction of the tracks remain (and these are primarily at very high redshift). 

\textcolor{black}{After applying PZFlow, we perform an additional cut to clean the CosmoDC2 rest-frame colour distribution. We limit our sample to galaxies with $-0.18<i-z<0.5$, which maintains the vast majority of the CosmoDC2 colour distribution but removes two unrealistic clumps that were visible on the red and blue extremes after re-sampling.}

\subsection{Selecting a CosmoDC2 subsample}\label{subsec:CosmoDC2_subsample}
Ideally, we would sample directly from the CosmoDC2 galaxies, \textcolor{black}{with our selection} informed by real data, to populate a catalogue of synthetic host galaxies for each transient class. Although the GHOST dataset includes \textcolor{black}{many} SN classes, for most classes \textcolor{black}{(e.g., SNe IIn, Ic-BL, and IIb)} the sample size is too small to contain representative information. \textcolor{black}{To combat this issue, before matching we group the hosts of archival GHOST SNe} into three broader categories:

\begin{enumerate}
    \item \textcolor{black}{6,284} hosts of Type~Ia SNe.
    \item \textcolor{black}{1,973} hosts of core-collapse SNe with hydrogen lines in their spectra, including those classified as Type~II, IIn, and IIP.
    \item \textcolor{black}{443} hosts of hydrogen-poor SNe, including stripped-envelope core-collapse SNe and hydrogen-poor superluminous SNe (SLSNe-I).
\end{enumerate}

For each group, we select a many-times larger sample of synthetic CosmoDC2 galaxies that approximates the distribution in GHOST galaxy properties while extending to higher redshifts. We construct a \texttt{HOSTLIB} for each of these 3 broad categories.

With insufficient high-$z$ data to model any cosmic evolution of transient-host correlations, a reasonable approach to high-redshift HOSTLIB sampling is to find high-$z$ CosmoDC2 galaxies with similar intrinsic properties as low-$z$ galaxies; e.g., rest-frame absolute magnitudes, rest-frame colours, and morphology. This approach ignores the possible evolution of transient-host correlations over cosmic time.  

\begin{figure*}
\includegraphics[width=\textwidth]{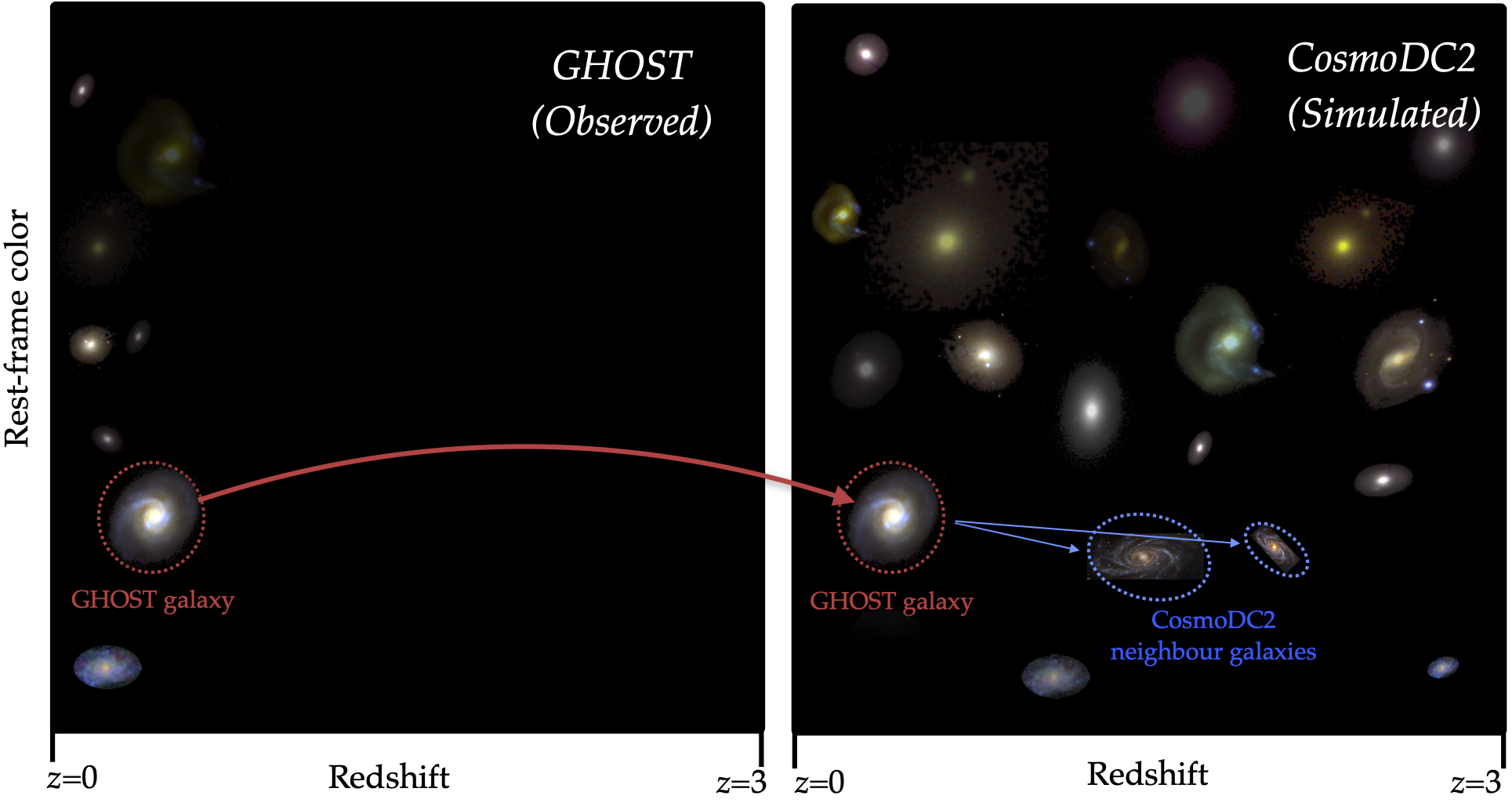}
\caption{
Illustration of CosmoDC2 sampling for \texttt{HOSTLIB} generation. \textbf{Left:} \textcolor{black}{We identify GHOST galaxies which hosted transients belonging to a particular category (the three categories are defined in the text). The GHOST sample is observationally biased toward low-redshift, but we assume the rest-frame properties (like color, shown on the $y$ axis) are representative of the true full sample of hosts.} \textbf{Right:} For every GHOST galaxy, we select $k$ synthetic galaxies from CosmoDC2 nearby in \textcolor{black}{rest-frame color, brightness, and redshift} (with small $d_{\rm ANNOY}$ values; see Appendix A from the supplementary materials for details) but extending to $z=3$. For simplicity, the diagram shows $k=2$ and a \textcolor{black}{2}-dimensional parameter space; in practice, however, $k=800 - 18,000$ and \textcolor{black}{5} parameters are used for matching.}
\label{fig:nn_matching_diagram}
\end{figure*}

We employ a nearest-neighbour (NN) technique, illustrated in Fig.~\ref{fig:nn_matching_diagram}, to achieve this matching. NN algorithms deteriorate in high-dimensional space due to decreased data density (the so-called `curse of dimensionality'), so we select a small number of galaxy properties that most strongly correlate with transient properties. 
%
We train our NN matching on absolute rest-frame $r$- and $i$-band magnitudes, rest-frame $g-r$ and $i-z$ colours, and \textcolor{black}{the PZFlow-shifted redshifts.} This combination of properties incorporates information from all $griz$ bands ($u$ and $y$ being unavailable after converting GHOST data to SDSS filters) without redundancy. \textcolor{black}{Although a similar matching could be achieved by simply using the $griz$ magnitudes and allowing colour to be implicitly included, doing so would down-weight the importance of colour. We choose to explicitly make two of the four dimensions colours, with an even weighting as the magnitudes, because \textcolor{black}{colour correlates strongly with a galaxy's star-formation rate, and consequently the rate of core-collapse SNe; while brightness correlates strongly with stellar mass, and consequently the rate of SNe~Ia and SNe~II (see \textsection~\ref{sec:host_correlations}).}}  We manually down-weight redshift with respect to the colours and magnitudes to allow for neighbour matching across a wide range of redshifts. We considered also matching on morphological information such as galaxy radius and ellipticity, but found the difference between observational and simulated estimates to be prohibitive given the necessary corrections to systematics in the GHOST data (e.g., deconvolving the PSF from ellipticity and radius measurements) and the strong redshift-dependent observational bias towards intrinsically large and face-on galaxies in targeted surveys.

Although exact NN algorithms (e.g., \texttt{scikit-learn}'s \texttt{KNeighbours}) are sufficient for samples of $N\sim10^5$, they become increasingly slow when scaling up to $\sim10^7$ galaxies. Instead, we make use of a rapid approximate NN method called \texttt{ANNOY}, which leverages Locality-Sensitive Hashing \citep[LSH;][]{lsh}, to \textcolor{black}{rapidly} cross-match datasets in parallel. We provide additional information on our use of this method in Appendix A in the supplementary materials accompanying this paper, available online. We select $k$ neighbours in CosmoDC2 with \textcolor{black}{low} Euclidean distance ($d_{\mathrm{ANNOY}}$) to each GHOST galaxy. Rather than impose a limit on $d_{\mathrm{ANNOY}}$, we set $k$ to a value that produces several million matched CosmoDC2 galaxies. We then remove synthetic galaxies that were matched to multiple GHOST galaxies, creating \texttt{HOSTLIBs} with $\sim3$ million objects each. This requires $k$ to range from 800-18,000 across the three host categories.  

\begin{figure*}
    \centering
    \includegraphics[width=0.31\textwidth]{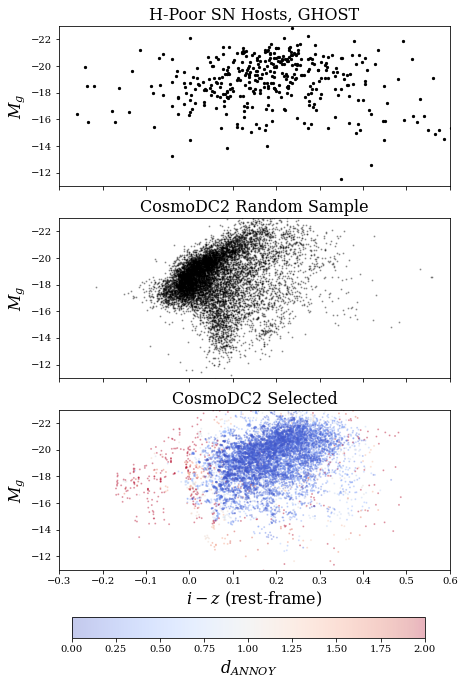}
    \includegraphics[width=0.32\textwidth]{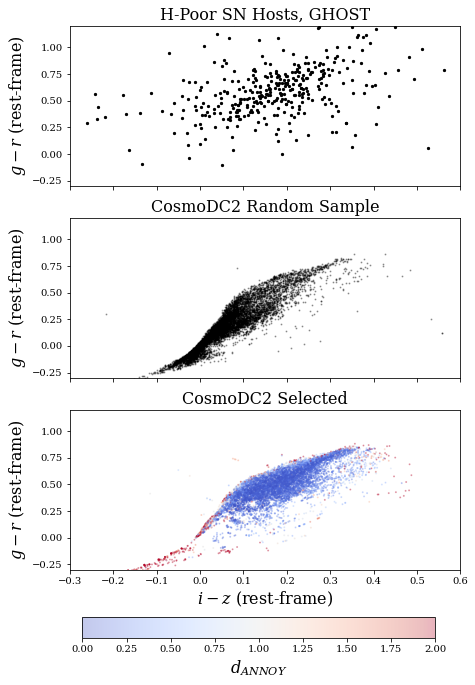}
    \includegraphics[width=0.32\textwidth]{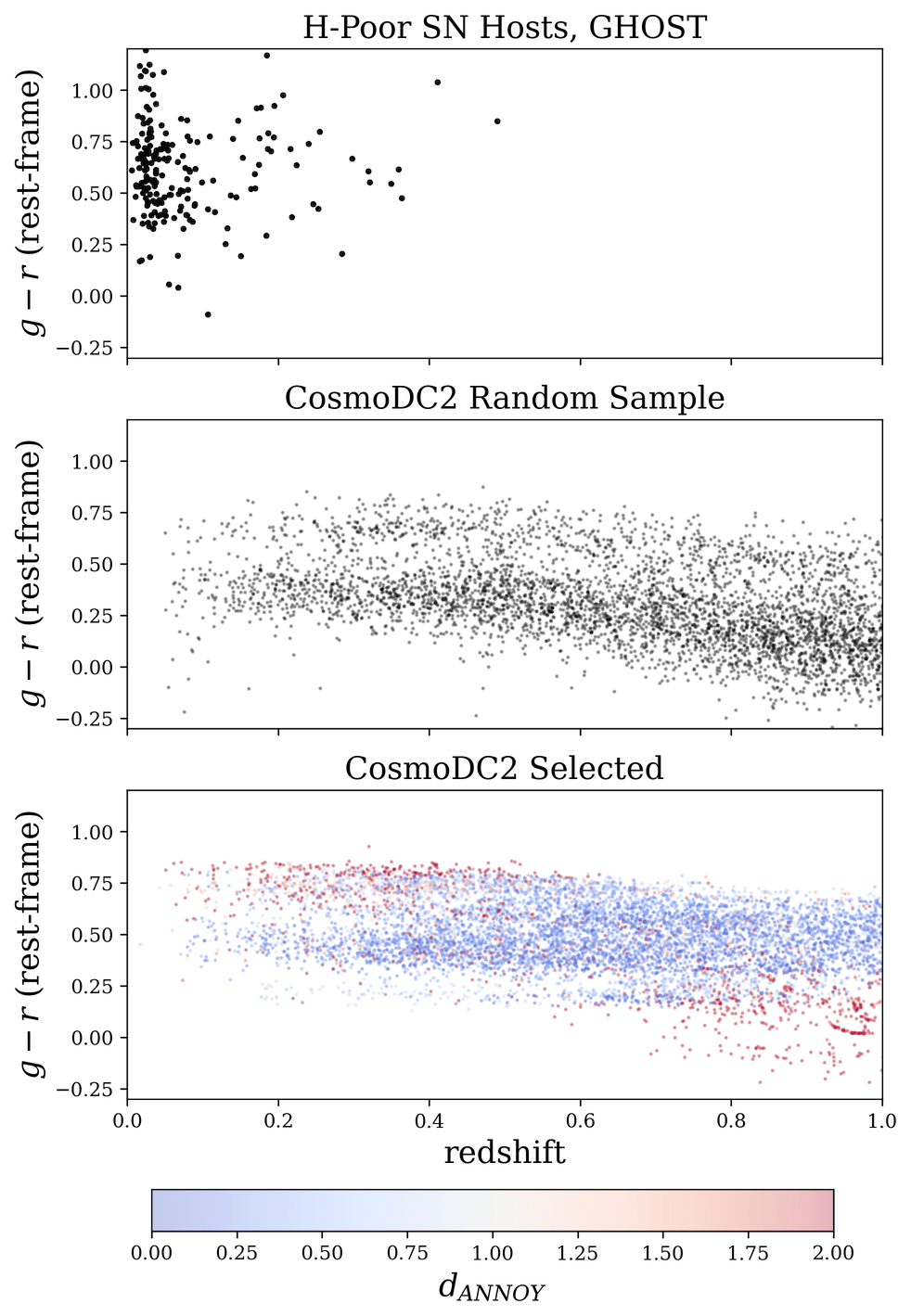}
    \caption{A representation of \textcolor{black}{how the GHOST-CosmoDC2 matching process produces a larger sample of GHOST-like galaxies from CosmoDC2. Each set of three panels (stacked vertically) show the same steps of the process which creates the H-poor host library. The left three plots show a magnitude-colour distribution, the middle show colour-colour space, while the right show colour vs. redshift. \textbf{Top}: the GHOST H-poor event host galaxies.  \textbf{Center}: a random subset of CosmoDC2 galaxies. \textbf{Bottom}: nearest-neighbour galaxies selected from CosmoDC2 to match each GHOST galaxy in a multidimensional parameter space.} The colour bar, applying only to the bottom three plots, signifies the nearest-neighbour \textcolor{black}{Euclidean} distance and is truncated at 2 for visual purposes. Red points are CosmoDC2 galaxies that are far from their nearest GHOST neighbour. \textcolor{black}{From the left and center plots, it is apparent that the limited colour ranges in CosmoDC2 also limit the effectiveness of matching, as some GHOST galaxies lie beyond the available DC2 colour extent. The right plots demonstrates how down-weighting redshift as a matched property allows us to select CosmoDC2 galaxies across a wide range (extending to $z=3$, but only shown up to $z=1$).}}
    \label{fig:annoy_plots}
\end{figure*}

Using this pipeline, we generate \texttt{HOSTLIB}s of CosmoDC2 galaxies matching the host-galaxy distributions in GHOST for the three groups of transient classes described above. We show the distribution of observed GHOST galaxies, a random subset of CosmoDC2 galaxies, and the final matched sample for the Hydrogen-poor host group in Fig~\ref{fig:annoy_plots}. In addition to the three \texttt{HOSTLIB}s described above, we create a \texttt{HOSTLIB} consisting of galaxies randomly sampled from CosmoDC2 for transient classes not in the GHOST catalogue. The host correlations for events using this unmatched \texttt{HOSTLIB} originate entirely from the transient-specific \texttt{WGTMAP}. 

\textcolor{black}{The CosmoDC2 simulation contains many ultra-faint galaxies that were injected into the simulation to account for its finite mass resolution \citep{CosmoDC22019}. These galaxies were not assigned properties as rigorously as the normal galaxies in the simulation; as a result, we exclude them from our \texttt{HOSTLIB}s. Our initial $r<28$ cut on CosmoDC2, as well as the matching to GHOST galaxies (which are generally bright), remove many of these galaxies from our sample. We identify any remaining ultra-faint galaxies by their negative \texttt{halo\_id} after \texttt{HOSTLIB} creation and cut them from the samples.} 


\section{Transient-Host Galaxy Correlations (WGTMAP) }
\label{sec:host_correlations}

The GHOST database contains only basic photometric properties for SN host galaxies. Additional properties like masses and star formation rates, which can be estimated from galaxy imaging and/or spectra, are not reported in GHOST due to the lack of availability. Thus, the \texttt{HOSTLIB}s of CosmoDC2 galaxies matched to GHOST are realistically distributed in colour and brightness, but only contain an implicit dependence on derived properties. There is motivation to fold in explicit dependence on mass and star formation rate: many studies have empirically verified this dependence for multiple transient classes (using smaller samples than GHOST) and found statistically significant differences between classes. To more precisely simulate these correlations, we construct a series of \texttt{WGTMAP}s that explicitly encode the class-specific dependence of host-transient association on a galaxy's derived properties. These \texttt{WGTMAP}s enable us to fine-tune the host populations of the three matched \texttt{HOSTLIB}s and add in class-specific correlations for events whose hosts will be drawn from the unmatched, randomly-sampled \texttt{HOSTLIB}.

\textcolor{black}{Each \texttt{WGTMAP} consists of a multidimensional grid of host properties, where values in the grid correspond to the \textcolor{black}{probability} of a galaxy with those properties being selected to host a given transient event in \texttt{SNANA}. The weightmap values can be scaled arbitrarily, so long as the \textit{relative} weights between galaxies of different properties are preserved. When the \texttt{SNANA} simulation generates a transient at $z$, a sample of candidate host galaxies within $\delta z$ are retrieved from the \texttt{HOSTLIB}. \texttt{SNANA} then assigns a corresponding weight to each potential host using the \texttt{WGTMAP}. The simulation creates a cumulative distribution function (CDF) for all galaxies, where the horizontal axis orders the galaxies arbitrarily, and the function is augmented by each galaxy's weight. The CDF is normalized to have a maximum value of one. The simulation selects a random number for the CDF vertical axis, and picks the corresponding galaxy along the horizontal axis, such that highly weighted galaxies have a higher CDF slope and are therefore more likely to be selected. Because the normalization is computed internally within \texttt{SNANA}, only the \textit{proportionality} linking the probability $P$ to the host galaxy parameters must be defined for the  \texttt{WGTMAP}s.}

\textcolor{black}{We emphasize that the transient rate models, described in \textsection~\ref{sec:transient_models} and extensively detailed in K19, are entered into the \texttt{SNANA} simulation \textit{independently} of host properties. In other words, the final transient event rate in the simulation is not affected by the associated host galaxies. Transients are generated according to the rate function and the total number desired (Table~\ref{tbl:class-models-numbers}), and only then assigned a host. The \texttt{WGTMAP} plays a role only in determining the relative likelihood of one host being selected over another.}

The following subsections review the observations that motivate the \texttt{WGTMAP}s of each class. We caution that for several classes \textcolor{black}{(TDEs, KNe, SNe Ic-BL, and SLSNe-I)}, the data statistics remain limited, and thus the resulting simulations may not capture the full diversity of host galaxies. Fig.~\ref{fig:elasticc_SFRvsMsol} provides some intuition for the \textcolor{black}{probability} equations that follow by showing the SFR-mass distribution of hosts selected by \texttt{SNANA} from these equations for six classes.



\subsection{SNe~Ia}\label{subsec:SNIa}
SN~Ia explosions are thought to occur when a white dwarf accretes material from a companion star (which may also be degenerate) in a binary system, approaches the Chandrasekhar mass, and ignites in a runaway chain reaction of carbon fusion. SNe~Ia occur more frequently in high-mass and highly star-forming galaxies \citep{Sullivan2006, smith2012,wiseman2021}.
Their light curves are parameterized in the \texttt{SNANA} simulation using the Spectral Adaptive Lightcurve Template (SALT2/SALT3) frameworks \citep{2007A&A_SALT2,2021Kenworthy}.  
\newcommand{\RKDE}{P_{\mathrm{Ia}}^{\mathrm{KDE}}}  

The \textcolor{black}{probability} of SNe~Ia as a function of galaxy properties can be written as the product of two components:
\begin{equation}
     P_{\mathrm{Ia}}(M_*, \mathrm{SFR}, x_1) = \RKDE  (M_*, \mathrm{SFR}) \times P_{\mathrm{Ia}}^*(x_1,M_*)
\end{equation}
where $M_*$ is the host galaxy stellar mass, SFR is its star formation rate, and \textcolor{black}{$x_1$} is the SALT2 stretch parameter of the SN light curve \citep{SALT22007}.

The first component of the \textcolor{black}{probability}, $P^{\mathrm{KDE}}_{\mathrm{Ia}}$, encodes a dependence on galaxy stellar mass and star formation rate. It has been constructed to replicate the distribution of 200 SN~Ia host galaxies from the Dark Energy Survey Supernovae Program \citep[][]{Smith2020}. \textcolor{black}{Using the \texttt{statsmodel} Python package,} we perform kernel density estimation on the DES \textcolor{black}{$\log(M_*)$ and $\log(\mathrm{SFR})$ values} to estimate a smooth probability density function in $\log(M_*)$ vs. $\log(\mathrm{SFR})$ space. \textcolor{black}{We use a Gaussian kernel for both parameters and determine the best-fit bandwidths from the KDE fit, finding 0.49 and 0.47 for $\log(M_*)$ and $\log(\mathrm{SFR})$, respectively.} Because  $\RKDE$ is a non-parametric estimate for the probability density of observed events, it has no closed form.


The second component represents the anti-correlation between $M_*$ and \textcolor{black}{$x_1$} as shown in Fig. 4 of \citet{Smith2020}. Based on the fact that few events have been observed with $M_*<10^{10}$  and negative \textcolor{black}{$x_1$}, we implement the following model from \citet{Vincenzi2021}:


\begin{eqnarray}
 P_{\mathrm{Ia}}^*(x_1, M_*) \propto 
  \begin{cases}
    e^{-x_1^2}, & \text{for } x_1<0 \; \mathrm{and} \; M_*<10^{10} \\
     1, & \text{for } x_1>0 \; \mathrm{and} \; M_*<10^{10}\\ 
     1, & \text{for } \forall x_1 \; \mathrm{and} \; M_*>10^{10}\\   
  \end{cases}
\end{eqnarray}

\subsection{Peculiar SNe~Ia}
In addition to `normal' SNe~Ia, our simulated sample includes two classes of peculiar transients: SN2002cx-like SNe \citep[SNe~Iax;][]{2013Foley_SNIax}, which are underluminous and red at peak light relative to their Branch-normal counterparts; and SN1991bg-like SNe~Ia, which are more rapidly evolving \citep{1992Filippenko_91bg}. Although observed samples are small, there are indications that SNe~Iax occur almost exclusively in late-type, star forming galaxies \citep{2017Jha_Iax, 2020Takaro_SNIax} and SNe~91bg-like preferentially occur in early-type, passive galaxies \citep{2005vandenBergh_91bg, 2020Hakobyan_PecSNeIa}. Accordingly, \cite{Vincenzi2021} set the SN~Iax \textcolor{black}{probability} to the SN~Ia \textcolor{black}{probability} in active hosts and 0 in passive hosts; conversely, they define the SN~91bg-like \textcolor{black}{probability} as equal to the SN~Ia \textcolor{black}{probability} in passive hosts and 0 in active hosts. We adopt a similar approach, and assume that the peculiar SNe~Ia originate from similar progenitors as normal SNe~Ia, but that their hosts represent opposite ends of a continuum in galaxy age and morphology. However, we exponentially taper the \textcolor{black}{probability} of SNe~Iax in passive galaxies and 91bg-like events in star-forming galaxies. This is done in order to avoid encoding an unrealistically discrete boundary in our catalogue, and to account for the possibility that a sub-population of these events does occur in hosts distinct from those previously discovered. 

As in \citet{Sullivan2006} and \citet{Vincenzi2021}, we define a threshold between active and passive galaxies at specific star formation rate (sSFR) values of $\log_{10}(\mathrm{sSFR/Gyr^{-1}})=-11.5$; active galaxies lie above this threshold and passive galaxies lie below it. For SNe~Iax, our combined \textcolor{black}{probability} function adopts the normal SN~Ia \textcolor{black}{probability} (without \textcolor{black}{$x_1$} dependence) for active galaxies and tapers the \textcolor{black}{probability} below the passive/active threshold as follows:
\thickmuskip=0mu
\begin{eqnarray}
  P_{\mathrm{Iax}}(M_*, \mathrm{SFR}) \propto 
  \begin{cases}
    P_{\mathrm{Ia}}(M_*, \mathrm{SFR}), & \text{for } x_s \geq -11.5 \\
      P_{\mathrm{Ia}}(M_*, \mathrm{SFR}) \times e^{(x_s+11.5)}, & \textrm{otherwise,}\\ 
  \end{cases}
\end{eqnarray}

\textcolor{black}{where $x_s = \rm{log_{10}(sSFR/Gyr^{-1})}$}. The SN~91bg-like \textcolor{black}{probability} does the opposite, tapering the \textcolor{black}{probability} symmetrically above the passive/active threshold:
\thickmuskip=0mu
\begin{eqnarray}
 P_{\mathrm{91bg}}(M_*, \mathrm{SFR}) \propto 
  \begin{cases}
    P_{\mathrm{Ia}}(M_*, \mathrm{SFR}), & \text{for } x_s < -11.5 \\
     P_{\mathrm{Ia}}(M_*, \mathrm{SFR}) \times e^{-(x_s+11.5)} , & \textrm{otherwise.}\\
  \end{cases}
\end{eqnarray}

Because $P_{\textrm{Ia}}$ is \textit{not} symmetric about $x_s$ (it increases with galaxy mass and star formation), we note that this formalism results in a higher probability for 91bg-like events to occur in star-forming host galaxies than SNe~Iax to occur in passive host galaxies. \catname\; contains $\sim0.1$\% of SNe~Iax in passive hosts and $>25$\% of SNe~91bg-like in active hosts. This implementation is motivated by previous work \citep[e.g.,][]{2020Hakobyan_PecSNeIa} showing a larger tail of 91bg-like hosts toward spiral-type morphologies than of SNe~Iax toward elliptical galaxies.

\subsection{Types II, Ib, and Ic Supernovae (SNe~II, SNe~Ib/c)}\label{subsec:SNII/Ibc}
Type~II, Type~Ib, and Type~Ic are three subclasses of core-collapse SNe. Type II events are characterized by strong Hydrogen emission lines, indicating that an outer Hydrogen shell was present in the progenitor star. Types Ib and Ic lack Hydrogen features, indicating that their outer shell was stripped from the progenitor star prior to explosion. Type Ic also lacks Helium features, indicating that the Helium envelope was also lost. 
Although SNe~IIb are typically grouped with these `stripped-envelope' events, they exhibit weak hydrogen features at early times. As the explosion progresses, the hydrogen emission quickly fades and the light curves begin to resemble those of SNe~Ib. 
Because of their short-lived progenitor systems (typically $<$50 Myr), core-collapse explosions are predominantly observed in galaxies that are actively forming massive stars. The \textcolor{black}{probability} of these events occurring in passive galaxies, in contrast, is suppressed. Active and passive galaxies can be distinguished by their sSFR, defined as the star formation rate of the galaxy per unit of stellar mass. 
 
The \textcolor{black}{probability} of these classes of core-collapse supernovae are set to \citep{Vincenzi2021}:


\begin{eqnarray}
 P_{\mathrm{II}}(M_*, \mathrm{SFR}) \propto 
  \begin{cases}
    (M_*/M_{\odot})^{0.16}, & \text{for } x_s \geq -11.5 \\
    0, & \text{otherwise}\
  \end{cases}
\end{eqnarray}

\begin{eqnarray}
 P_{\mathrm{Ib/c}}(M_*, \mathrm{SFR}) \propto 
  \begin{cases}
    (M_*/M_{\odot})^{0.36}, & \text{for } x_s \geq -11.5 \\
    0, & \text{otherwise}\
  \end{cases}
\end{eqnarray}
Additionally, we distinguish the hosts of SNe~Ic-BL events from the lower-energy class of SNe~Ic explosions through the properties of their host galaxies. To do so, we embed a further correlation with host galaxy metallicity. 
Additional details concerning this correlation are described in Sec.~\ref{subsec:SNIc-BL}.

\subsection{Type~Ic-BL Supernovae (SNe~Ic-BL)}\label{subsec:SNIc-BL}
It is hypothesized that long-duration gamma ray bursts (lGRBs, lasting longer than $\sim$2 seconds) are produced by relativistic jets originating in an SN explosion and interacting with surrounding circumstellar material \citep{2021Roy_LGRBs}. Broad-Lined SNe Ic (SNe~Ic-BL) are the only SNe that have been unambiguously associated with lGRBs, and these explosions have been found to occur in galaxies of lower average metallicity than the less-energetic SNe~Ic \citep{2020Modjaz_IcBL}. A thorough understanding of the host-galaxy correlations of SNe~Ic-BL could permit a reconstruction of the physical conditions that give rise to relativistic jets in SNe.  
To simulate a distinction between SNe~Ic and SNe~Ic-BL hosts, the \texttt{WGTMAP} probabilities for both SNe Ic and SNe Ic-BL are constructed as the product of two terms. 
The first term embeds the preference for core-collapse events to occur in active galaxies, 
which is parameterized by equation 6. 
For the second term, we consider the metallicity of each galaxy and
model the distribution of SN~Ic and SN~Ic-BL host galaxy metallicities as a set of logistic 
distributions:

\begin{eqnarray}
 P_{\mathrm{Ic}}^{\mathrm{MZR}} & \propto  &  f(9(Z-8.9))
   \\
P_{\mathrm{IcBL}}^{\mathrm{MZR}} & \propto &  f(10(Z-8.5))
\end{eqnarray}
%
\textcolor{black}{where $Z$ is the metallicity of the galaxy measured as $\textrm{log}_{10}(12 + O/H)$ and estimated using the modified Mass-Metallicity Relation (MZR) given by Eq.~2 in \citet{2010Mannucci_FMR} (galaxy metallicity is not included in the CosmoDC2 catalogue)}. The functions $f$ are Gaussians which roughly reproduce both the probability density functions and the cumulative distributions of Ic and Ic-BL hosts shown in 
Fig.~5 of \citet{2020Modjaz_IcBL}.

We note that the SN~Ic and Ic-BL light curve models are photometrically similar, and therefore these classes provide a unique opportunity to test the value of host galaxy information in classification.

\subsection{Type~I Superluminous Supernovae (SLSNe-I)}\label{subsec:SLSN}
Superluminous SNe (SLSNe) are unusually bright explosions potentially powered by exotic engines \citep[e.g., pair instability or magnetar spin-down;][]{2015Kozyreva_PISN, 2017Nicholl_MagnetarSLSN}. Type-I SLSNe are hydrogen-poor and have been found to occur predominantly in low-mass, metal-poor, and highly star-forming galaxies \citep{Lunnan2014, 2016Perley_SLSNe, Schulze2018}. 
We model their rate based on the observations of $\sim50$ SLSN-I hosts from \citet{Lunnan2014,2016Perley_SLSNe,Schulze2018}.  
All studies report SLSN-I hosts with specific star formation rates greater than $10^{-10}$ yr$^{-1}$; 
thus, we model the SLSN-I probability to be exponentially suppressed below this threshold.

Based on the observed distribution of host galaxy masses and star formation rates, we further suppress the \textcolor{black}{probability} of \textcolor{black}{events in galaxies that are both massive and highly star-forming:}

\begin{equation}
 P_{\mathrm{SLSNI}} \propto 
  \begin{cases}
   e^{(x_s+10)}, & \text{for } x_s<-10,\; x_M<10\; \text{ and }\; x_f<-0.5 \\
   e^{-(x_M-3)}, & \text{for } x_M>10\; \text{ and }\; x_f>-0.5 \\
    1 & \text{otherwise}\,
  \end{cases}
\end{equation}
\textcolor{black}{where $x_M = \rm{log_{10}(M_*/M_{\odot})}$, \textcolor{black}{$x_s = \rm{log_{10}(\mathrm{sSFR}/Gyr^{-1})}$}, and $x_f=\mathrm{log_{10}(\mathrm{SFR}/Gyr^{-1})}$}. We normalize these \textcolor{black}{probabilities} by eye such that hosts in the `otherwise' category all have an equal probability of being selected, hosts beyond the defined boundaries are increasingly less likely to be selected, and the discontinuity at the boundaries is minimized.


\subsection{Using short-duration gamma-ray bursts as a proxy for kilonova correlations}
\label{subsec:KN}
The discovery of gravitational-wave event GW170817 by the LIGO and Virgo collaborations, followed by short-duration ($\lesssim2$ sec) Gamma-Ray Burst (sGRB) GRB170817A a few seconds later and the optical kilonova (KN) AT2017gfo several hours later, confirmed the association of these signals during a likely neutron star - neutron star merger \citep{2017Abbott_BNSMerger,2017Abbott_GRB170817, Goldstein2017}. The multi-band photometry of AT2017gfo, an event which took place in the nearby ($z\approx0.00972$) galaxy NGC~4993, provided direct evidence for compact mergers as a site where r-process elements are forged \citep{Drout2017}. Nevertheless, many questions remain. Although the colour of the emission reveals information on the fraction of lanthanides and actinides produced from the neutron-rich ejecta, more events are needed to understand the relationship between an event's emission and its precise nucleosynthetic yield. Further, the differences between KNe generated from the merging of a neutron star-black hole system and those originating in a neutron star-neutron star collision remain unknown. Though there have been no observations of KNe from a neutron star-black hole system yet, numerical simulations show differences between these types of KNe \citep{Kawaguchi_KNeNumericalSim,2014Tanaka_anotherKNeNumSim,2019Shibata_neutronStarBinaries,2021Bulla_polarizedKNe}. The strength of the subsequent electromagnetic signal also encodes information regarding the neutron-star equation of state \citep{2017Bauswein_NSEoS,2021Li_NICER,2021Raaijmakers_NICER}.
Because our knowledge of these systems starts and ends at this prototypical event, owing to the large sky localization area of gravitational wave observations -- especially when events are observed only in some of the existing detectors -- and the low intrinsic event rate, additional KN discoveries in upcoming surveys would vastly improve our understanding of these systems.

Because our aim is to improve the ability to accurately identify a diversity of KN events in upcoming survey streams, we do not want to limit our KN host-galaxy model solely to galaxies matching the properties of NGC~4993 (the host of AT2017gfo). 
Indeed, whereas NGC~4993 is a quiescent galaxy \citep{2017Levan_GW170817}, sGRBs have been discovered in a variety of both star-forming and quiescent systems \citep{2006Prochaska_sGRBHosts,2009Berger_sGRBHosts}. 
Based on recent evidence that these signatures share a common astrophysical site \citep{2017Abbott_BNSMerger}, for this work, we assume that sGRBs and KNe inhabit host galaxies with similar properties. This assumption has the added benefit that GRBs are regularly observed across the full sky \citep{DAvanzo_GRBs}, and as a result we have amassed hundreds of GRB detections to date. 

We begin constructing the KN \texttt{WGTMAP} by retrieving the positions of \textcolor{black}{reasonably well-localized sGRBs (uncertainty radius $r_{e} < 0.01^{\circ}$ and $t_{90} \lesssim 2$ s)} from the NASA Fermi GBM Burst catalogue \citep{2020vonKienlin_GRBCat,2014Gruber_GRBCat,2014vonKienlin_GRBCat,2016Bhat_GRBCat} \textcolor{black}{and from the GRB Host Studies catalogue \citep{GHostScatalogue2006}.} 
Next, we use the \texttt{astro\_ghost}\footnote{\url{https://pypi.org/project/astro-ghost/}} software to identify the most-likely host galaxy associated with these events \textcolor{black}{and retrieve photometric data from these galaxies}. Whereas the original implementation of the package considered only northern-hemisphere galaxies within the Pan-STARRs source catalogue \citep{2020Flewelling_PanStarrs}, for this work we have extended the code to query the SkyMapper catalogue \citep{2019Onken_SkyMapper} for southern-hemisphere sources as well. After visually verifying each association, we arrive at a sample of \textcolor{black}{11} sGRB-hosts, one of which is NGC~4993. Apparent $gri$-band magnitudes are reported for all hosts except two, which lack \textcolor{black}{high-SNR brightness estimates in} the $r$ band. \textcolor{black}{We then convert apparent magnitudes to absolute magnitudes. We adopt the galaxies' spectroscopic redshifts as reported in the NASA/IPAC Extragalactic Database\footnote{\url{https://ned.ipac.caltech.edu/}} where possible, and use a photo-$z$ estimator\footnote{\url{https://github.com/awe2/easy_photoz/tree/main}} for the galaxies without a reported redshift.} We generate a kernel density estimate (KDE) to approximate the \textcolor{black}{joint} distribution of \textcolor{black}{absolute} magnitudes in $gri$ for these hosts using the available data. \textcolor{black}{From this KDE and using a Gaussian kernel, we determine best-fit bandwidths of $0.93$, $0.62$, and $0.57$ for $gri$-band absolute magnitudes, respectively.}

We \textcolor{black}{then} sample from this KDE along a uniformly-spaced grid spanning the apparent magnitudes of a random sample of CosmoDC2 galaxies. This sampling forms the basis of our KN \texttt{WGTMAP}. 

To select hosts for our KN simulations, we initially applied this \texttt{WGTMAP} to the \texttt{HOSTLIB} of randomly-sampled cosmoDC2 galaxies. The resulting simulations exhibited an unrealistically small scatter in the host $r-i$ vs. $g-r$ colours. We identified this as a result of tight colour correlations in the \texttt{HOSTLIB} coming from the cosmoDC2 colour assignment, unrelated to the \texttt{WGTMAP} construction. To correct for this artifact, we shift the apparent magnitudes of cosmoDC2 galaxies in each LSST band by a random number less than or equal to the observational uncertainty in that band. This sample of magnitude-shifted cosmoDC2 galaxies is stored as a distinct \texttt{HOSTLIB} and used only for the KN class. The shifts are effective at smearing the tight colour relationship. The KN \texttt{WGTMAP} is unchanged.

\subsection{Active Galactic Nuclei}\label{subsec:AGN}
The emission from Active Galactic Nuclei (AGN) originates from black holes at the centers of host galaxies in one of various states of active accretion \citep{Lynden-Bell_1969}. Multiple theoretical and observational studies have lent support for the co-evolution of supermassive black-holes and the galaxies they inhabit \citep{2011Ellison_AGN, 2015Rosario_AGN}, making these bright transients powerful probes both for galaxy formation and black-hole evolution across cosmic time \citep{2007Alonso_AGN}. \citet{2003Kauffmann_AGN} found by analyzing a sample of 22,623 AGN within $z<0.3$ that these transients occur predominantly in massive ($>10^{10} \; M_{\odot}$) galaxies. The properties of an AGN, including its overall luminosity and the characteristic timescale of its optical variability, have also been linked to the luminosity and stellar mass of its host, respectively \citep{1980Heckman_AGN, 1997Ho_AGN, 2021Burke_AGN}. 



\textcolor{black}{To simulate the properties of AGN host galaxies, we model the properties of 2,204 AGN hosts from \cite{2020Stemo_AGNSample}. While not a volume-limited sample, the catalogue provides publicly-available stellar masses and star-formation rates for AGN hosts from SEDs derived using \textit{HST}, \textit{Spitzer}, and \textit{Chandra} data and spans $0.2 < z < 2.5$, nearly the same redshift range considered in this work.}

\textcolor{black}{As was done for SNe~Ia, we model the joint probability distribution $P_{\mathrm{AGN}}(M_*, \mathrm{SFR})$ by performing kernel density estimation on the observed sample, using a Gaussian kernel for both parameters and the best-fit bandwidth of 0.18 for both parameters.}

\subsection{Tidal Disruption Events}\label{subsec:TDEs}
In a Tidal Disruption Event (TDE), material from a star is stripped during close passage to a supermassive black hole \citep[SMBH;][]{2000Ayal_TDE,2009Strubbe_TDE}. The stripped material circularizes into a debris disk around the SMBH and accretes onto it, resulting in a luminous flare spanning the electromagnetic spectrum. This occurs when the separation between the star and the SMBH falls within the tidal disruption radius of the star and, in some cases, the star is completely destroyed. Early detection and characterization of these events facilitates photometric and spectroscopic follow-up that illuminates the dynamic accretion physics of SMBHs across a range of mass scales \citep[particularly for quiescent black holes;][]{1989Evans_TDE}. To date, however, only a few these events have been observed in detail. 
TDEs are found predominantly in post-starburst galaxies \citep{2017French_TDEs, 2020French_TDEs}, and these correlations can be incorporated into targeted searches in upcoming surveys \citep{2021Zabludoff}. We encode these correlations into a TDE \texttt{WGTMAP} by constructing the following functional approximation, which roughly reproduces the distribution of events as a function of host galaxy SFR and $M_*$ as shown in Fig.~3 of \citet{2020French_TDEs}. 

\begin{equation}
    P(M_*, SFR) \propto
\begin{cases}
         e^{-x_f},& \text{if } \mathrm{SFR} \geq 10^{-9} M_{\odot}\; \textrm{yr}^{-1}\\
         &\textrm{ and } M_{*} < 10^{11} M_{\odot}\\
         e^{-x_M},& \text{if } \mathrm{SFR} < 10^{-9} M_{*}\; \textrm{yr}^{-1} \\
         &\textrm{ and } M_{*} \geq 10^{11} M_{*}\\
    e^{-(x_f+x_M)},              & \text{otherwise.}
\end{cases}
\end{equation}
We use the \texttt{HOSTLIB} of randomly-selected cosmoDC2 galaxies for this class.

\begin{figure*}
    \centering
    \includegraphics[width=\linewidth]{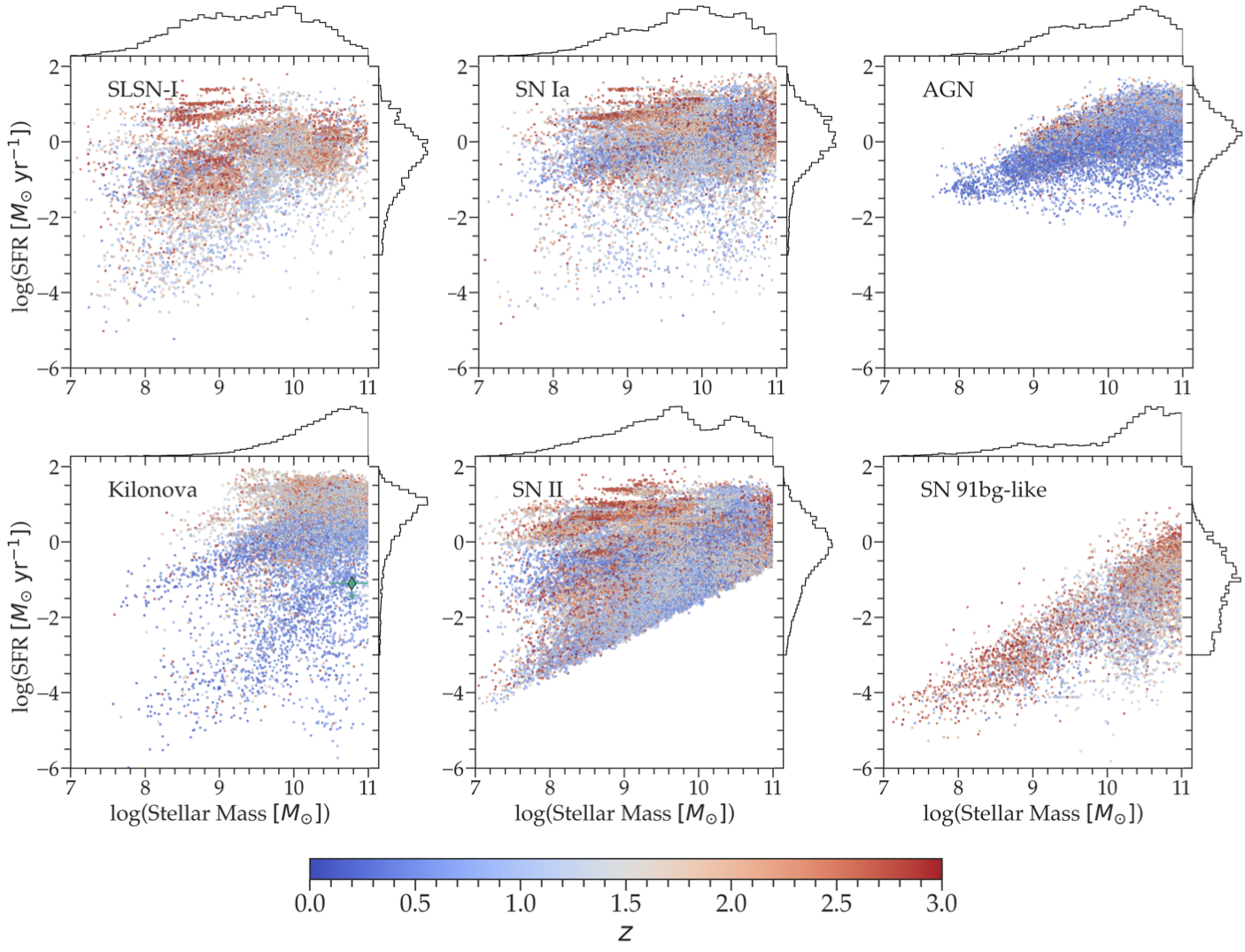}
    \caption{Distribution of derived host-galaxy properties for six simulated classes of transients. Galaxies are coloured by redshift (with red points for high-redshift galaxies).  Marginal distributions are given at the top and right of each sub-plot. \textcolor{black}{NGC~4993, the host of GW170817, is shown as the green diamond.} 
    }
    \label{fig:elasticc_SFRvsMsol}
\end{figure*}

\section{Offsets from the Host Centre}\label{sec:offsets}
\subsection{Transient offset model and validation}
\textcolor{black}{Within the \texttt{SNANA} simulation, we select the position of a transient relative to its host galaxy centre by randomly sampling from a probability distribution defined by the galaxy's surface brightness profile. This is motivated by prior studies that, averaging over class, SN surface density profiles roughly trace the surface brightness distribution of the spiral disks that host them \citep{2009Hakobyan_radialCCSNe}. Nevertheless, differences exist between the observed distributions of radial offsets for multiple transient classes. While we have aimed primarily at reproducing \textit{global} transient-host galaxy correlations in this work, we also integrate a basic local model for three classes of transients: AGN, KNe, and TDEs. When post-processing \catname\, to place events within a survey footprint, care should be taken to preserve the listed offsets.}

\textcolor{black}{Because AGN are nuclear transients by definition, we place them closer to their host centres than any other class. AGN are placed within a radius that encompasses 10\% of the total galaxy flux; within this radius, their distribution traces the surface brightness profile. We caution that AGN \textit{have} been detected that are more offset relative to their galaxy centres, but these are rare \citep[<5\%;][]{2014Comerford_OffsetAGN} and may be traced back to specific merger events, which is challenging to model with our existing framework.}

\textcolor{black}{TDEs, typically discovered as the optical signal from the tidal interaction between a star and a central supermassive black hole, are placed at small offsets in our simulations. However, the recent discovery of a TDE from a candidate intermediate-mass black hole (IMBH) in the nucleus of a dwarf galaxy \citep{Angus2022} suggests that future observations may reveal TDEs from IMBHs offset from the nucleus of massive hosts \citep[see][for a discussion of wandering IMBHs]{Bellovary2019}. To allow for this possibility, we model the TDE offsets slightly more broadly than AGN, with probability profiles truncated beyond the radius that encompasses 30\% of the host galaxy flux. }

\textcolor{black}{Conversely to TDE and AGN, we place KNe at broad offsets: their positions are sampled from distributions in which the innermost 20\% of a host galaxy is truncated. This is done to encode the prediction that natal kicks prior to explosion eject KNe from the inner regions of their host galaxies \citep{2020Gompertz_Kilonovae}. }

\begin{figure}
\begin{center}
\includegraphics[width=\columnwidth]{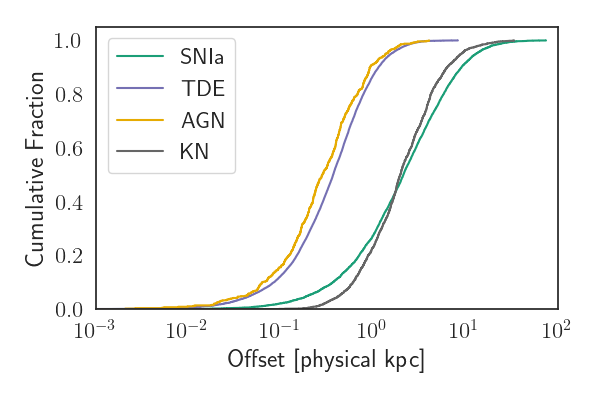}
\includegraphics[width=\columnwidth]{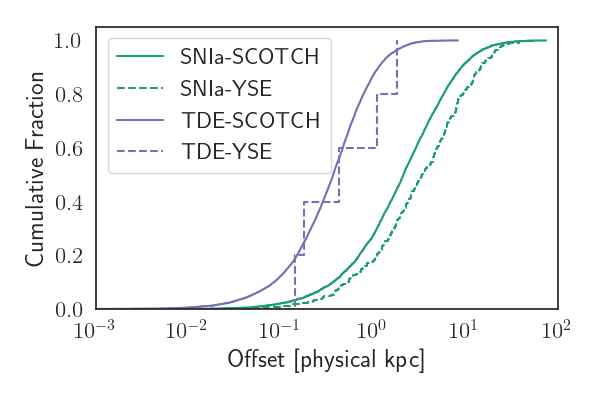}
\includegraphics[width=\columnwidth]{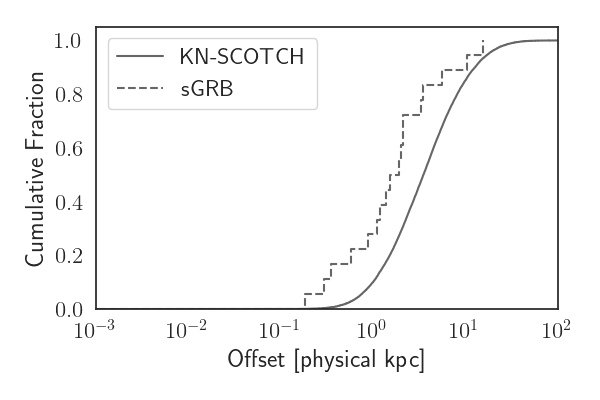}
\caption{\textcolor{black}{The CDF for transient-nucleus offsets for several classes in \catname\, only (top) and a \catname\, comparison to observational data from YSE (middle) and Swift data (bottom). \textbf{Top:} qualitative differences in our treatment of several classes are visible; AGN and TDE are closest to the center of galaxies while SN Ia and KN are distributed further out. KN are not allowed to exist within the centers of galaxies to simulate the kick effect. Here, \catname\; data is limited to the range $z<0.25$, where most YSE data lies, to facilitate a comparison in the middle plot. \textbf{Middle:} SCOTCH TDE and SNIa CDFs are duplicated and YSE data is added. The simulated TDE fit the YSE data well; the simulated SNIa are distributed closer to the centre than in YSE. \textbf{Bottom:} SCOTCH KN data is shown for $z<1$ and are distributed somewhat further than the distribution of sGRBs from Swift data, which also extend to $z\sim1$ \citep{Fong2013}.}}
\label{fig:separations}
\end{center}
\end{figure}

\textcolor{black}{We present the offsets for AGN, KNe, TDEs, and SNe~Ia in our catalogue in Fig.~\ref{fig:separations}, and compare the SNe~Ia and TDE offsets to observed SNe from the Young Supernova Experiment 
(YSE) Data Release 1 \citep{2022Aleo_YSEDR1} and the KN offsets to \textit{Swift} satellite sGRB data \citep{Gehrels2004} that was previously collected and analyzed in \cite{Fong2013, 2017Pan_Offsets}. The qualitative differences in our simulation clearly follow our model prescription: AGN lie closest to their host nucleus, TDE are slightly more offset, KN are evacuated from the central region of the host, and SNe~Ia (representative of the treatment for all SN classes) appear throughout the host galaxy with most within the inner few kpc. The observational data shows reasonable agreement to the simulations, despite the fact that our approach for offset prescriptions was approximate, and we did not explicitly aim to replicate observations. We leave a more detailed prescription for individual SN classes to future work.}

\subsection{Hostless events}
\textcolor{black}{Some transient events are reported without a host in \catname\;. This occurs when the distance of the transient from its host exceeds a particular threshold. After generating the position as described above, \texttt{SNANA} calculates the $d_{\mathrm{DLR}}$, the angular separation from the transient and nucleus (in arcseconds) normalized by the DLR, which is a measure of the galaxy radius in the direction of the transient. If $d_{\mathrm{DLR}}>4$, the host is rejected to simulate the fact that a realistic survey would be unlikely to match that transient with its host (specifically, the rate of mismatch within $d_{\mathrm{DLR}}<4$ is expected to be small, $\sim4\%$ according to \citealt{gupta2016host}). Hostless events due to $d_{\mathrm{DLR}}$ rejection do not occur for TDE or AGN due to their inner-galaxy placement; for all other classes, the rate of hostless events is at the few-percent level ($\sim 3-8\%$).}

\textcolor{black}{In \catname\,, these transients appear without host association in the final catalogue. A more realistic treatment would be to provide an incorrect host when another galaxy lies closer to the transient than the true host. However, because \catname\, removes all on-sky position information for hosts, including galaxy neighbors as false hosts in this manner is infeasible.}

\section{Validation of Host Galaxy Correlations}\label{sec:validation}
\begin{figure*}
\begin{center}
\includegraphics[width=\textwidth]{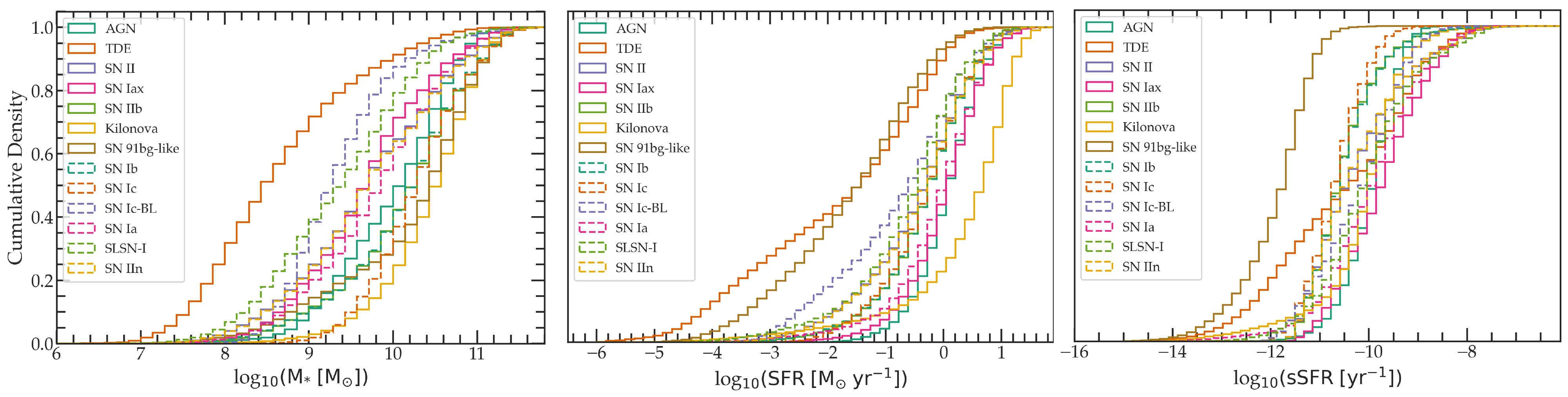}
\caption{CDFs in host galaxy stellar mass (left), star formation rate (\textcolor{black}{middle), and specific star formation rate (right)} for a representative sample of transient classes in the catalogue. \textcolor{black}{Differences between transient classes reflect a combination of \texttt{WGTMAP} and \texttt{HOSTLIB} effects.}}
\label{fig:class_cdfs}
\end{center}
\end{figure*}

To validate the \catname\; catalogue,
we first confirm that the correlations described in \textsection\ref{sec:host_correlations}
appear in the final sample as expected. Fig.~\ref{fig:elasticc_SFRvsMsol} displays the SFR vs. $M_*$ relationships for a representative sample of hosts associated with six transient classes. Each matches qualitative trends in host types: SN~Ia hosts are massive and highly star-forming, \textcolor{black}{SLSNe-I occur mainly in galaxies with high specific star formation rates \textcolor{black}{(and trace the cosmic star formation rate)}}, and SNe~91bg-like occur in passive, massive galaxies. The generated samples, which used \textcolor{black}{\texttt{WGTMAP} probabilities} from \citet{Vincenzi2021}, match the correlations therein, but extend to lower masses and star formation rates due to our use of CosmoDC2 versus their use of observed DES galaxies.
To more directly compare the host-galaxy derived properties for each class, we plot the CDFs for $M_{*}$ and SFR in Fig.~\ref{fig:class_cdfs}. 

\textcolor{black}{The CDFs show that, of all transients simulated in \catname\;, TDEs are simulated in the lowest-mass host galaxies,} followed by SLSNe-I. Simulated \textcolor{black}{KNe} are hosted by the highest-mass galaxies, followed by the host galaxies of stripped-envelope systems (particularly SNe~Ic). The star formation rate CDFs indicate an overabundance of TDEs in galaxies of low star formation rate, which is consistent with observations of them in post-starburst galaxies. SNe~91bg-like events in our simulation occur on average in the galaxies with the lowest star formation rates, a feature of our exponential suppression in active galaxies for this class. Simulated SNe~Iax and \textcolor{black}{KNe} occur in galaxies with the highest average star formation rate. For SNe~Iax, this reflects the exponential suppression of these events in passive hosts. 

\textcolor{black}{Consistent with the fact that we simulated nearly all core-collapse supernovae (besides SNe~Ic-BL) with the same \texttt{WGTMAP} and \texttt{HOSTLIB}, the CDFs do not show any clear differences between the host galaxy SFR of different core-collapse subclasses. The exception is the Ic-BL class, which appears more often in the simulation in galaxies of low mean SFR due to the different implemented \texttt{WGTMAP}}. 
We note that, in our models, we can distinguish between statistical samples of SNe~Ic and SNe~Ic-BL by the mass of their host galaxy, as SNe~Ic-BL occur in lower-metallicity host galaxies \citep{2020Modjaz_IcBL} and therefore in lower-mass hosts on average. A detailed analysis of the typical observational uncertainties associated with these derived properties and how they might obfuscate these differences in \textcolor{black}{observed} data is beyond the scope of this work.

\begin{figure*}
    \centering
    \includegraphics[width=\linewidth]{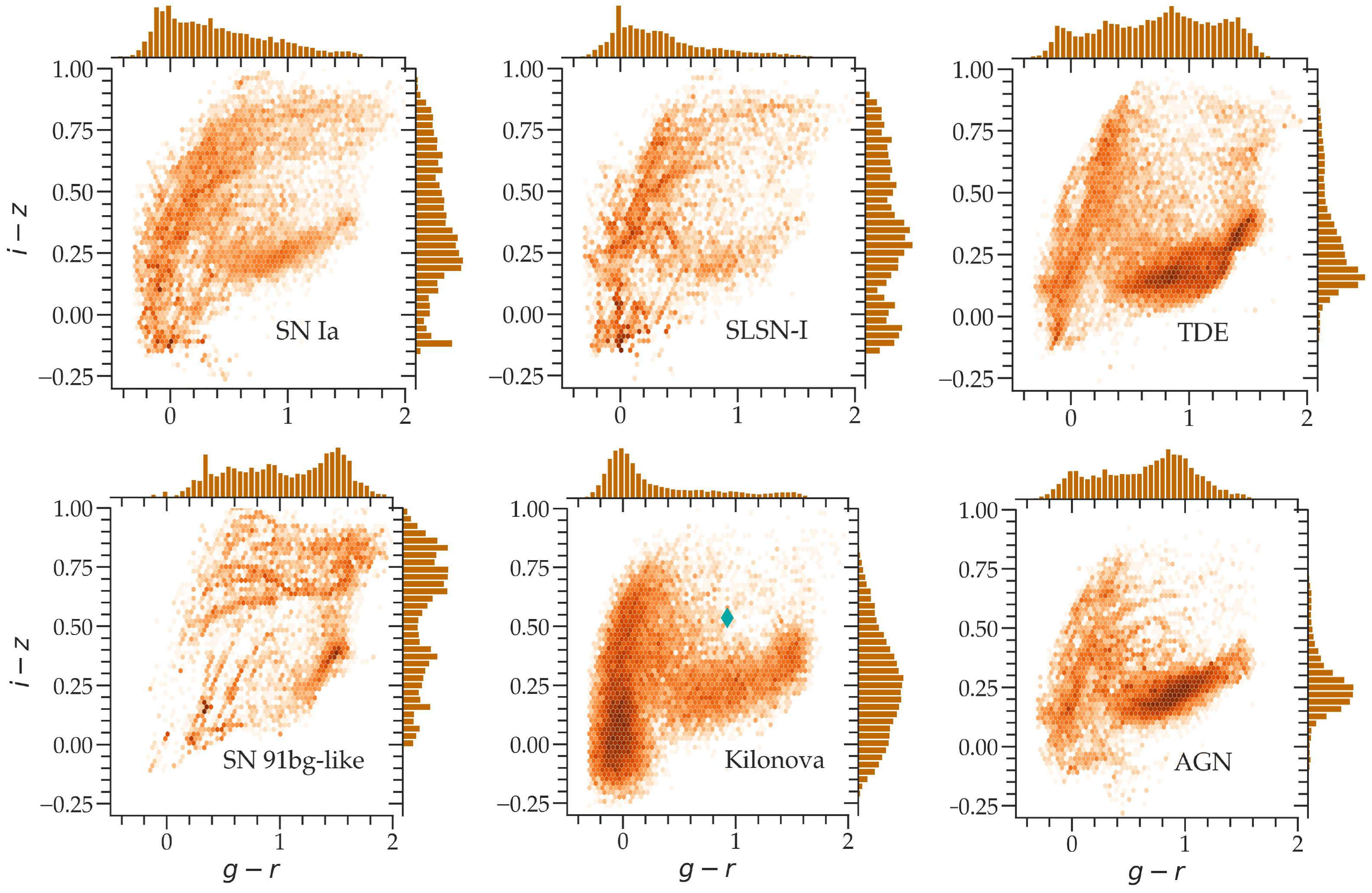}
    \caption{Density distribution of \textcolor{black}{observed} colours for simulated transient classes \textcolor{black}{(combining \texttt{WGTMAP} and \texttt{HOSTLIB} contributions)}. Marginal histograms are shown at top and at right of each sub-panel. The green diamond indicates NGC~4993, the host of the single kilonova observed. \textcolor{black}{Overdensities near $g-r = 0$ reflect remaining discrete tracks in CosmoDC2 galaxy distribution.}
    }
    \label{fig:elasticc_colorcolor}
\end{figure*}

Fig.~\ref{fig:elasticc_colorcolor} shows the distributions of host colours within each class. Redder galaxies appear to the upper-right and bluer to the lower-left. We did not directly encode colour correlations into the \texttt{WGTMAPS} for any of these classes; therefore, the differences between distributions stem from the initial GHOST matching in conjunction with the \texttt{WGTMAPs} in which transient \textcolor{black}{probabilities} are dependent on galaxy mass and star formation rate. \textcolor{black}{We caution that we have not converted galaxy colours to the rest-frame for Fig.~\ref{fig:elasticc_colorcolor}, and so the redshift-dependent rates for each class affect this comparison.}

Qualitatively, the \textcolor{black}{observed color} plots display many of the \textcolor{black}{intended comparative} attributes. Simulated SLSN-I host galaxies are bluer than other SN hosts, reproducing observations that show that these events occur in highly star-forming galaxies \citep{Schulze2018}. 
SN~91bg-like host galaxies are redder than normal SN~Ia, as intended. The KN hosts \textcolor{black}{are surprisingly blue; given the broad redshift distribution of simulated events (Fig.~\ref{fig:elasticc_colorcolor}), this may reflect the preference of observed sGRBs for late-type hosts \citep{2013Fong_sGRBs}. As GW170817 (diamond in Fig.~\ref{fig:elasticc_colorcolor}) occurred in a lenticular galaxy, its host galaxy is redder than the majority of the simulated sample.} 
The simulated $g-r$ distribution for TDE host galaxies is broader than all others, which is consistent with observations \citep{2020French_TDEs};
however, the $i-z$ span may be unrealistically restricted.

Next, we quantitatively compare the colours of SNe~Ia (including peculiar) hosts with the observed distributions reported in \citet[][ H20]{2020Hakobyan_PecSNeIa}. H20 examined $u-r$ colours (SDSS photometry) for a low-redshift sample and found statistically significant differences between the $<u-r>$ of `normal' SNe~Ia, 91bg-type, and Iax-type host galaxies. The authors reported that galaxies hosting SNe~Iax are bluer than normal, while SNe~91bg are redder than normal, 
with $\Delta <u-r>{\sim}0.4$ in each case. To investigate this trend in our own data, we calculate the average $u-r$ SDSS rest-frame colour among 10,000 random members of the SCOTCH host sample for the same three classes. Although H20 examined hosts with $z<0.04$, for the comparison we draw the SCOTCH host sample from the entire $z<3$ redshift range because our simulation methodology treats high-$z$ host selection equivalently to low-$z$. The results are shown in Table~\ref{tab:u-r_color_comparison}. Our average mean simulated host colour values are significantly bluer than those of the H20 sample for two of the three classes. \textcolor{black}{Our mean SN~Iax host colour is also bluer, but consistent within 2$\sigma$ with the H20 observations.} The blue trend may be a consequence of unrealistic galaxy colours in CosmoDC2, as demonstrated in Fig.~3 of \citet{dc2val}: the CosmoDC2 $u-g$ distribution is significantly bluer than observational data while $g-r$ is consistent with observations, suggesting that $u-r$ tends to be nonphysically blue. \textcolor{black}{The colour \textit{differences} between the Ia classes in SCOTCH are in the same direction as the observed differences and are statistically significant. The magnitudes of the differences are somewhat mismatched from observations.} \textcolor{black}{Although qualitatively imperfect,} the fact that SCOTCH reproduces the general trend in colour differences between the SN~Ia subtypes should enable catalogue users to test algorithms which include host information for classification. 

\begin{table*}
    \centering
    \begin{tabular}{c||cc||cc}
    \hline
        & \multicolumn{2}{c}{H20} & \multicolumn{2}{c}{This work} \\
         & $<u-r>$ & $\Delta$-norm & $<u-r>$ & $\Delta$-norm\\
         \hline
         \hline
        SN~Ia (normal) & $1.86 \pm 0.03$ & & $1.372 \pm 0.005$ & \\
        SN~Iax & $1.47 \pm 0.10$ & $-0.39 \pm 0.10$ & $1.255 \pm 0.006$ &$-0.117\pm0.008$ \\
        SN~Ia-91bg & $2.23 \pm 0.05$ & $0.37 \pm 0.06$ & $1.908 \pm 0.003$ & $0.536 \pm 0.007$\\
    \hline
    \end{tabular}
    \caption{A comparison of the mean rest-frame $u-r$ colour of observed SN~Ia host galaxies (all $z<0.04$) in \citealt{2020Hakobyan_PecSNeIa}~(H20) with the simulated catalogue (spanning all redshifts) presented in this work. Each reported uncertainty is the standard error of the mean. $\Delta$-norm is the difference between $<u-r>$ of the peculiar SN~Ia sample and the normal sample. \textcolor{black}{All simulated hosts are bluer than the observed hosts; SN~Ia and SN~Ia-91bg-like hosts are significantly bluer, while SN~Iax are consistent within 2$\sigma$.} The qualitative relationship between the classes (SN~Iax hosts are bluer, SN 91bg-like hosts are redder) is similar.}
    \label{tab:u-r_color_comparison}
\end{table*}

\begin{figure}
    \centering
    \includegraphics[width=\columnwidth]{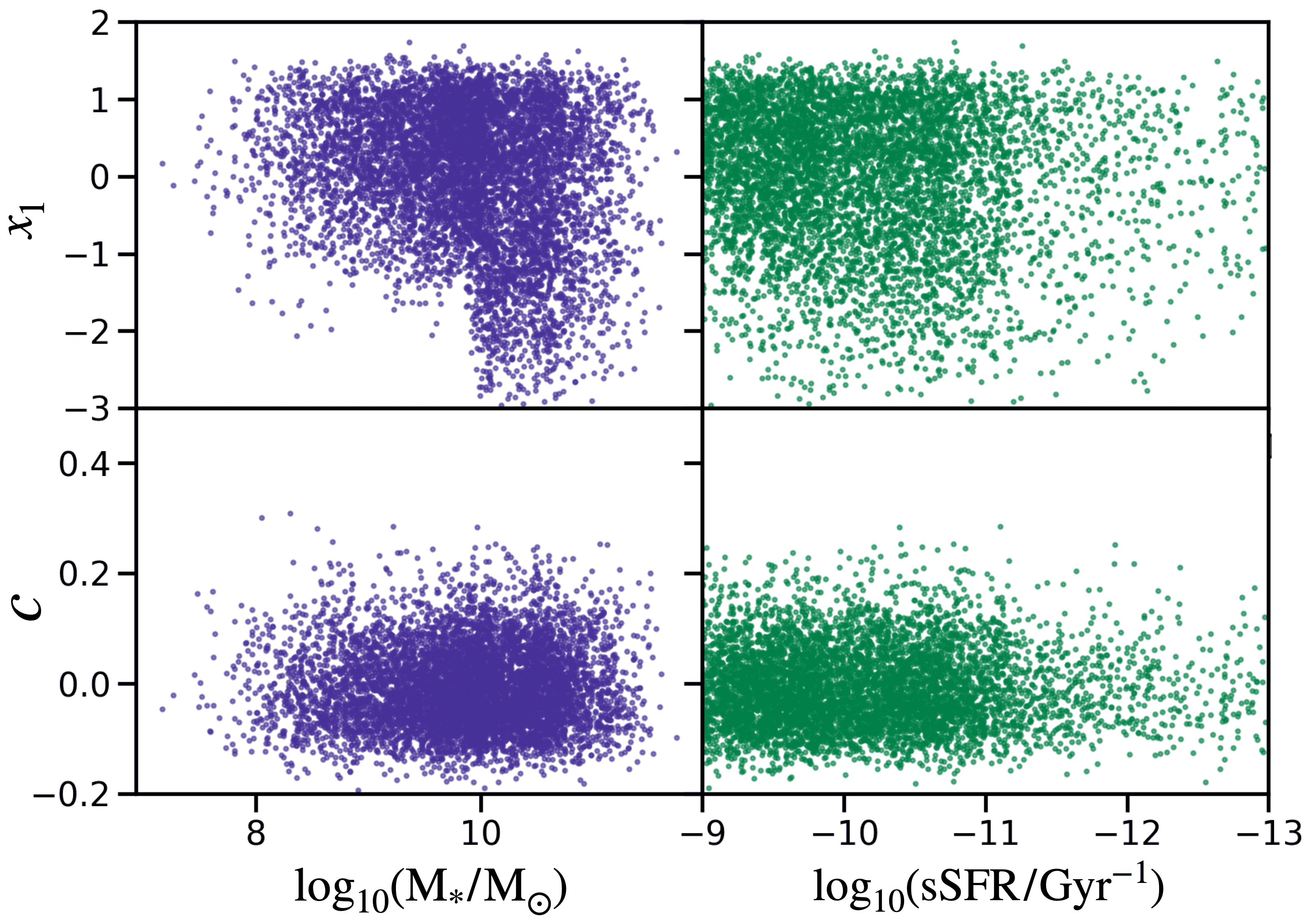}
    \caption{SALT2 \citep{SALT22007} fitted parameters \textcolor{black}{$x_1$} and $c$ as a function of host galaxy $M_*$ and sSFR. Our simulations reproduce the distribution of events observed in \protect\cite{Childress2013}.}
    \label{fig:SALT2}
\end{figure}

\textcolor{black}{We also aim to reproduce observed correlations between the photometric light curves of SNe~Ia and their host galaxies. These have been extensively studied in the literature due to the use of these events as cosmological distance indicators. We present the SALT2 \citep{SALT22007} fitted parameters \textcolor{black}{$x_1$} and $c$ (describing the stretch and colour of an SN~Ia light curve, respectively) as a function of the intrinsic properties of our associated hosts in Fig.~\ref{fig:SALT2}. We find similar distributions to those presented in \cite{Childress2013}, including a mass-step at $10^{10} \; M_{\odot}$ (this has been explicitly encoded in our simulations, so its presence is not surprising), and weak anti-correlations between \textcolor{black}{$x_1$} and both host galaxy $M_*$ and sSFR. We observe a few SALT2 $c$ values above $c=0.2$, and multiple are identified in \cite{Childress2013}; although these are anomalous, our model their presence indicates that our correlation model is able to reproduce both the central distribution and the tails of the observed data.}

\begin{figure*}
    \centering
    \includegraphics[width=0.7\linewidth]{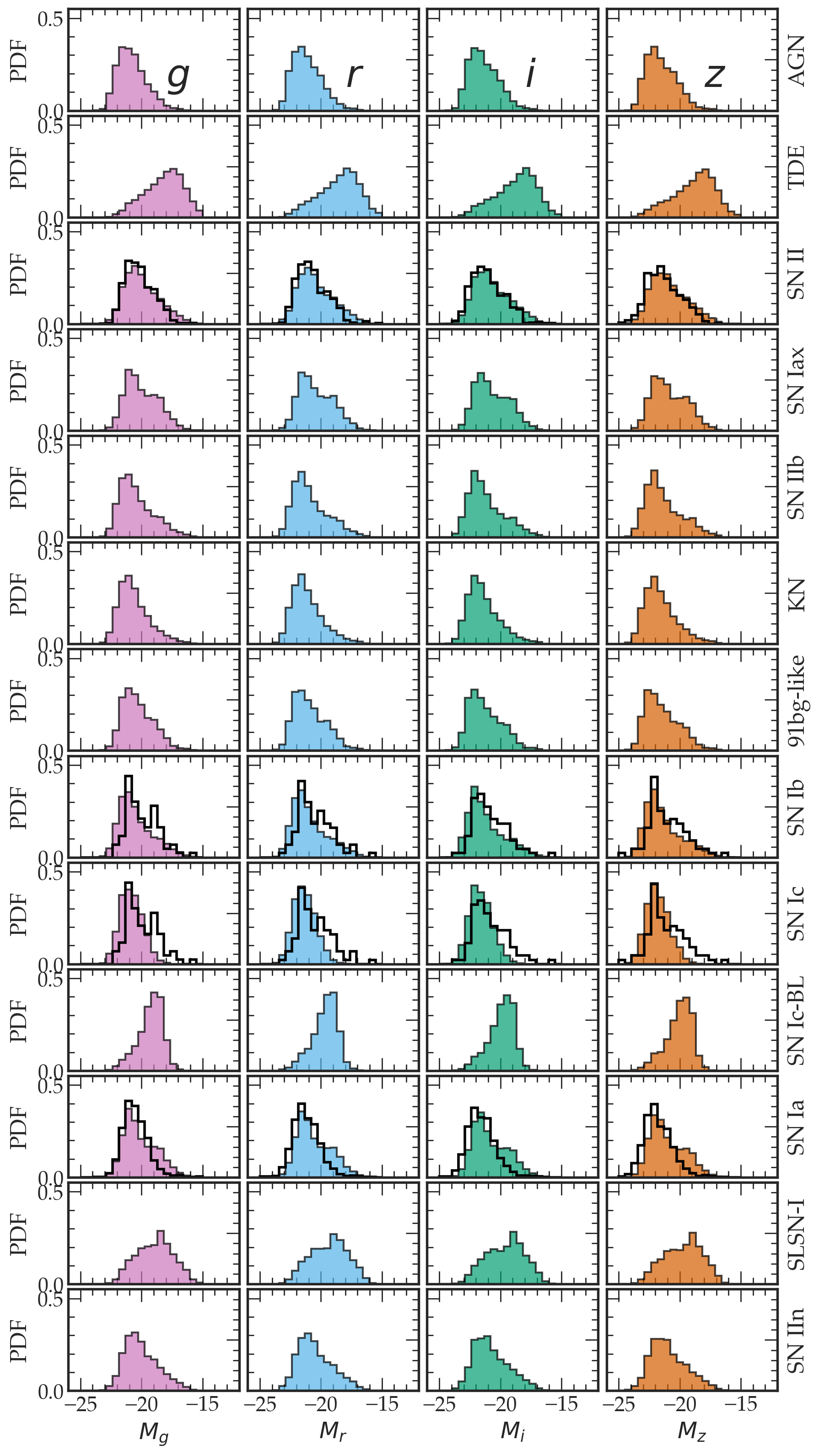}
    \caption{Distribution of host-galaxy absolute magnitudes for a representative sample of \textcolor{black}{SCOTCH} transients using both \texttt{HOSTLIB}s and \texttt{WGTMAP}s. \textcolor{black}{Host galaxy photometry from the GHOST catalogue for classes with statistical samples is shown as a black outline. Photometry is presented in the SDSS filter system.}}
    \label{fig:magnitudes}
\end{figure*}

We present \textcolor{black}{SDSS-$griz$ histograms} of the absolute magnitudes of host galaxies associated with each transient in Fig.~\ref{fig:magnitudes}. \textcolor{black}{GHOST data with $z>0.05$ are shown for several classes with black histograms; the cut is motivated by the fact that \catname\, has very few events below this redshift. The data reasonably fit the simulation.} The impact of our encoded correlations is readily apparent by the variation in median and skew between distributions. \textcolor{black}{However, we caution that the host magnitude differences between classes are not only due to encoded correlations, but also due to} two important redshift-dependent effects within \catname\;: the intrinsic rate of different transients across cosmological distances and the $r<28$ magnitude cut imposed prior to matching in order to cross-match with the 5-yr DC2 image catalogue. The confluence of these effects results in a Malmquist-like bias in the sample, with the distribution of host galaxies for a given class tending toward intrinsically brighter galaxies with increasing redshift (fainter hosts being unavailable to draw from the \texttt{HOSTLIB}s at those redshifts). For events occurring in galaxies ranging in brightness and observed only at low and intermediate redshifts (e.g., TDEs), we observe a realistic skew toward fainter galaxies that dominate the \texttt{HOSTLIB}s at these redshifts. For bright classes (e.g., SLSNe-I) and common ones spanning the full simulated redshift range (SNe~II), the majority of events will be simulated at high-redshift where this bias is strongest. As a result, the histograms in Fig.~\ref{fig:magnitudes} reveal a distribution for SNe~Ia, SNe~II, and SLSNe-I skewed toward bright hosts. \textcolor{black}{This effect is especially visible when examining absolute magnitude as a function of redshift for both GHOST data and \catname\, results (not shown): the distributions match well across the overlapping redshift range, but both the simulated and observed hosts taper rapidly from including a broad range of magnitudes at low-$z$ to a tighter, brighter range of hosts as $z$ increases.}

\begin{figure*}
    \centering
    \includegraphics[width=2\columnwidth, trim={3cm 3cm 3cm 3cm}, clip]{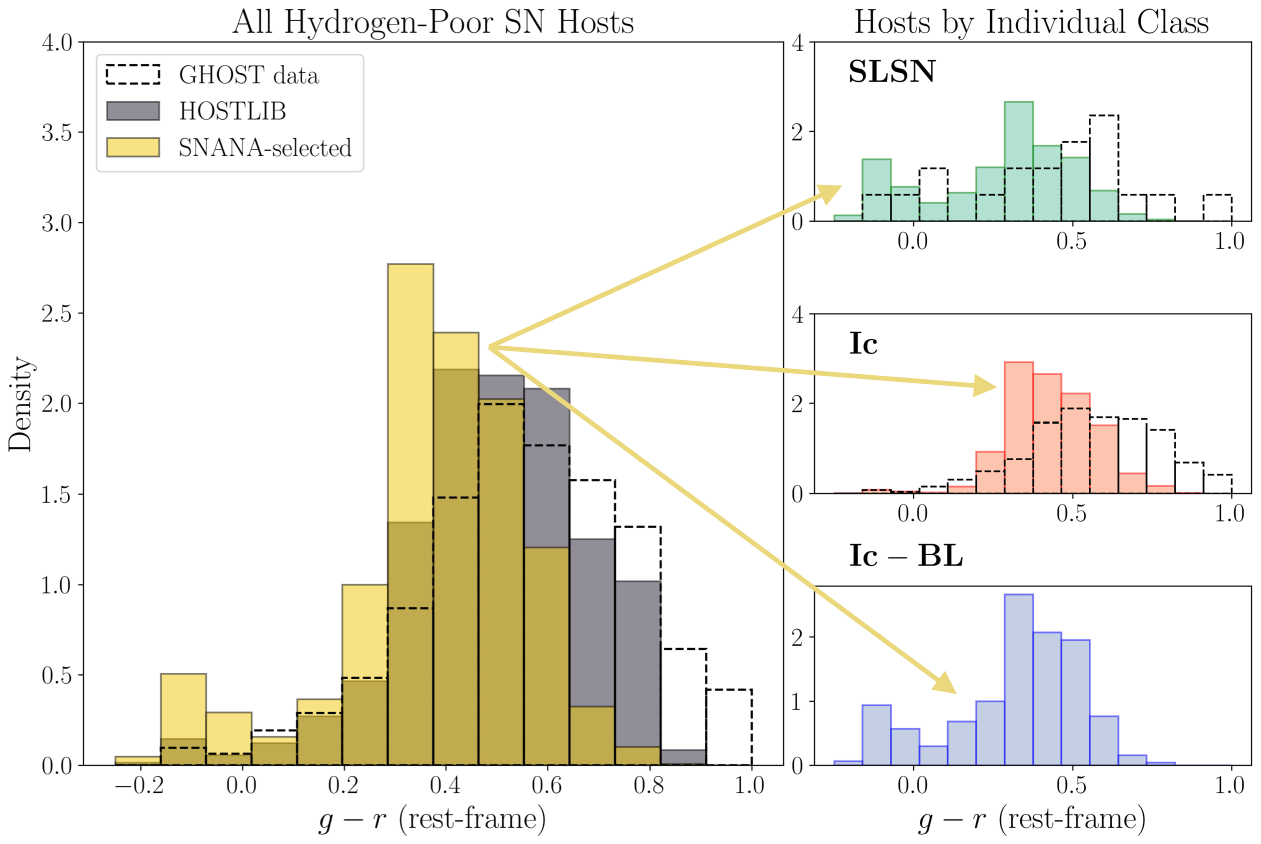}
    \caption{Histogram of $g-r$ colour for synthetic Hydrogen-poor SN hosts before and after running the SNANA simulation. \textbf{Left:} The \texttt{HOSTLIB} of all H-poor (Ib/c, SLSN, IIb) hosts (gray),  which was tailored to match the GHOST data (dashed line). The hosts selected after running the \texttt{SNANA} simulation for all H-poor classes with their respective \texttt{WGTMAP}s are shown in yellow. \textbf{Right:} SNANA-selected hosts of SLSN, Ic, and Ic-BL events. SLSN hosts are bluest, and Ic-BL hosts are bluer than the rest of the Ic class due to SFR and metallicity correlations. GHOST data are shown with a dashed outline, and we avoid plotting SNe~Ic-BL hosts in GHOST due to the small sample size.}
    \label{fig:all_steps}
\end{figure*}

We next compare the distributions of various properties of our final sample to GHOST to examine the combined effects of our two-step host selection process. Fig.~\ref{fig:all_steps} shows the galaxy colour distribution for the hydrogen-poor event \texttt{HOSTLIB} (consisting of SN~Ib, Ic, IIb, and SLSN host matches to GHOST) and the distribution of hosts selected by SNANA after \texttt{WGTMAPS} are applied. The \texttt{WGTMAPS} cause the overall $g-r$ distribution to become slightly bluer and more peaked than the GHOST data. Examining the colours for individual host classes reveals how the different WGTMAPs cause the distributions to stratify. The final SLSN-I host sample is bluer than the others, as encoded via SFR. The Ic hosts separate from the Ic-BL hosts in colour space due to the implemented metallicity dependence. Overall, the final combined host grouping remains similar to observed data, but the individual host classes within that group are subtly different, demonstrating the combined power of the \texttt{HOSTLIB} + \texttt{WGTMAP} formalism.

\begin{figure*}
\begin{center}
    \includegraphics[width=.48\textwidth]{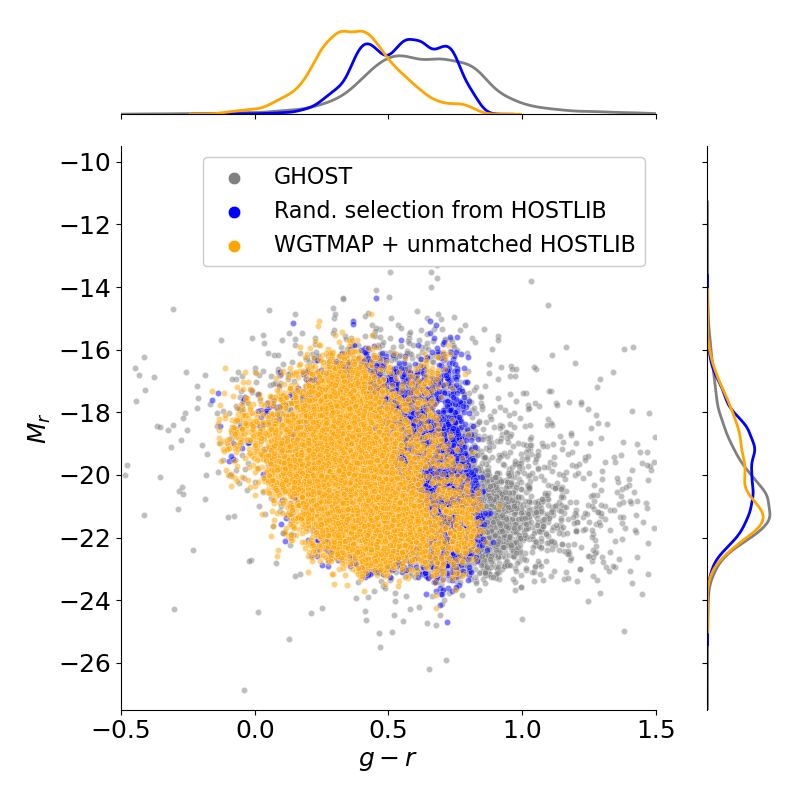}
\includegraphics[width=.48\textwidth]{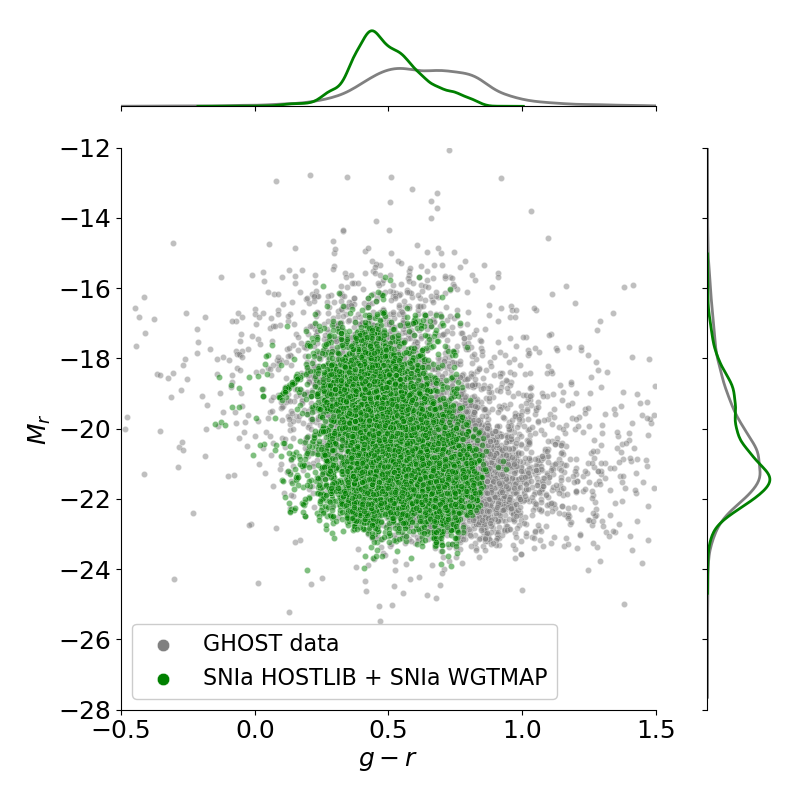}
\includegraphics[width=.48\textwidth]{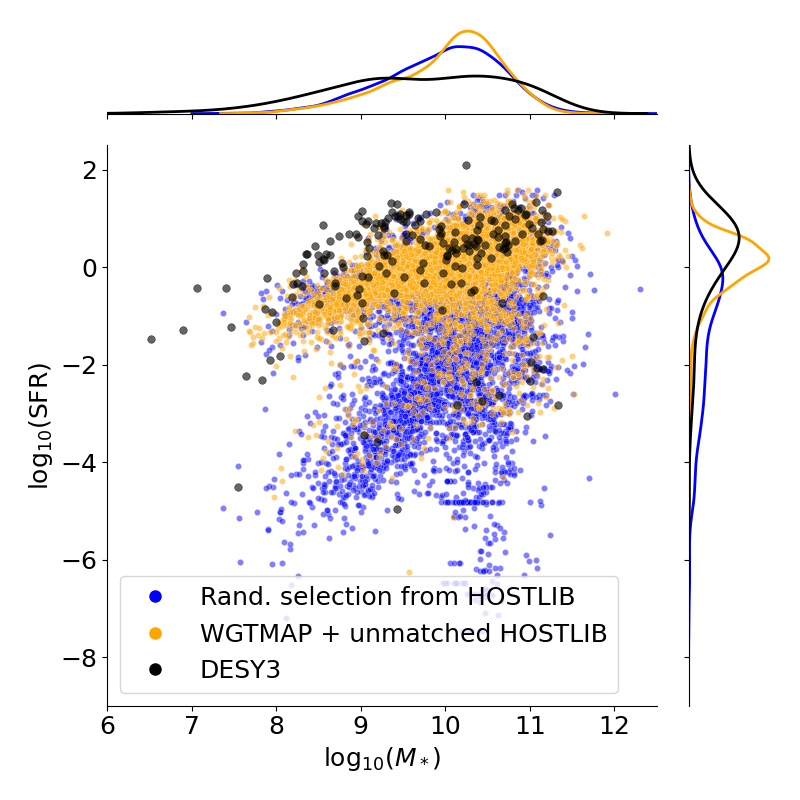}
\includegraphics[width=.48\textwidth]{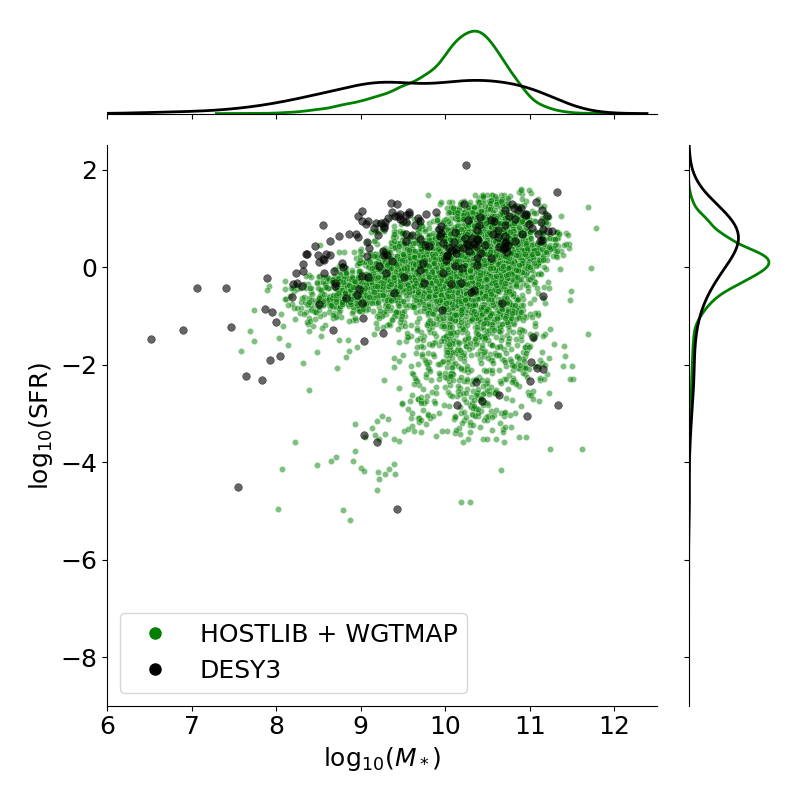}
\caption{Simulated SN~Ia hosts compared to observational data for simulations using either a \texttt{HOSTLIB} \textit{or} a \texttt{WGTMAP} (left), and simulations using both (right). Marginal kernel density estimates are shown at top and at right in each panel. In the upper left panel, the \texttt{HOSTLIB}-only simulation (blue) decently reproduces the colour-magnitude spread of SN~Ia hosts in GHOST (grey), \textcolor{black}{although it is worse in $g-r$ than in $M_r$ due to the CosmoDC2 colour limitations. The blue points, however,} are a poor match to DES data from \citet{Smith2020} (black) in SFR-stellar mass (SFR-$M_*$) space (lower left), \textcolor{black}{especially in the SFR distribution}. A random sampling of CosmoDC2 galaxies which is associated with \texttt{SNANA} transients via a realistic \texttt{WGTMAP} (yellow) yields more realistic SFR-$M_*$ results but does worse in colour-magnitude space. \textcolor{black}{Notably, the yellow distribution in $g-r$ is in strong disagreement with GHOST.} Simulations using both \texttt{HOSTLIB} and \texttt{WGTMAP} (green at right) result in decent matches to both data sets in both parameter spaces. \textcolor{black}{In all plots, the redshift range is limited to $z<0.85$ to enable fairer comparison with observational data. In the lower plots the simulations are further limited to $r<23.8$, the 10-$\sigma$ magnitude limit for the DES3YR sample, for optimal comparison.}}
\label{fig:hostlib_wgtmap_combos}
\end{center}
\end{figure*}

To further demonstrate the utility of including both a \texttt{HOSTLIB} and \texttt{WGTMAP} in our simulation, we compare our final results for SN~Ia host galaxies with (1) a sample simulated \textit{only} using the GHOST-matched \texttt{HOSTLIB} and no \texttt{WGTMAP}, and (2) a sample simulated using a randomly-selected (unmatched) CosmoDC2 \texttt{HOSTLIB} and the SN~Ia \texttt{WGTMAP}. Figure~\ref{fig:hostlib_wgtmap_combos} demonstrates how the implementation of either a \texttt{HOSTLIB} \textit{or} a \texttt{WGTMAP} results in reasonable distributions for some, but not other, host properties. Implementing both the \texttt{HOSTLIB} and \texttt{WGTMAP} leads to \textcolor{black}{an improved} match with data from both 
DES \citep{Smith2020} and GHOST in multidimensional parameter space.


\section{Catalogue Access and Intended Use}\label{sec:cat_access}
The \catname\; catalogues \textcolor{black}{are presented in HDF5 format and} can be accessed on Zenodo at \url{https://zenodo.org/record/7563623#.Y8_2my9h2yA}. There are two separate catalogues, one for transients and one for hosts, which can be cross-matched to access the complete associated host-transient information. The properties of the transient and host catalogue are listed in Tables~\ref{tbl:transient_cat_schema} and \ref{tbl:host_cat_schema}, respectively. Each transient in the catalogue is assigned a unique transient ID (\texttt{TID}) and each host a unique galaxy ID (\texttt{GID}), which are reported in both catalogues to enable cross-matching. If an event is considered hostless using our pipeline, we include an associated entry in the host catalogue with a unique \texttt{GID}, but populate the host properties with filler values ($999$).

We additionally provide \texttt{dc2ID}, the original ID of each galaxy within the CosmoDC2 catalogue, so that users can query the CosmoDC2 \textcolor{black}{image} simulations\footnote{\texttt{CosmoDC2\_v1.1.4\_image} at \url{https://data.lsstdesc.org/doc/cosmodc2}.} for additional properties. For example, although SCOTCH erases the realistic position information of host galaxies due to the necessary repeat use of \texttt{HOSTLIB} galaxies (see \textsection \ref{sec:SNANA}), if a user is interested in the original position and large-scale environment of a galaxy, e.g. to check whether the galaxy was a cluster member, they may retrieve the relevant information from CosmoDC2. A user in need of host galaxy SED information may access the original CosmoDC2 SED, find the amount that the galaxy's redshift has changed from CosmoDC2 to SCOTCH, and red- or blue-shift the SED accordingly. However, we warn that for \textit{kilonova} hosts in particular, the magnitude and colours of SCOTCH  hosts are \textit{not} directly traceable to the CosmoDC2 native values, as the kilonova host galaxy magnitudes in SCOTCH have been shifted (\textsection\ref{subsec:KN}). In the case of hostless events, the \texttt{dc2ID} of the galaxy which was rejected by the $d_{\mathrm{DLR}}$ threshold is included in order to enable fetching information about the rejected host.

With top-of-the-galaxy observing conditions, users aiming to simulate optical surveys with similar passbands to LSST have a straightforward path to forecasting survey data.  The path would include converting the magnitudes accordingly, subselecting events according to expected survey densities, updating transient and host positions to span the survey footprint, and implementing Galactic extinction and atmospheric noise. The ideally-sampled light curves can be degraded by adding noise and by sub-sampling as needed. \textcolor{black}{We provide a series of basic tutorials for querying the catalogue and post-processing the data in Appendix~B, available in the online supplementary materials.}

\textcolor{black}{Due to the increased volumes at higher redshifts in the universe and the fact that we simulate a fixed number of transient events per class, the full $z<3$ catalogue is sparsely populated at very low redshifts. Users wishing to access a larger amount of low-$z$ data may alternatively make use of an additional $z$-limited catalog that we release. In this catalog, we restrict the maximum $z$ to 0.8 for each transient class so that the same number of events $N$ as the full catalogue, as given in Table~\ref{tbl:class-models-numbers}, are generated at lower redshifts. The $z<3$ catalogue is named \texttt{SCOTCH\_Z3} and the additional catalogue is named \texttt{SCOTCH\_ZLIM}}.

For users wishing to simulate significantly different surveys, we also provide the \texttt{HOSTLIBs} and \texttt{WGTMAPs} in the Zenodo release for re-running the \texttt{SNANA} simulation code.

\begin{table*}
    \centering
    \begin{tabular}{c|l}
      \hline
        Variable & Summary \\
    \hline
    \hline
         \texttt{TID} & Transient ID\\
         $z$ & True Redshift (of host and transient) \\
        \texttt{GID} & Host-galaxy ID  \\
         MJD & Array of Modified Julian Dates of light curve observations (only spacing is meaningful) \\
         $m_{<\mathrm{band}>}$ & Apparent brightness in LSST $ugrizY$ bands (AB magnitudes) \\
         Class & Transient class\\
         Model & Simulation model from Table \ref{tbl:class-models-numbers} \\
         Cadence & Time-spacing of light curve samples [days] \\
         
         $\rm RA_{\rm off}$ & Transient offset from host nucleus in R.A. [$''$]\\
         $\rm \delta{\rm off}$ & Transient offset from host nucleus in Dec. [$''$]\\
         Sep & Total great-circle distance between transient and host nucleus [$''$]\\ 
    \hline
    \end{tabular}
    \caption{Summary of the transient information in the \catname\; catalogue.}
    \label{tbl:transient_cat_schema}
\end{table*}

\begin{table*}
    \centering
    \begin{tabular}{c|l}
      \hline
        Variable & Summary \\
    \hline
    \hline
         \texttt{GID} & Host-galaxy ID  \\
         \texttt{TID} & Transient ID\\
         \texttt{dc2ID} & ID of corresponding CosmoDC2 galaxy  \\
         $m_{<\mathrm{band}>}$ & Apparent AB magnitudes in LSST $ugrizY$ bands \\
         $\sigma_{m,<\mathrm{band}>}$ & 10-year estimated apparent AB magnitude errors for LSST \\
         $e$ & Shear ellipticity $(1-q)/(1+q)$, where $q$ is the axis ratio\\
         $R_d$ & Disk half-light radius in physical kpc\\
         $R_s$ & Spheroid half-light radius in physical kpc\\
         $\log{(M_*)}$ & Log stellar mass $\left[M_{\odot}\right]$\\
         $\log{(\mathrm{SFR})}$ & Log star formation rate
         $\left[M_{\odot}/\textrm{yr}\right]$ \\
         $n_i$ & Sersic index for $i=[0,1]$; $n_0=1$ (exponential disk) and $n_1=4$ (deVaucouleurs bulge)  \\
         $w_i$ & Weight of $i=[0,1]$ Sersic components (bulge and disk) \\
         $a_i$ & Major-axis half-light size $\left[''\right]$ for $i=[0,1]$ Sersic components \\
         $b_i$ & Minor-axis half-light size $\left[''\right]$ for $i=[0,1]$ Sersic components \\
         $e_i$ & Ellipticity of $i=[0,1]$ Sersic components \\
         $e_{\rm tot}$ & Luminosity-weighted sum of bulge and disk ellipticities \\
         $a_{\mathrm{rot}}$ & Rotation angle of major axis with respect to the +RA coordinate [$^{\circ}$] \\

    \hline
    \end{tabular}
    \caption{Summary of the host galaxy information in the \catname\; catalogue. \textcolor{black}{We note that the original CosmoDC2 catalog contains inaccurate ellipticities due to a bug; in SCOTCH, we report the corrected ellipticity and minor axis values following the corrections from \url{https://github.com/LSSTDESC/gcr-catalogs/blob/ellipticity\_bug\_fix/GCRCatalogs/cosmodc2.py}}.}
    \label{tbl:host_cat_schema}
\end{table*}

\section{Conclusions and Discussion}\label{sec:conclusions}

We have presented our methodology for constructing the \catname\; catalogue, a dataset consisting of multiple classes of simulated transients and their host galaxies. The final catalogue includes 5 million light curves for 10 distinct classes of supernovae and 3 additional transient classes, as well as the observed and derived properties of their host galaxies and \textcolor{black}{physically-motivated} offsets between the galaxy and transient positions. The cadence of these simulated light curves is class-specific in order to resolve rapidly-evolving features while limiting total data volumes. This catalogue significantly advances the framework for simulating the time-domain sky initially constructed for PLAsTiCC, which included minimal host-galaxy information. 

This catalogue can be used for a broad range of scientific studies. First, teams developing transient classification algorithms for upcoming surveys can test the value of large samples of realistic host information. This added contextual information may help classify transients earlier in their light curves, facilitating rapid follow-up of interesting objects. Host information can also increase the precision of classifications, of particular value for collecting large uncontaminated samples of Type~Ia supernovae for precision dark energy measurements. However, users should take caution when training classifiers on this catalogue, as overconfidence in the correlations represented therein could limit classifiers from finding transients in hosts not described by extant samples. We have not attempted to include any anomalies in the catalogue to mitigate this problem. Additionally, this catalogue is missing some known transient classes, and does not include any mock new classes of transients. Classifier trained for anomaly detection should be able to flag a totally unknown class of event in an upcoming survey, and this will be an important aspect of upcoming surveys.

Beyond classification, this catalogue can be used to forecast future population-level studies of transients -- e.g., for progenitor studies or cosmological applications -- using upcoming survey data. Besides the LSST-specific passbands, the catalogue is not tailored to fit the observational limitations of any particular survey. \textcolor{black}{With empirically-derived transient rates extrapolated} to $z\;=\;3$ and host galaxy magnitudes as faint as $r\;=\;28$, users can post-process the catalogue to reflect the expected footprint, depth, and observational biases of planned surveys. The catalogue can then be used to generate a proof-of-concept for science ideas, forecast results, estimate expected biases, and test new methods to mitigate them.

Although not the focus of this work, the validated correlations between simulated SNe~Ia and their host galaxies are encouraging. Because we have linked the properties of observed SN~Ia host galaxies to synthetic galaxies in CosmoDC2, myriad intrinsic host properties are available. \textcolor{black}{The authors encourage future catalogue users} to evaluate the use of synthetic SNe~Ia for cosmological analyses, and explore additional correlations between Hubble Residuals and the CosmoDC2 properties of host galaxies.

The realism of this simulated catalogue is limited by approximations made during the simulation pipeline and the limitations of pre-existing transient-host galaxy catalogues. \textcolor{black}{Here we re-emphasize several points and discuss opportunities for future improvements.}

\textcolor{black}{Generally, we allow the catalogue to reflect the trends seen in the existing data, so that developers of classification algorithms can better incorporate host galaxy information where available. The classes with the sparsest currently available data will have the least accurate host property distributions. Thus the transient-host correlations learned by a classifier trained on \catname\, may be revealed to be too strong, or even fully inaccurate, with future data influxes.}

\textcolor{black}{Future updates to the catalog should seek to augment the small samples with recent data. For example, the Transient Host Exchange \citep{2022Qin_THEx} could be ingested into our simulation framework to incorporate additional data into the existing \catname\, classes as well as expand to more classes. 
During the writing of this work, the first long-duration GRB (lGRB) was discovered with evidence for coincident kilonova emission \citep{2022Rastinejad_lGRB_KN}. This result underscores our limited understanding of these events. Should additional future events be discovered with joint lGRB-KN emission, upcoming iterations of host galaxy simulations should include the properties of lGRB host galaxies as possible KN sites. The authors will consider incorporating this in future \catname\, releases given sufficient demand. Finally, while preparing this manuscript, \cite{2022OConnor_sGRBhosts} also investigated a series of high-redshift ($z<2$) sGRB host galaxies and found a potential increase in the number of hostless events with redshift. Given the small number of observed sGRBs considered in this study, additional work should be devoted to identifying and integrating correlations among high-$z$ samples.}


In addition, a major approximation in \catname\, is that the hosts and transients are \textit{not} distributed in a realistic large-scale-structure. Although the relative offsets between transients and their host galaxies are \textcolor{black}{physically motivated}, we have avoided placing these systems on the sky so that the catalogue may be of general use for generating synthetic transient catalogues for both northern and southern hemisphere surveys by the user. Therefore, the \catname\; catalogue cannot be used to forecast cosmological applications of surveys that rely on realistic positions or velocities within the large-scale structure. \textcolor{black}{Previous studies have found SNe to be highly clustered within the large-scale structure of the universe \citep{2022Tsaprazi_LSS}, so at the very least, retaining the locations of the synthetic SNe hosts within their simulated cosmic web environment would be a major improvement in realism. To retain large-scale structure information while matching empirically-determined transient rates, it is necessary to have very large \texttt{HOSTLIB}s. The current computational time required to read the large files into memory many times during the simulation is a bottleneck. In addition, retaining realistic position information would limit the sky area to the 440 sq. deg. currently simulated in CosmoDC2, which is less helpful for creating survey mocks. The future release of 5,000 sq. deg. of CosmoDC2 will present an opportunity for expansion, while other larger-volume synthetic galaxy catalogues could also be explored \citep[e.g., Buzzard,][]{deRose2019}.}

Another possible extension of this work is to implement correlations between a transient and its location within a galaxy, as multiple studies have revealed local transient correlations (\textcolor{black}{for example, with the amount of host-galaxy extinction for SNe~Ia; see \citealt{2008MNRASHolwerdaDust,2015Holwerda_SNIaDust,2021Popovic_Dust}).} \textcolor{black}{For some classes (e.g., TDEs), this local information may be more valuable in early classification than global host galaxy information \citep{2022Gomez_Fleet}.}  

In addition to these catalogue-level corrections, our methods could be expanded to produce simulated images. This is a non-trivial issue because the transient image must be overlaid on the host galaxy image as well as surrounded by a realistic background which accurately reflects the galaxy density at the location in the simulation (which would require the preservation of large-scale structure information). \textcolor{black}{Machine learning-based algorithms such as Generative Adversarial Networks (GAN) have shown significant promise in generating galaxy images that are both survey-specific and realistic \citep[][among others]{2019Fussell_GAN, 2020Dia_GalaxyImages, 2021Buncher_Survey2Survey}, and conditional networks allow for the encoding of model-specific parameters in generated images \citep{2016DES_Generative}. A generative model conditioned on transient and host galaxy correlations, along with local galaxy density, could be used to rapidly generate realistic images for the systems in the SCOTCH sample.} This framework would represent a significant milestone toward complete end-to-end validation of survey infrastructure. 

\section*{Acknowledgements}\label{sec:acknowledgements}
We thank the referee for their invaluable comments, which improved the clarity of the manuscript and quality of the simulations.

This paper has undergone internal review in the LSST Dark Energy Science Collaboration. The internal reviewers were Eve Kovacs, Christopher Frohmaier, Benjamin Rose, and Douglas Clowe. The authors thank the internal reviewers for their thorough reviews of this work, which have improved both its content and presentation.

The DESC acknowledges ongoing support from the Institut National de Physique Nucl\'eaire et de Physique des Particules in France; the  Science \& Technology Facilities Council in the United Kingdom; and the Department of Energy, the National Science Foundation, and the LSST Corporation in the United States.  DESC uses resources of the IN2P3 Computing Center (CC-IN2P3--Lyon/Villeurbanne - France) funded by the Centre National de la Recherche Scientifique; the National Energy Research Scientific Computing Center, a DOE Office of Science User  Facility supported by the Office of Science of the U.S.\ Department of Energy under Contract No.\ DE-AC02-05CH11231; STFC DiRAC HPC Facilities,  funded by UK B\textcolor{black}{E}IS National E-infrastructure capital grants; and the UK 
particle physics grid, supported by the GridPP Collaboration.  This work was performed in part under DOE Contract DE-AC02-76SF00515.

M.L. acknowledges the support of the Natural Sciences and Engineering Research Council of Canada (NSERC) [PGSD3 - 559296 - 2021]. M.L. also partially conducted this research while supported by the Queen Elizabeth II/Graduate Scholarships in Science and Technology (QEII-GSST).

A.G. is supported by the National Science Foundation Graduate Research Fellowship Program under Grant No.~DGE–1746047. A.G. further acknowledges funding from the Center for Astrophysical Surveys Fellowship at UIUC/NCSA and the Illinois Distinguished Fellowship. 

The Pan-STARRS1 Surveys (PS1) and the PS1 public science archive have been made possible through contributions by the Institute for Astronomy, the University of Hawaii, the Pan-STARRS Project Office, the Max-Planck Society and its participating institutes, the Max Planck Institute for Astronomy, Heidelberg and the Max Planck Institute for Extraterrestrial Physics, Garching, The Johns Hopkins University, Durham University, the University of Edinburgh, the Queen's University Belfast, the Harvard-Smithsonian Center for Astrophysics, the Las Cumbres Observatory Global Telescope Network Incorporated, the National Central University of Taiwan, the Space Telescope Science Institute, the National Aeronautics and Space Administration under Grant No. NNX08AR22G issued through the Planetary Science Division of the NASA Science Mission Directorate, the National Science Foundation Grant No.\ AST-1238877, the University of Maryland, Eotvos Lorand University (ELTE), the Los Alamos National Laboratory, and the Gordon and Betty Moore Foundation.

The Dunlap Institute is funded through an endowment established by the David Dunlap family and the University of Toronto. 
The authors at the University of Toronto acknowledge that the land on which the University of Toronto is built is the traditional territory of the Haudenosaunee, and most recently, the territory of the Mississaugas of the New Credit First Nation. 
R.H is a CIFAR Azrieli Global Scholar in the Gravity and the Extreme Universe Program, and acknowledges funding from the Alfred P. Sloan Foundation, the National Science and Engineering Research Council of Canada and the Connaught Fund.

This research made use of the ``K-corrections calculator'' service available at \url{http://kcor.sai.msu.ru/}. 

This work was partially enabled by funding from the UCL Cosmoparticle Initiative.

L.S. is supported by the Data Science in Multi-Messenger Astrophysics program at the University of Minnesota.

J.F.C is supported by the U.S. Department of Energy, Office of Science, under Award DE-SC0011665.

A.I.M. acknowledges support from the Max Planck Society and the Alexander von Humboldt Foundation in the framework of the Max Planck-Humboldt Research Award endowed by the Federal Ministry of Education and Research.

This project uses models developed for the Photometric LSST Astronomical Time-series Classification Challenge. PLAsTiCC was a joint project between the Dark Energy Science Collaboration and Transient and Variable Stars Science Collaboration, and was supported by a LSST Enabling Science Grant, as well as funding from Kaggle, a subsidiary of Google.

\textcolor{black}{This research has made use of the GHostS database (www.grbhosts.org), which is partly funded by Spitzer/NASA grant RSA Agreement No. 1287913.}
\par
\noindent Author contributions are listed below. \\
M.~Lokken: Data curation, SCOTCH simulations and validation, writing - editing, tutorial creation\\
A.~Gagliano: Data curation, SCOTCH simulations and validation, writing - editing, tutorial creation\\
R.~Hlo\v{z}ek: supervision, PLAsTiCC data curation and validation, project administration for PLAsTiCC, writing - editing \\
G.~Narayan: supervision, ELAsTiCC data curation and validation, project administration for ELAsTiCC, writing - editing \\
R.~Kessler: supervision, updating code for SNANA simulations, setting up SNANA simulations, writing - editing\\
J.~F.~Crenshaw: Supervising use of PZFlow, writing - editing\\
L.~Salo: Data curation of kilonova hosts, writing - editing \\
C.~Alves: SNANA simulation setup, writing - editing  \\
D.~Chatterjee: Updating kilonova light curve model, writing - editing\\
M.~Vincenzi: consulting on WGTMAP and HOSTLIB setup, updating core collapse SN light curve model, writing - editing\\
A.I.~Malz: Discussions and infrastructure for follow-up work\\
\section*{Data Availability}

The simulated data are available at \url{https://zenodo.org/record/7563623#.Y8_2my9h2yA} \citep{scotch2022}. Any updates after the publication of this manuscript will be given a unique Zenodo identifier and link, and accompanied by text describing the changes. The most up-to-date version can always be accessed using \url{https://doi.org/10.5281/zenodo.6601210}. All code used for this project is publicly visible at \url{https://github.com/LSSTDESC/transient-host-sims}. Users are encouraged to report any catalog problems or bugs in the code via GitHub Issues or by e-mail to the authors. The supplementary materials to this paper, available online, include a description of the nearest-neighbors matching from Sec.~\ref{subsec:CosmoDC2_subsample} (Appendix~A) and a description of the catalogue walk-through tutorials (Appendix~B).



\bibliographystyle{mnras}
\bibliography{sn_host_simcat}

\appendix

\section{Approximate Nearest-Neighbours Matching of GHOST and CosmoDC2 Galaxies with ANNOY}\label{sec:ANNappendix}

In LSH methods,  a hashing function $h$ is constructed that maps a high-dimensionality observation to a scalar such that, for any three observations $a$, $b$, and $c$ for which some distance metric $s$ is defined in the original space, the following criterion is met: \begin{equation}
    d(a, b) < d(b, c)  \Longrightarrow  h(a,b) < h(b,c)
\end{equation}
In this example, the two observations $a$ and $b$ with a small relative distance will retain a small relative distance (high relative similarity) after being mapped to a scalar value. Because $d \neq h$, the nearest neighbours returned by LSH methods may vary from those returned by a brute-force distance calculation between all observations; but the `approximate' solution is achieved in significantly less time.

A multitude of methods exist both for constructing the hashing function $h$ and the distance measure $d$ \citep[see][for additional details]{2014Wang_Hashing}. In the random projection method employed by \texttt{ANNOY}, a catalogue of observations is provided and two are selected at random. A maximum-margin hyperplane is then constructed that separates the two observations \citep{charikar2002similarity}. This process is repeated and a decision tree is constructed with each node in the tree corresponding to a hyperplane. The hashing function encodes the side of each separating hyperplane on which the observations fall. Similar to random forest algorithms, multiple decision trees are constructed in \texttt{ANNOY}. The nearest-neighbours returned by an observation query to the space are the indexed observations that fall on the same side of all separating hyperplanes in a combination of decision trees. 

In contrast with previous implementations of LSH methods, \texttt{ANNOY} decouples the construction of the indexed space and the querying of the space for individual vectors. Because the indexed space can be saved as a static file and then loaded into shared memory, matching can be achieved in parallel to dramatically reduce matching time. 

We construct our indexed space from the following properties of 30M galaxies in CosmoDC2: absolute rest-frame $r$- and $i$-band magnitudes, $g-r$ and $i-z$ colours, and redshift. We normalize all properties by the StandardScaler algorithm, after which the distribution for each parameter has a mean of 0 and a variance of 1. Finally, we weight the redshift by 5\%, enabling a wider distribution of redshifts in the matched sample than the low-$z$ GHOST data. We select the same parameters from observed galaxies within the GHOST catalogue, normalize them, and query the indexed space for $k$ approximate nearest-neighbours in CosmoDC2. For a transient class $i$, we set $k = 2\times10^6/N_{i, \rm Ghost}$, where $N_{i, \rm Ghost}$ is the number of galaxies of class $i$ in the GHOST catalogue. This allows our matched sample to contain $\sim$2M galaxies for each class.

\section{Using the SCOTCH Catalogue}\label{sec:walkthroughs}
\begin{figure*}
    \centering
    \includegraphics[width=2\columnwidth, trim={2cm 4cm 2cm 4cm},clip]{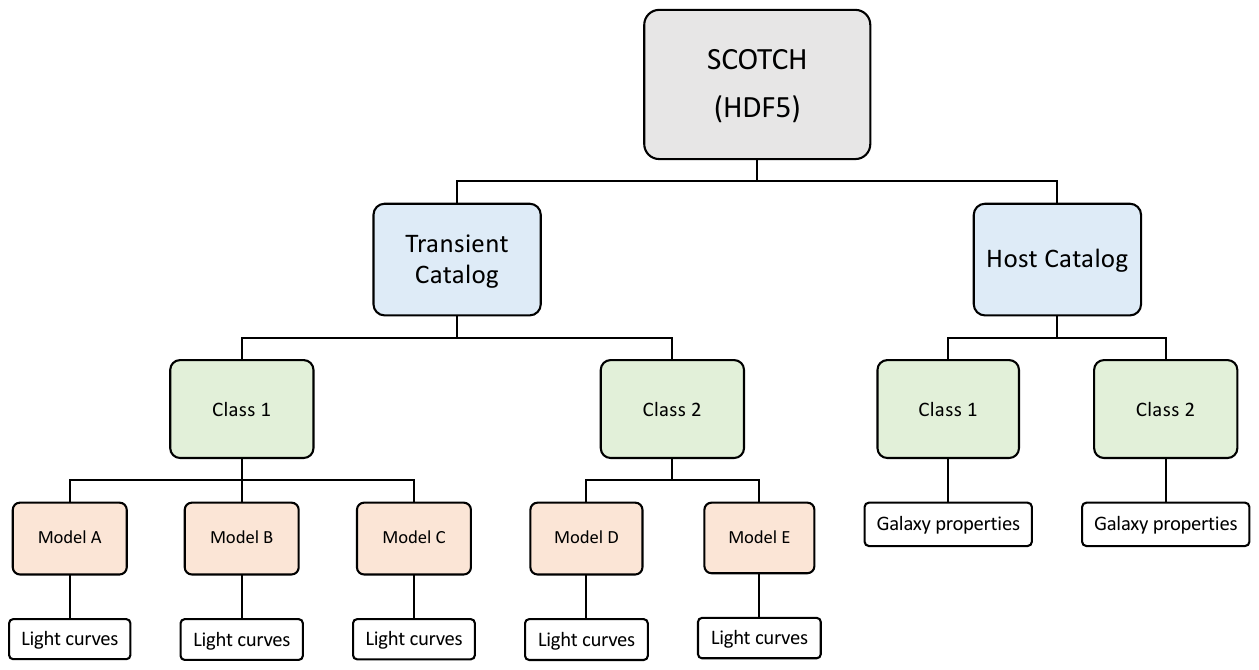}
    \caption{\textcolor{black}{A schematic diagram showing the hierarchical organization of the SCOTCH catalogue. Only two transient classes are shown for illustrative purposes. The Hierarchical Data Format (HDF5) allows users to read only the desired components of the catalogue into memory.} }
    \label{fig:hdf5_diagram}
\end{figure*}
\textcolor{black}{We have released the SCOTCH catalogue as an HDF5 file consisting of a transient table and a host galaxy table. We present an overview of the SCOTCH catalogue hierarchy in Fig.~\ref{fig:hdf5_diagram}, and provide Python code for retrieving basic information about transients and host galaxies below. Full tutorials can be found in the repository associated with this release.\footnote{\url{https://github.com/LSSTDESC/transient-host-sims/blob/main/notebooks/SCOTCH_walkthroughs.ipynb}}}

\subsection{Properties of Transient Host Galaxies}
To load in the SCOTCH catalogue, we use the Python package \verb|h5py|\footnote{\url{https://www.h5py.org/}}:
\begin{verbatim}
>> scotch = h5py.File("./scotch.hdf5", "r")
>> print(scotch.keys())

<KeysViewHDF5 ['HostTable', 'TransientTable']>
\end{verbatim}
The host galaxy table is further broken down by transient class: 
\begin{verbatim}
>> scotch_hosts = scotch['HostTable']
>> print(scotch_hosts.keys())

<KeysViewHDF5 ['AGN', 'KN', 'SLSN-I',  'SNII', 
'SNIIb', 'SNIa', 'SNIb', 'SNIc', 'TDE']>

\end{verbatim}
The parameters linked to each host galaxy are given by retrieving the hosts of a given class: 
\begin{verbatim}
>>    print(scotch_hosts['AGN'].keys())

<KeysViewHDF5 ['GID', 'T', 'TID', 'a0', 'a1', 
'a_rot', 'b0', 'b1', 'dc2ID', 'ellipticity', 
'logMstar', 'logSFR','mag_Y', 'mag_g', 'mag_i', 
'mag_r', 'mag_u', 'mag_z', 'magerr_Y', 'magerr_g', 
'magerr_i', 'magerr_r', 'magerr_u', 'magerr_z', 
'n0', 'n1', 'w0', 'w1', 'z']>
\end{verbatim}
Values for each parameter can then be retrieved for each event: 
\begin{verbatim}
>> print(scotch_hosts['AGN']['logSFR'][0])

-0.4994781
\end{verbatim}

\subsection{Recovering Transient Light Curves}

We now consider the table of transient properties in SCOTCH, and focus on SNe~II in the catalogue.
\begin{verbatim}
>> scotch_transients = scotch['TransientTable']
>> SNII = scotch_transients["SNII"]
\end{verbatim}

Each transient table is further broken down by transient model.
\begin{verbatim}
>>    print(SNII.keys())

<KeysViewHDF5 ['SNII+HostXT_V19', 'SNII-NMF', 
'SNII-Templates',  'SNIIn+HostXT_V19', 
'SNIIn-MOSFIT']>    
\end{verbatim}
The details for these models are provided in \textsection~4.2 of the main text as well as \cite{Plasticcv12019}.  

\begin{figure}
    \centering
    \includegraphics[width=0.8\linewidth]{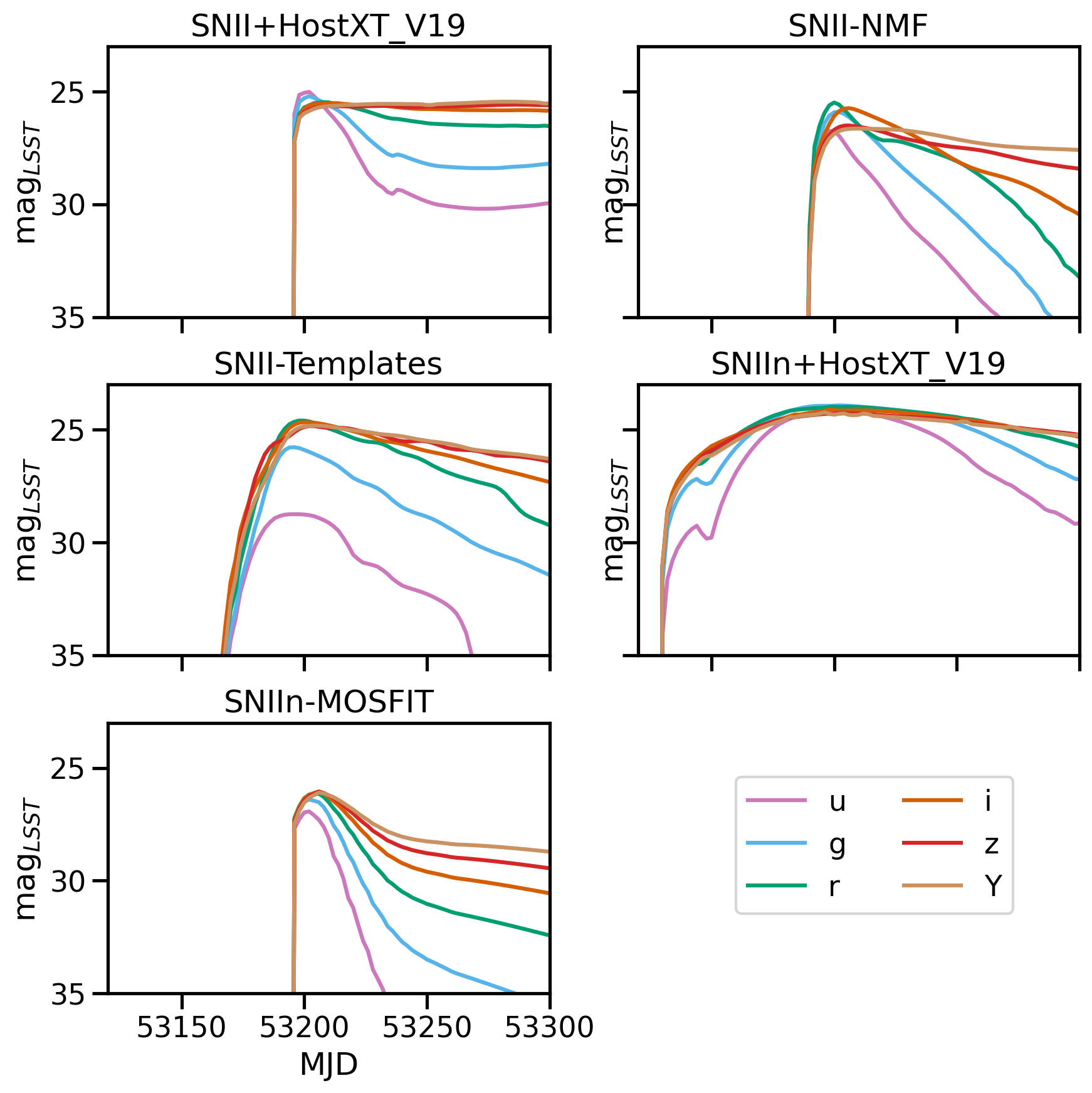}
    \caption{Sample light curves from the SCOTCH catalogue for each SN~II model used.}
    \label{fig:SNeII_tutorial}
\end{figure}

Below, we retrieve these models and plot the light curves of a single event for each model:

\begin{verbatim}
>>    SNII_models = list(SNII.keys())
>>
>>    bands = 'ugrizY'
>>    cols_sns = sns.color_palette("colorblind", 10)
>>    cols = [cols_sns[4], cols_sns[9], cols_sns[2],
>>          cols_sns[1], 'tab:red', cols_sns[5]]

>>    fig, axs = plt.subplots(nrows=3, 
>>          ncols=2, figsize=(10, 10), 
>>          sharex=True, sharey=True)
>>    axs[2, 1].set_axis_off()
>>    plt.subplots_adjust(hspace=0.25, wspace=0.2)
>>
>>    for i in np.arange(len(SNII_models)):
>>        model = SNII_models[i]
>>        SNII_oneModel = SNII[model]
>>       ax = axs.ravel()[i]
>>       for j in np.arange(len(bands)):
>>           ax.plot(SNII_oneModel['MJD'][0], 
>>                  SNII_oneModel['mag_%s'
                    % bands[j]][0], 
>>                  c=cols[j], label=bands[j])
>>        if i==0:
>>            ax.legend(bbox_to_anchor=(2.05, -1.7), 
>>                  borderaxespad=0, ncol=2)
>>        ax.set_ylim((35, 23))
>>        ax.set_xlim((53120, 53300))
>>        ax.set_ylabel("mag$_{LSST}$")
>>        ax.set_title(model)
>>    axs.ravel()[4].set_xlabel("MJD");
\end{verbatim}
This code produces Fig.~\ref{fig:SNeII_tutorial}.

\subsection{Survey-Specific Analyses}\label{sec:ssa_appendix}
\textcolor{black}{
SCOTCH is intended as a truth catalog, but for many purposes the goal is to simulate observations taken by real surveys with specific footprints and limited efficiency. In the associated tutorials for this work, we provide a basic example of how to emulate observations matching the depths and sky footprint of the LSST Deep Drilling Fields (DDFs), using PLAsTiCC data\footnote{\url{https://zenodo.org/record/2539456\#.YmHSu5LMKNF}} as a comparison.}

\begin{figure}
    \centering
    \includegraphics[width=\columnwidth]{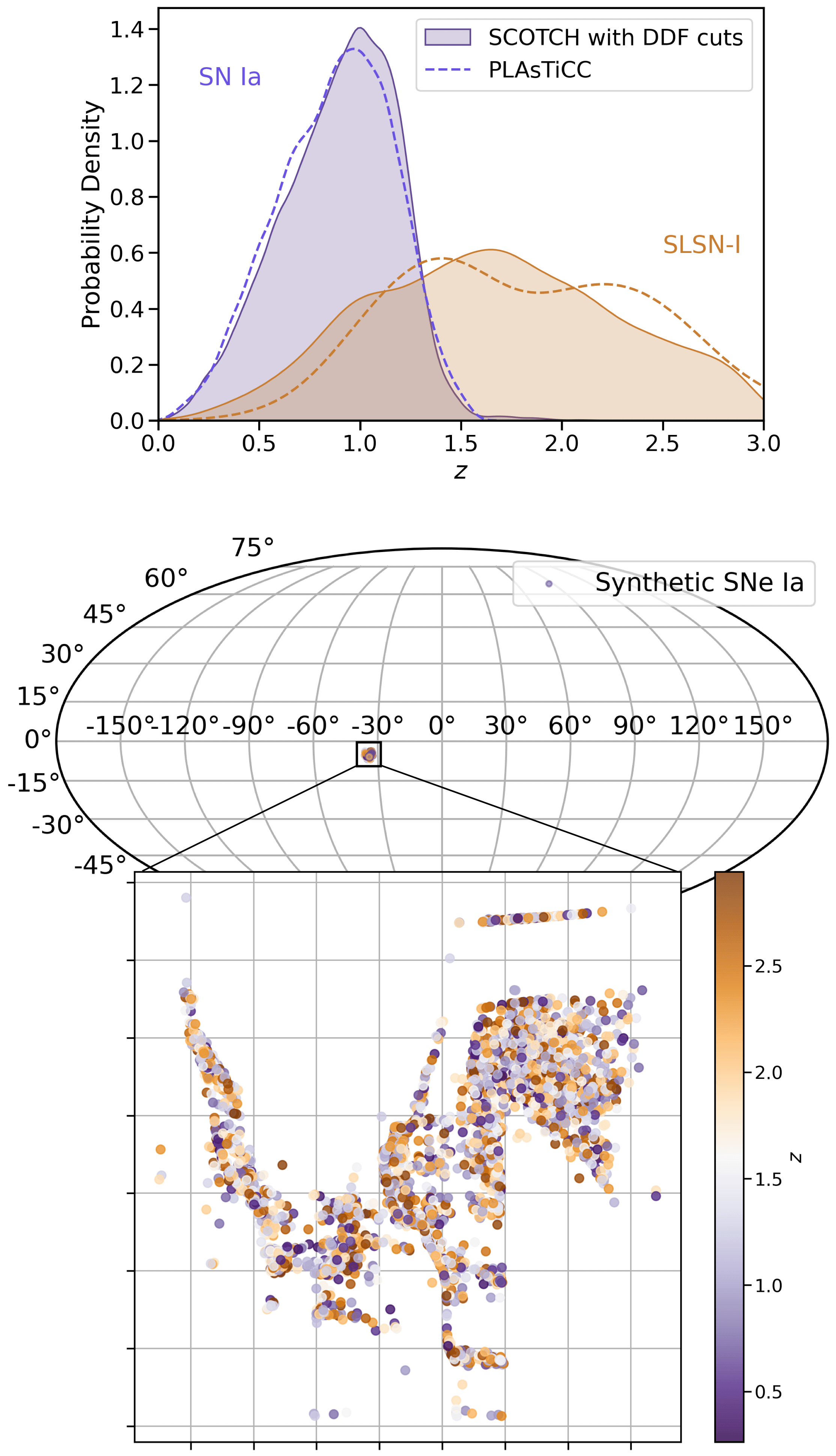}
    \caption{\textbf{Top:} Redshift distributions for SNe simulated within the LSST DDFs. PLAsTiCC data (dashed lines) are compared to SCOTCH SNe after a cut to match the expected 5$\sigma$ depth of the DDFs. \textbf{Bottom:} SCOTCH SNe~Ia placed within an LSST DDF field on the sky and colored by redshift. \textcolor{black}{SN Locations were generated by oversampling observations from within a simulated library of DDF observations with PZFlow \citep{PZFlow_JFC}. The structure observed for SN locations is not physical, but reflects  the limited number of observations used for oversampling. Additional details can be found at the repository for this work.}}
    \label{fig:DDF_tutorial}
\end{figure}

\textcolor{black}{We plot the results of our magnitude limit cuts in Fig.~\ref{fig:DDF_tutorial}, where we have also placed SCOTCH SNe realistically on the sky. The redshift distributions for SNe~Ia and SLSNe-I now show reasonable agreement to those from the test set of the PLAsTiCC challenge. The remaining discrepancy reflects Milky Way extinction, which is considered in PLAsTiCC and not in SCOTCH; and the difference in how observational limits are incorporated into the SNANA simulated framework (as an efficiency versus SNR curve, and not as an explicit SNR cut as we have imposed here).}

\begin{figure}
    \centering
    \includegraphics[width=\columnwidth]{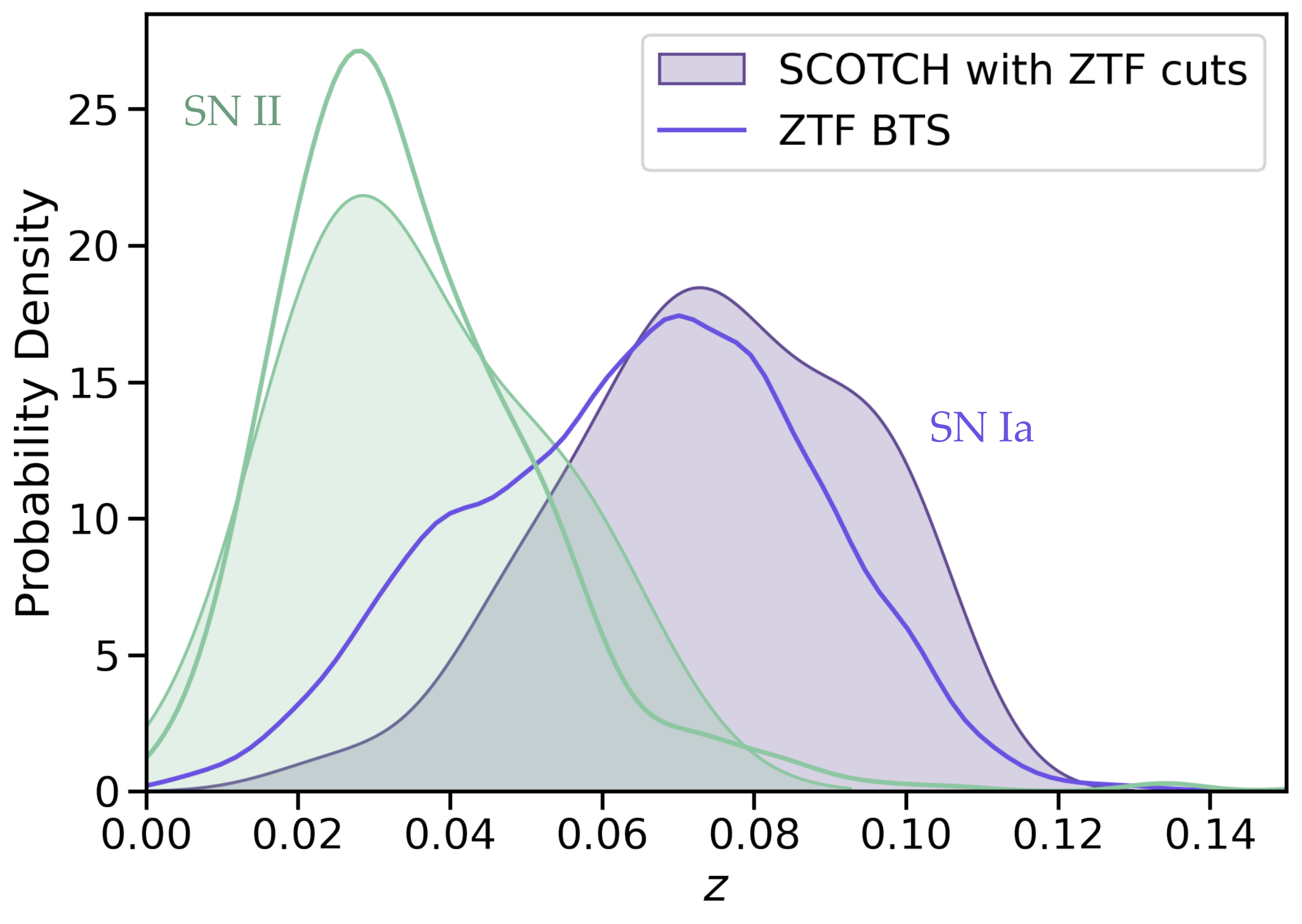}
    \caption{\textcolor{black}{Distribution for \texttt{SCOTCH} SNe post-processed to match the detection criteria of the ZTF BTS. Observed samples are shown as solid lines.}}
    \label{fig:ZTFBTS_match}
\end{figure}

\textcolor{black}{Because \texttt{SCOTCH} does not simulate Milky Way extinction or instrument systematics, some additional work is required to more accurately reflect the observed distribution of transients from a survey. We can roughly estimate these observational effects for an observational catalogue such at the Zwicky Transient Facility Bright Transient Sample \citep[ZTF BTS;][]{2020Fremling_ZTFBTS}. We download the ZTF BTS, model the distribution of Galactic $A_V$ values across the sample, and draw new values for \texttt{SCOTCH} transients. After mapping the given trigger criterion to LSST-$gr$ bands, we arrive at the distribution of SNe~Ia and SNe~II in Fig.~\ref{fig:ZTFBTS_match}. We find reasonable agreement, despite the facts that the photometric systems used are different and that the the precise observing strategy of the survey is unknown.}

The number of objects per class in SCOTCH do not represent realistic volumetric rates. However, \texttt{SNANA} provides calculations of the expected number of events per solid angle per survey time given the input volumetric rate models. These are shared in Table~\ref{tbl:rates_for_scaling} to enable re-sampling the simulation to create realistic volumetric populations.

\begin{table}
    \centering
    \begin{tabular}{c|c|c}
    \hline
    Class & Model name & $n$ [yr$^{-1}$]\tabularnewline
    \hline 
    \hline
    \multirow{3}{*}{Ia } & SALT2-Ia & 21,773,710 \tabularnewline
     & Iax & 15,933,345\tabularnewline
     & 91bg-like & 5,035,905 \tabularnewline
    \hline
    \hline
    \multicolumn{3}{c}{Hydrogen-Rich Core-Collapse (2M total)}\tabularnewline
    \hline
    \multirow{3}{*}{SN II} & SNII-Templates & 15,652,660 \tabularnewline
    & SNII-NMF & 15,652,660 \tabularnewline
    & SNII+HostXT\_V19 & 31,402,045 \tabularnewline
    \hline
    \multirow{2}{*}{SN IIn } & SNIIn-\texttt{MOSFiT} & 2,270,665 \tabularnewline
    & SNIIn+HostXT\_V19 & 2,270,665 \tabularnewline
    
    \hline
    \hline
    
    \multicolumn{3}{c}{Stripped Envelope / H-Poor Core-Collapse}\tabularnewline 
    \hline
    
    \multirow{2}{*}{Ib} & SNIb-Templates & 5,217,675 \tabularnewline
     & SNIb+HostXT\_V19 & 5,217,675 \tabularnewline
    \hline 
    \multirow{2}{*}{Ic} & SNIc-Templates & 3,623,355 \tabularnewline
     & SNIc+HostXT\_V19 & 3,623,355  \tabularnewline
    \hline 
    IcBL & SNIcBL+HostXT\_V19 & 1,062,880 \tabularnewline
    \hline 
    IIb & SNIIb+HostXT\_V19 & 10,531,710 \tabularnewline
    \hline 
    \hline 
    SLSN-I & SLSN-I-\texttt{MOSFiT} & 53,290 \tabularnewline
    \hline 
    \hline 
    \multirow{2}{*}{KN} & Kasen 2017 & 122,640 \tabularnewline
     & Bulla 2019 & 122,640\tabularnewline
     \hline
     \hline
    TDE & TDE\_\texttt{MOSFiT} & 38,690\tabularnewline
    \hline
    \hline
    \end{tabular}
\caption{Number of transients per year, $n$, predicted by each model for the entire sky and for $z<3$. The relative amounts of core collapse SNe follow the relative rates measured by \protect\cite{2017ShivversLickObservatory}. }
\label{tbl:rates_for_scaling}
\end{table}

\textcolor{black}{We provide additional tutorials for comparing the properties of transient light curves and their host galaxies in the associated repository.}


{scotch2022}{[dataset]}


\bsp	
\label{lastpage}
\end{document}